\documentclass[a4paper,11p]{article}

\pdfoutput=1

\usepackage[T1]{fontenc} 
\usepackage{graphicx}
\usepackage{epsfig}
\usepackage{amsmath, amsthm, amsfonts, amssymb, latexsym}
\usepackage{euscript,array,mathrsfs}
\usepackage{bbold}
\usepackage{epsf}
\usepackage{slashed}
\usepackage{comment}
\usepackage{colortbl}
\usepackage{cite}

\usepackage{xcolor}
\usepackage[a4paper,left=2.7cm,right=2.7cm,top=3cm,bottom=3.5cm]{geometry}
\usepackage[utf8]{inputenc}
\usepackage{longtable}
\usepackage{bm}
\usepackage{breqn}

\usepackage[caption=false]{subfig}
\usepackage[colorlinks=true,linktocpage=true,linkcolor=blue,citecolor=blue]{hyperref}

\usepackage{tikz}
\usetikzlibrary{decorations.markings,decorations.pathmorphing}

\usepackage{float}
\usepackage{subfig}

\usepackage{authblk}

\usepackage[font=small,labelfont=bf]{caption}
\graphicspath{ {./images/} }

\numberwithin{equation}{section}

\title{\LARGE \bf Field redefinitions and evolutions in relativistic 
Navier-Stokes}

\author[1]{Yago~Bea}
\author[2]{Pau~Figueras}

\affil[1]{Departament de F\'\i sica Qu\`antica i Astrof\'\i sica and Institut de
  Ci\`encies del Cosmos (ICC), Universitat de Barcelona, Mart\'\i\ i Franqu\`es
  1, ES-08028, Barcelona, Spain}

\affil[2]{School of Mathematical Sciences, Queen Mary University of London, Mile
	End Road, London E1 4NS, United Kingdom}

\date{\today}

\begin{document}

\maketitle

\thispagestyle{empty}

\begin{abstract}

In recent years the equations of relativistic first-order viscous hydrodynamics, that is, the relativistic version of Navier-Stokes, have been shown to be well posed and causal 
under appropriate field redefinitions, also known as hydrodynamic frames. 
We perform real-time evolutions of these equations for a conformal fluid and explore, quantitatively, the consequences of using different causal frames  for different sets of initial data. 
By defining specific criteria, we make precise and provide evidence for the statement that the arbitrarily chosen frame does not affect the physics up to first order, as long as the system is in the effective field theory regime. 
Motivated by the physics of the quark-gluon plasma created in heavy-ion collisions we also explore systems 
which are marginally in the effective field theory regime, finding that even under these circumstances the first order physics is robust under field redefinitions. 

\end{abstract}

\newpage

\setcounter{page}{1}
\tableofcontents

\section{Introduction}


Relativistic hydrodynamics is an effective field theory that provides the real-time description of microscopic theories when 
the system is locally in thermal equilibrium and gradient corrections are small. 
If the characteristic macroscopic scale of the system is much larger than the microscopic scale of the underlying theory, then ideal hydrodynamics should  provide a good description. However, if these two scales are not well separated, then viscous effects are expected to become relevant.

Relativistic viscous hydrodynamics is well known to 
play a fundamental role 
in the experimental description of the quark gluon-plasma (QGP) created in heavy-ion collision experiments at RHIC and LHC \cite{Romatschke:2017ejr,Busza:2018rrf}. In this case the size of the QGP droplet is comparable to the microscopic scale of quantum chromodynamics (QCD). The data from the Beam Energy Scan experiment at RHIC \cite{Aparin:2023fml} is currently being analysed, and more experiments are planned for the near future at FAIR and NICA.
Relativistic viscous hydrodynamics also plays a relevant role in
astrophysical systems. 
For instance, in neutron star mergers,
weak processes during the post merger dynamics can give rise to an effective bulk viscosity relevant for the scales of the system \cite{Alford:2017rxf}. 
Moreover, magnetorotational
instabilities may act as the onset of turbulence, whose dynamics can be effectively captured by introducing an effective viscosity \cite{LandauBook}.
Evaluating the effect of viscosity in numerical evolutions of neutron star mergers is an active area of research \cite{Chabanov:2021dee,Shibata:2017jyf,Fujibayashi:2017puw,Fujibayashi:2020jfr,Chabanov:2023blf,Chabanov:2023abq,Ripley:2023lsq}. 
 
Both the QGP and neutron star mergers provide two experimental windows into the hot and dense properties of the strongly interacting matter. They may contribute to provide access to uncharted regions of the QCD phase diagram, 
and relativistic viscous hydrodynamics is a key element to obtain the relevant real-time dynamical evolutions. Moreover, the transport properties might be useful to distinguish between different phases of matter. 
\newline

The equations of relativistic first-order viscous hydrodynamics have their origins in the works of Eckart \cite{Eckart:1940te} and Landau and Lifshitz \cite{LandauBook}. 
From a more modern perspective in the language of effective field theory, the two different sets of equations proposed by Eckart and Landau and Lifshitz can 
be obtained by using different field redefinitions of the fundamental variables (temperature, velocity...) 
in a gradient expansion of the constitutive relations of the stress tensor up to first order in derivatives.\footnote{In the Landau frame the local fluid velocity is chosen such that there is no energy flow in the local rest frame of the fluid, and in the Eckart frame the local fluid velocity is chosen in such a way that there is no charge flow in the local rest frame of the fluid. The fundamental variables are not uniquely defined away from thermodynamic equilibrium and in this case the field redefinitions of the local fluid velocity correspond to different extensions of these quantities away from equilibrium with different physical meanings.}
Different choices of field redefinitions are referred in the literature as `hydrodynamic frames'. 
Since then, working in a specific hydrodynamic frame such as Eckart's frame or Landau's frame became the usual approach.
Written in this form, the equations of first-order viscous hydrodynamics were observed to present causality issues and instabilities \cite{Hiscock:1985zz}. 
That is, the relativistic generalization of the Navier-Stokes equations, a theory that is meant to provide the effective description of any relativistic viscous fluid, would seem to be unphysical. 
This puzzling  theoretical question remained unsolved for many years.

Meanwhile the experimental analysis of the QGP created in heavy-ion collisions required of some effective viscous relativistic hydrodynamical description. 
%
An approach that provides such a description was introduced by M\"uller, Israel and Stewart (MIS) \cite{Muller:1967zza,Israel:1976tn,Israel:1979wp}, who proposed a set of equations for which the causality and stability problems are absent. 
The strategy consists in including second order terms in the gradient expansion and introduce extra variables (and their corresponding evolution equations), in such a way that the global set of equations has good causality and stability properties.
There are several formulations based on the same idea that we denote by MIS-type theories, for example: BRSSS \cite{Baier:2007ix}, DNMR \cite{Denicol:2012cn}, divergence-type theories \cite{Geroch:1990bw,Lehner:2017yes,Montes:2023dex}, etc.. 
This approach has been very successful describing the experimental data in heavy-ion collision experiments. In fact, Geroch famously showed that all these different theories of relativistic dissipative hydrodynamics provide the same physical predictions and should have the same physical content as the Navier-Stokes equations as long as they are in the regime of validity of effective field theory \cite{Geroch:1995bx}.
In spite  of this success, the question of why the relativistic generalization of Navier-Stokes appeared to be unphysical remained unsolved.
A solution was proposed in recent years.

Bemfica, Disconzi, Noronha
\cite{Bemfica:2017wps,Bemfica:2019knx,Bemfica:2020zjp}
and Kovtun \cite{Kovtun:2019hdm}  proposed a formulation of the relativistic Navier-Stokes equations 
which is well-posed, causal and stable. 
Such a formulation has become known as the  `BDNK equations'.
The key insight was to realize that by performing field redefinitions up to first order in the hydrodynamic gradient expansion, one can change the highest derivative terms in the equations of motion (i.e., the principal part) and achieve 
local well-posedness of the initial value problem (Cauchy problem), relativistic causality and dynamical stability of the homogeneous thermal states.
Thus, the solution to obtain a mathematically and physically well behaved relativistic version of the  Navier-Stokes equations is to perform an appropriate field redefinition, or choice of hydrodynamic frame.
Traditionally used frames such as the Landau or Eckart's frame do not belong to this set of allowed frames.
\newline

Having at our disposal this novel theory, we  may wonder: why do we need another formulation of relativistic viscous hydrodynamics if the MIS-type theories were successful describing the experimental data? We provide arguments why first-order viscous hydrodynamics might be a good alternative to MIS-type theories.

An important advantage of first-order viscous hydrodynamics theories over MIS-type theories are the well posedness properties of the initial value problem. 
Local well posedness of the initial value problem has been well established and studied for many physically relevant sets of equations such as the Maxwell equations, ideal hydrodynamics or the Einstein equations of general relativity. However, there are limited results  in the context of relativistic viscous hydrodynamics. It was not until very recently that a study was performed for a set of MIS-like equations \cite{Bemfica:2020xym}. The result is that these theories present some limitations, for example,
local well posedness may depend on the specific state under consideration. Thus, the conditions must be checked pointwise in spacetime for each solution. These conditions were shown to fail in realistic simulations of the QGP created in heavy-ion collisions experiments \cite{Plumberg:2021bme}. 
In contrast, for the relativistic Navier-Stokes equations, once a causal frame has been chosen, local well posedness is satisfied independently of the specific state. Thus, this is a clear advantage of relativistic Navier-Stokes over MIS. 

Related to the previous point there is another argument regarding the principles of relativity. For the MIS equations the ability to propagate information (more precisely, the characteristic velocities of the PDEs) might be larger than the speed of light, and again this depends on the specific state and must be checked pointwise in every evolution. 
In contrast, for the relativistic Navier-Stokes equations the characteristic velocities are fixed once the frame has been chosen, and they do not depend on the specific state. Choosing a frame for which the characteristic velocities are not larger than the speed of light ensures that every evolution performed with those equations respects the principles of relativity. In MIS it cannot be ensured that the solution respects the principles of relativity, and this has to be checked for every solution.

Another important advantage of relativistic Navier-Stokes is that it is able to capture arbitrarily strong shockwave solutions, while MIS-theories may present limitations evolving large shocks; this may be an advantage in astrophysical applications. 
The ability to evolve large shocks is related to the characteristic velocities, and in MIS-type theories, these depend on the state and might be less than the speed of light. A shock with velocity larger than the characteristic velocities cannot be evolved with the MIS equations and hence their predictive power is lost in these situations. 
In relativistic Navier-Stokes, once the frame has been fixed, also the characteristic velocities are fixed. Moreover, frames can be chosen such that these characteristic velocities are equal to the speed of light. Mathematical results in this direction were obtained in \cite{Freistuhler:2021lla} and numerical studies were performed in \cite{Pandya:2021ief}, providing evidence that arbitrarily large shocks can be evolved with the relativistic Navier-Stokes equations.

An advantage of MIS-type formulations is that the numerical infrastructure is already implemented in the context of the QGP. For relativistic Navier-Stokes this infrastructure is still under construction and this paper aims to contribute towards this goal. 
\newline

The above arguments suggest that first-order viscous hydrodynamics might be a very good alternative to MIS-type theories in aspects which are relevant for an appropriate description of physical systems of interest like neutron star mergers or the QGP created in heavy-ion collisions.  However, for a practical implementation of the first-order viscous hydrodynamics equations it is relevant to address the following question. Now that we are not using traditionally preferred frames like the Landau frame, what frame should we use in actual numerical evolutions? In principle, one can use any frame within the set of causal frames. 
Actually, if the system is within the effective field theory regime, then it should not matter which specific frame is being used, and using different frames should lead to an equivalent physical description. Equivalently, if the system is within the effective field theory regime, the physics to first order should be independent of the arbitrarily chosen frame. If not, the equations would not have predictive power. Numerical studies of the first-order viscous hydrodynamics equations were performed in \cite{Bantilan:2022ech,Pandya:2021ief,Pandya:2022pif,Pandya:2022sff}.
These papers constitute a proof of principle that first-order viscous hydrodynamics produces physically sensible real-time evolutions. Moreover, numerical techniques have been studied, with extensions to non conformal and charged systems.  
\newline

In this paper we go beyond the state of the art by performing a detailed, quantitative study of the effect of field redefinitions in numerical evolutions of the relativistic  Navier-Stokes equations for a conformal theory.
For this purpose we consider different sets of initial data: a small amplitude perturbation of a homogeneous thermal state, a large amplitude Gaussian perturbation of a homogeneous state and shockwave solutions. 
We define a specific set of criteria to assess different conditions on the numerical solutions.
By using these criteria, we make precise and provide evidence for the statement that the arbitrarily chosen frame does not affect the physics to first order, as long as the system is in the effective field theory regime. 
Motivated by the physics of the QGP, we also study the effect of field redefinitions in situations where the system is marginally in the effective field theory regime. Furthermore, we also study the effect of changing the frame in the initial data. 

In this paper we use equivalently the terminology `relativistic first-order viscous hydrodynamics equations', `relativistic Navier-Stokes equations' and `BDNK equations'. From the perspective of effective field theory it is natural to use `relativistic Navier-Stokes equations' \cite{Kovtun:2019hdm,Hoult:2020eho}.
We define it as a collection of sets of partial differential equations (PDEs), where each set of PDEs is obtained by choosing a hydrodynamic frame and truncating the hydrodynamics gradient expansion of the constitutive relations up to first order in derivatives and plugging it into the conservation equation for the stress tensor. By using different frames the truncations are different, obtaining a different set of PDEs for each frame.
Each set of PDEs can be considered as a different version of relativistic Navier-Stokes, like Eckart's or Landau's versions; it is expected that they all describe equivalent first-order physics as long as the solutions are in the effective field theory regime. The main results of this paper provide a quantitative assessment of  this point of view.\footnote{See \cite{Reall:2021ebq} for a detailed and rigorous study of the validity of effective field theory in the context of the Abelian Higgs model, where similar issues arise. } The reader interested in a summary and conclusions may jump to Section \ref{sec:Discussion} Discussion.

\section{First-order hydrodynamics and effective field theory}
\label{First_order_hydrodynamics_and_effective_field_theory}

We consider a conformal fluid in 3+1 dimensions in Minkowski spacetime.  Conformal symmetry fixes the equation of state
\begin{equation}
	p = \frac{\epsilon}{3}\,,  
	\label{conformal_equation_of_state} 
\end{equation}
where $\epsilon$ is energy density and  $p$ is pressure. It also imposes that the energy is proportional to $T^4$,  $\epsilon=\mathcal{C}\,T^{4}$,  where $T$ is the temperature and  $\mathcal{C}$ is a constant that we choose $\mathcal{C}=\frac{3}{4} \pi^4$.\footnote{We choose this constant motivated by holographic fluids.} 
The conservation of the stress-energy tensor
\begin{equation}
	\nabla_{\mu}T^{\mu\nu} = 0\,, 
	\label{conservationideal} 
\end{equation}
provides the evolution equations for the relativistic fluid. Considering the 4-velocity of the fluid $u^\mu$, normalized such that $u^2=-1$,  we define $\Delta^{\mu\nu} \equiv g^{\mu\nu} + u^\mu\,u^\nu$, $\dot{\epsilon}\equiv u^{\mu}\nabla_{\mu}\epsilon$, $\nabla_{\perp}^{\mu}\equiv \Delta^{\mu\nu} \nabla_{\nu} $, $\nabla \cdot  u \equiv \nabla_{\rho} u^{\rho}$  and 
$\sigma^{\mu\nu} \equiv 2  \nabla^{\langle\mu} u^{\nu\rangle}$, where
\begin{align}
	&A^{\langle\mu\nu\rangle} \equiv \frac{1}{2} \Delta^{\mu\alpha} \Delta^{\nu\beta}(A_{\alpha \beta}+A_{\beta \alpha})-\frac{1}{3}\Delta^{\mu\nu} \Delta^{\alpha\beta}A_{\alpha \beta}  \,, 
\end{align}
is symmetric, traceless and transverse to $u^{\mu}$.

We start by presenting the equations of ideal hydrodynamics. The constitutive relations of ideal hydrodynamics are 
	\begin{equation}
	T^{\mu\nu} = \epsilon \left(u^\mu\,u^\nu + \frac{1}{3}\,\Delta^{\mu\nu}\right)\,.
	\label{constitutive0} 
\end{equation}

The conservation of the stress-energy tensor \eqref{conservationideal} with \eqref{constitutive0}
provides evolution equations for the dynamical variables, namely  $\epsilon$ and the independent components of the velocity vector $u^\mu$, 
\begin{subequations}
	\begin{align}
		\frac{3}{4}\frac{\dot{\epsilon}}{\epsilon}+\nabla \cdot u &=0 \,\,, 	\label{conservationidealexplicita} \\
		\dot{u}^{\mu}+\frac{1}{4} \frac{\nabla_{\perp}^{\mu} \epsilon}{\epsilon} &=0\,\,.	\label{conservationidealexplicitb} 
	\end{align}
	\label{conservationidealexplicit} 
\end{subequations}

 The constitutive relations of relativistic hydrodynamics up to first order in the gradient expansion in the Landau frame\footnote{The Landau frame is defined by a local fluid velocity that satisfies $ u_{\mu}T^{\mu \nu} =\epsilon \, u^{\nu}$,
where $\epsilon$ is the energy density in the local rest frame of the fluid.
}
can be written as 
	\begin{equation}
	T^{\mu\nu} = \epsilon \left(u^\mu\,u^\nu + \frac{1}{3}\,\Delta^{\mu\nu}\right) -\eta\, \sigma^{\mu\nu}+ O(\partial^2)\,,
	\label{constitutivefirstorder} 
\end{equation}

In the spirit of effective field theory we consider the most general field redefinition of the fundamental variables $\{\epsilon,u^{\mu}\}$ by first order terms compatible with Poincar\'e and conformal symmetries. There are two Lorentz scalars $\dot{\epsilon}$, $\nabla \cdot u$ and two Lorentz vectors $\dot{u^{\mu}}$, $\nabla_{\perp}^{\mu} \epsilon$ that we can construct with the first derivatives of the dynamical fields. However,
conformal symmetry uniquely determines the most general field redefinition \cite{Kovtun:2019hdm}
\begin{subequations}
	\begin{align}
		\epsilon &\rightarrow \epsilon+ a_1\,\eta  \left(\frac{3}{4}\frac{\dot{\epsilon}}{\epsilon}+\nabla \cdot u \right) + O(\partial^2)\,\,, \label{changetoLandau0a}\\
		u^\mu  &\rightarrow u^\mu +\frac{3\, a_2\,\eta }{4\epsilon}\left(\dot{u}^{\mu}+\frac{1}{4} \frac{\nabla_{\perp}^{\mu} \epsilon}{\epsilon}\right) + O(\partial^2)\,\,.
	\end{align}
	\label{changeframeconformal}
\end{subequations}
In these equations, the overall factor multiplying the terms in brackets has dimensions fixed by conformal symmetry, and we choose it to be proportional to the shear viscosity $\eta$ times a constant, which we denote by $a_1$ or $a_2$ respectively.
By performing a general field redefinition 	\eqref{changeframeconformal} of the constitutive relations in the Landau frame \eqref{constitutivefirstorder} and neglecting higher order terms, we obtain the stress tensor up to first order in the gradient expansion in a general hydrodynamic frame

\begin{equation}
	\begin{aligned}
		T^{\mu\nu}=& ~ \epsilon \left(u^\mu\,u^\nu + \frac{1}{3}\,\Delta^{\mu\nu}\right)-\eta\,\sigma^{\mu\nu} \\
		&+ a_1 \eta\left( \frac{3}{4}\frac{\dot{\epsilon}}{\epsilon}+\nabla \cdot u \right)\left(u^\mu\,u^\nu + \frac{1}{3}\,\Delta^{\mu\nu}\right)\\  
		&+a_2 \eta \left[\left(\dot{u}^{\mu}+\frac{1}{4} \frac{\nabla_{\perp}^{\mu} \epsilon}{\epsilon} \right)u^\nu + (\mu\leftrightarrow \nu)\right]+ O(\partial^2)\,.
	\end{aligned}
	\label{eq:tmunu11}
\end{equation}

The two real numbers $\{a_1,a_2\}$ in \eqref{eq:tmunu11} specify the hydrodynamic frame. The Landau frame constitutive relations \eqref{constitutivefirstorder} correspond to $\{a_1,a_2\}=\{0,0\}$. 
The stress tensor \eqref{eq:tmunu11} can be equivalently obtained by considering the most general stress tensor up to first order in the gradient expansion compatible with Poincar\'e and conformal symmetries \cite{Bemfica:2020zjp,Kovtun:2019hdm}.
The first order terms in \eqref{eq:tmunu11} are the shear term $-\eta\,\sigma^{\mu\nu}$, which is field redefinition independent, and the terms in second and third lines of \eqref{eq:tmunu11}, which we refer to as the  `$a_1$ term' and `$a_2$ term' respectively and depend on the choice of frame. 
We use the notation $T_{\mu\nu}^{(0)}$ to refer to the ideal part and $T_{\mu\nu}^{(1)}$ to the first order part in \eqref{eq:tmunu11}.
The equations of first-order viscous hydrodynamics are obtained by plugging \eqref{eq:tmunu11} into the conservation equation for the stress tensor \eqref{conservationideal}, and are of second order in derivatives.
\newline

Inspired by the works of Van and Biro \cite{Van:2011yn} and Freistuhler and Temple \cite{Freistuhler1,Freistuhler2,Freistuhler3}, Bemfica, Disconzi and Noronha \cite{Bemfica:2017wps,Bemfica:2019knx,Bemfica:2020zjp} and Kovtun \cite{Kovtun:2019hdm}, showed that in some frames the equations of first-order hydrodynamics present good causality and stability properties. The key observation is that different hydrodynamic frames give rise to different evolution equations 
at the level of two derivatives. 
More precisely,  the principal part of the equations, that is, the highest derivative terms,  depends on the constants $a_1$, $a_2$ and hence on the choice of frame.
Local well posedness and the causality properties of the PDEs are determined by the principal part of the equations and these references found that in some frames the equations have these good properties. More specifically, for the conformal fluid it was proven \cite{Bemfica:2017wps,Disconzi:2017qqm,Bemfica:2019hok} that if  
\begin{equation}
	a_1 \geq 4   \,, ~~~~   a_2\geq \frac{3a_1}{a_1-1}  \, , 
	\label{hyperbolicity_conditions_BDNK}
\end{equation}
then the PDEs \eqref{conservationideal} with \eqref{eq:tmunu11} are strongly
 hyperbolic and the initial value problem is locally well posed.\footnote{By `well posed' we mean that the solution locally exists, is unique and it depends continuously on the initial data, which should be in a suitable Sobolev space.}
 Also, the characteristic velocities can be chosen to be no larger than the speed of light, thus respecting the principles of relativity.

These results provide an answer to the longstanding question of why first-order viscous hydrodynamics had apparent ill causality properties: it is just a matter of choice of hydrodynamic frame and it can be solved by an appropriate field redefinition.
The Landau frame $\{a_1,a_2\}=\{0,0\}$ lies outside the region \eqref{hyperbolicity_conditions_BDNK}, and there is a gap between the causal frames and the Landau frame. 

Another important aspect of the first-order viscous hydrodynamics equations is that in the Landau frame  the equilibrium states are unstable \cite{Hiscock:1985zz}. Thus, apart from the local well posedness and causality properties, it is also important to discuss the stability of equilibrium states under the conditions \eqref{hyperbolicity_conditions_BDNK}, and this was shown in \cite{Bemfica:2017wps,Kovtun:2019hdm}. 
A relevant result in this direction was recently proven in \cite{Bemfica:2020zjp}, who showed that under the conditions \eqref{hyperbolicity_conditions_BDNK}, if a thermal equilibrium state is stable in one Lorentz frame, then it is stable in all Lorentz frames.
\newline

A natural question that arises if one is familiar with working in the Landau frame  is  the physical meaning of the `extra terms' $a_1$ and $a_2$ in \eqref{eq:tmunu11}. We now provide intuition about the significance of these terms.

We start by emphasizing that the field redefinition dependent terms in \eqref{eq:tmunu11}, that is, the $a_1$ and $a_2$ terms, are proportional to the lower order equations of motion, namely the ideal hydrodynamics equations \eqref{conservationidealexplicit}. Thus, upon using the equations of motion, these terms are 
equivalent to second  order terms \cite{Bemfica:2020zjp}. Actually, this is a general principle in 
effective field theory: when a term at a given order in the derivative expansion is proportional to the lower order equations of motion, it can be pushed to a higher order by a field redefinition. 
Therefore, on shell, the $a_1$, $a_2$ terms are effectively of second order and hence one should expect that their contribution to the first order physics  is negligible, as long as the system is in the effective field theory regime. In other words, in the effective field theory regime they are smaller than the shear viscosity term, which is of first order and field redefinition independent. Thus, changing frame should leave the physics to first order invariant, as it is a second order effect.\footnote{In charged fluids the physical interpretation of the local fluid velocity is different in the Landau and Eckart frames, and this is because the change of frame is first order but not proportional to the ideal equations of motion. In the conformal and uncharged fluids studied in this paper, all field redefinitions are second order on shell and so the physical interpretation of the fundamental variables does not change under field redefinitions.}
This is a formal statement in effective field theory, where gradients are assumed to be infinitesimally small. However, in realistic situations the gradients are finite,
and hence it would be desirable to have some specific quantitative criteria to assess whether the higher order gradients are indeed negligible compared to the first order ones.
In the following we suggest some criteria that capture different aspects relevant for this discussion; more precise definitions of these criteria will be given in Section \ref{sec:sinusoidal}.

\begin{itemize}
\item {\bf Criterion A}~ The `size' of each term evaluated on a solution of the viscous equations at given order in the hydrodynamic gradient expansion is smaller than any of the lower order ones. 
\item {\bf Criterion B}~ The `size' of the  $a_1$ and $a_2$ terms in  \eqref{eq:tmunu11} evaluated on a solution of the viscous equations are much smaller than the shear term. 
\item {\bf Criterion C}~ The difference between two solutions of the viscous equations obtained in two different frames is much smaller than the difference between the solution of the ideal equations and any of the solutions of the viscous equations. 
\end{itemize}

In practice, to check Criterion A we will evaluate the ideal and first order terms in \eqref{eq:tmunu11} on our numerical solutions. Criteria B and C are two different and complementary criteria to make precise when the physics to first order in the gradient expansion is independent of the choice of hydrodynamic frame. For Criterion C to be meaningful,  we need to require that the different causal frames are sufficiently distinct from each other, otherwise, by continuity and Cauchy stability,  two different frames corresponding to values of $a_1$ and $a_2$ that are sufficiently close will give solutions that are trivially close to each other even if they are not in the effective field theory regime. For this reason, in this paper we will compare situations in which the frames $\{a_1,a_2\}$ differ by a factor of 2 at least. Typically used values will be $\{a_1,a_2\}=\{5,5\}$ and $\{a_1,a_2\}=\{10,10\}$. 
\newline 

We will consider different sets of initial data to obtain solutions to the viscous equations  that we will use to analyse in detail the different criteria above. In particular, 
 we study situations in which the system is  marginally in the effective field theory regime.
This is motivated by the physics of the QGP created in heavy-ion collision experiments, 
where at the initial stages of the hydrodynamic evolution gradients might not be small and hence it is not clear that the system is within the effective field theory regime.  
We will explore the behaviour under change of frames by employing criteria B and C in this limiting situation in which Criterion A might be violated.  Also motivated by the physics of the QGP, we will address the following question: given some initial data, how do we change from the usual Landau frame to the chosen causal frame?  To do so, we will study the effect of the change of frame in the initial data.
\newline 

The equations of first-order relativistic hydrodynamics \eqref{conservationideal} with \eqref{eq:tmunu11} in a frame satisfying \eqref{hyperbolicity_conditions_BDNK} can be understood as a UV completion of hydrodynamics in the following sense. The equations are obtained after truncating the gradient expansion of the stress-energy tensor to first order in derivatives \eqref{eq:tmunu11}, and then plugging the latter into \eqref{conservationideal}.
In general, solutions to these truncated equations will not 
be in the effective field theory regime: they could be arbitrarily far from equilibrium, for example setting initial data with large gradients.\footnote{In particular, the theory  \eqref{eq:tmunu11} contains non-hydrodynamic modes that depend on the choice of frame. Similarly, MIS-like theories also contain non-hydrodynamic modes and can be thought of as different UV completions. }
The fact that there are solutions to the viscous hydrodynamics equations that are not in the effective field theory regime is a generic feature in truncated effective field theories. 
If we want that a solution of the truncated theory to correspond to the long distance description of a solution of the microscopic theory, the former must be in the effective field theory regime at all times.
Let us emphasize that this is the case not only for this particular set of equations; the same happens with the equations of ideal hydrodynamics and other theories of viscous hydrodynamics such as MIS-type equations. The fact that we construct solutions to the equations of (viscous) hydrodynamics does not necessarily mean that such solutions are consistent; for instance, there may be physical situations in which solutions explore the UV of theory, e.g., turbulence in 3+1 dimensions, and preventing a flow of energy to UV in an ad-hoc way in the truncated theory may be unphysical. 
We also emphasize that when we state that the system is in the `regime of hydrodynamics' it may sound trivial because we are solving the hydrodynamics equations, but it is not: we mean that the system is within the effective field theory regime, and this is in general not true for certain solutions of the equations of hydrodynamics.
\newline

To assess if a solution is in the regime of hydrodynamics one would typically compare ideal and first order terms evaluated on the solution, that is, Criterion A.
However, we can try to do better. One motivation is the following: there might be situations in which the system is away from the regime of hydrodynamics and still first order terms are smaller than ideal terms, so Criterion A would fail. One example is adding a perturbation to a homogeneous thermal state with a very small amplitude and large momentum: the small amplitude can make the first order terms small even if the momentum is large compared to the temperature. In that case the ratio of second order terms to first order terms would be large, indicating that the system is not in the regime of hydrodynamics, but the ratio of the first order terms to ideal terms is small. Another example is the Gaussian initial data that we explore in section \ref{Gaussian_section}: at $t=0$ gradients might be arbitrarily large (for sufficiently small width of the Gaussian) and yet the shear term vanishes. Then, can we improve our practical assessment of whether the system is in the regime of hdyrodynamics? We now provide one possible idea.

Nothing prevents us from measuring the size of the higher order gradients of our specific solutions; in particular, we can evaluate the second derivative terms in the gradient expansion of the stress-energy tensor on the solutions of the first order theory. Note that doing so is perfectly consistent  in our scheme since we are solving the classical equations of motion, which are of second order in derivatives; therefore, the second order terms in the gradient expansion are well-defined in our solutions.\footnote{The presence of viscosity smoothes out shock solutions, which therefore are continuous.}  For this purpose, we collect here the constitutive relations of relativistic viscous hydrodynamics up to second order in the derivative expansion in the Landau frame for a conformal fluid \cite{Baier:2007ix}
	\begin{align} 
		T^{\mu\nu} =&~ \epsilon\,u^\mu\,u^\nu + p\,\Delta^{\mu\nu} -\eta\, \sigma^{\mu\nu}\nonumber  \\ 
		 &+\eta\,\tau_{\pi} \left(\dot{\sigma}^{\langle\mu\nu\rangle}+\frac{1}{3} \,\sigma^{\mu\nu} \,\nabla \cdot  u\right)+{\lambda_1}\, {\sigma^{\langle\mu}}_{\rho}\sigma^{\nu\rangle\rho}+{\lambda_2}\, {\sigma^{\langle\mu}}_{\rho}\Omega^{\nu\rangle\rho}+\lambda_3\, {\Omega^{\langle\mu}}_{\rho}\Omega^{\nu\rangle\rho} \,,
    	\label{constitutive0sheartensor0} 
	\end{align}
 where $\eta$ is the shear viscosity, $\tau_{\pi}$, $\lambda_1$, $\lambda_2$, $\lambda_3$ are second order transport coefficients and $\Omega_{\mu\nu}=\Delta_\mu^{\phantom{\mu}\alpha}\Delta_\nu^{\phantom{\nu}\beta}\partial_{[\alpha}u_{\beta]}$ is the vorticity tensor.
 We will use the second line of  \eqref{constitutive0sheartensor0} to measure the size of the second order gradients in our solutions. Note that \eqref{constitutive0sheartensor0} is in the Landau frame, and our solutions will be in a different (causal) frame. 
We could change the hydrodynamic frame of our solutions to Landau frame and evaluate \eqref{constitutive0sheartensor0} in the Landau frame. However, the contribution of the terms corresponding the the change of frame to the second line in  \eqref{constitutive0sheartensor0} is of higher order, and hence negligible if the solution is in the hydrodynamic regime. 
 Having a measurement of the size of the second order terms allows for a more detailed assessment of when the solution is in the regime of hydrodynamics since we can compare these second order terms with the ideal and first order terms. Hence, we will state that the system is in the effective field theory regime when there is a clear 
 hierarchy between the `sizes' of the ideal, first order and second order terms respectively.
 In practice, when considering Criterion A, we will include the second order terms even though we do not use them to obtain the solutions.

\section{Real-time evolutions}
\label{Section3}

In this section we perform a detailed study of real-time numerical evolutions
of the relativistic first-order viscous hydrodynamics equations of a conformal fluid in 3+1 dimensions. These are obtained by plugging \eqref{eq:tmunu11} into the conservation equation \eqref{conservationideal}.
We perform evolutions in Minkowski spacetime and we restrict the dynamics in 1+1 dimensions, assuming homogeneity along the other two spatial dimensions. We use Cartesian coordinates $(t,x,y,z)$ and we choose the dynamics to take place along the $x$-direction. Our evolution variables are $\{\epsilon, u_x\}$, where $u_x$ is the $x$ component of the 4-velocity $u^{\mu}$. The numerical code that we use was presented in \cite{Bantilan:2022ech}, see Appendix A of that paper for details.\footnote{Even if our code evolves the equations in a 2+1 dynamical set up, in this paper we present evolutions with dynamics in 1+1 dimensions for presentation purposes: it is enough for the physics that we want to explore and easier to present. We do not expect that the qualitative conclusions obtained in this paper depend on the dimensionality of the dynamics of the problem.} In Appendix \ref{sec:AppendixA} we include some convergence tests for the runs presented in this paper.

In our evolutions we use hydrodynamic frames $\{a_1,a_2\}$ that satisfy the causality conditions \eqref{hyperbolicity_conditions_BDNK}. Typical values used in this paper will be $\{a_1,a_2\}=\{5,5\}$ or $\{a_1,a_2\}=\{10,10\}$, but we will also use other frames that will be specified in due course. We use the conformal equation of state \eqref{conformal_equation_of_state}
and the shear viscosity
\begin{equation}
	\eta = \frac{s}{4\pi} ~, 
	\label{shear_viscosity}
\end{equation}
where $s$ is the entropy density. We use the value of shear viscosity \eqref{shear_viscosity} of large-$N$ and large (infinite) coupling holographic theories, which is a universal result within holographic theories with an Einstein gravity dual \cite{Policastro:2001yc,Kovtun:2003wp}. 
Another motivation for choosing \eqref{shear_viscosity} is that this value is very close to the shear viscosity measured in the QGP created in heavy-ion collision experiments \cite{Schafer:2009dj}.

Even if in this paper we only perform evolutions of the first-order viscous hydrodynamics equations, in order to assess if our solutions are in the effective field theory regime we will evaluate pointwise in spacetime the second line of \eqref{constitutive0sheartensor0}, as explained above. 
For this purpose, we use the following second order transport coefficients \cite{Bhattacharyya:2007vjd,Baier:2007ix}
\begin{subequations}
	\begin{align}
		\eta \,\tau_{\pi} &= \frac{s}{8 \pi^2 T}\left(2-\ln 2 \right)  ~, \label{tau_pi}\\
		\lambda_1 &= \frac{s}{8 \pi^2 T}  ~, \label{lambda_1}\\
		\lambda_2 &= \frac{s}{8 \pi^2 T} \left(-2\ln 2 \right)   ~, \label{lambda_2}\\
		\lambda_3 &= 0   ~.\label{lambda_3}
	\end{align}
\label{N=4SYMcoeffs}
\end{subequations}
These are the coefficients of a set of 3+1 dimensional holographic conformal field theories, such as $\mathcal{N}=4$ Super Yang Mills, which admit a dual gravitational  description in terms the Einstein-Hilbert action plus a negative cosmological constant.
In planar 1+1 dynamics only $\tau_{\pi}$ and $\lambda_1$ are relevant, as the tensors multiplying $\lambda_2$, $\lambda_3$ vanish.
We use these coefficients because in these holographic theories these coefficients are known; in other frameworks, obtaining these coefficients from a microscopic quantum field theory may be challenging.
\newline

We will start by performing evolutions of a small perturbation of a homogeneous thermal state. We consider small amplitude sinusoidal perturbations with a large wavelength, and thus by construction this system is close to the linear hydrodynamic regime. We evolve the equations at the full non-linear level, but being close to the linear regime gives us good control of the system. This is particularly useful when studying  the effect of field redefinitions.  
More specifically, by continuously varying the momentum we can have a system that is well within the regime of hydrodynamics (low momentum), or far from equilibrium (large momentum), thereby allowing us to assess the effect of field redefinitions in situations where the system may be on the verge of exiting the regime of hydrodynamics.

We will continue our studies by considering initial data corresponding to a thermal state deformed  by a Gaussian with a large amplitude  to explore the non-linear regime of the theory. By tuning the width of the Gaussian profile, we can smoothly interpolate between  the regime of hydrodynamics (large widths) and the far from equilibrium regime (very small widths).

Finally, we continue exploring the non-linear regime of the theory in more extreme situations by studying shockwave solutions. By changing the amplitude of the shockwave we can take the system to be well in the regime of hydrodynamics (small shocks) or far from equilibrium (large shocks). 

\subsection{Small perturbation of a homogeneous thermal state}
\label{sec:sinusoidal}

We start by considering a physical system in which we can study the properties of the first-order viscous hydrodynamics equations under well controlled conditions:
a small amplitude sinusoidal perturbation of a homogeneous thermal state. By choosing large wavelength perturbations, the system is by construction in the regime of (linear) hydrodynamics. Choosing a small amplitude is useful because we can obtain some intuition from linear hydrodynamics; we use this intuition to analyse our solutions, that are obtained by solving the full non-linear equations.

We consider a thermal homogeneous state with temperature  $\overline{T}$ and energy $\overline{\mathcal{E}}=\frac{3}{4} \pi^4\,\overline{T}^{4} $, and a sinusoidal perturbation of amplitude $0.01 \overline{\mathcal{E}}$ and momentum $k/\overline{T} \simeq 0.184$.  
Specifically, the initial data is as follows:
\begin{subequations}
	\begin{align}
		\epsilon |_{t=0}&= \overline{\mathcal{E}} \left( 1+ 0.01 \cos(k x) \right) \, \,, \label{Initial_data_sinusoidala}\\
		\partial_t \epsilon|_{t=0}&= 0\,\,, \label{Initial_data_sinusoidalb}\\
		u_x|_{t=0}&= 0 \, \,, \label{Initial_data_sinusoidalc}\\
		\partial_t	u_x|_{t=0}&= \frac{0.01 k \sin(k x)}{4(1+ 0.01 \cos(k x)) } \, \,. \label{Initial_data_sinusoidald}
	\end{align}
	\label{Initial_data_sinusoidal}
\end{subequations} 
The time derivative of the velocity at $t=0$ is chosen so that the ideal hydrodynamics equations \eqref{conservationidealexplicit} are initially satisfied.
If a system is expected to be in the regime of hydrodynamics, the ideal equations should be nearly satisfied, up to higher order terms which are supposed to be small. Therefore, setting them to be exactly zero initially should be  a good approximation to this situation. Other options such as  setting to zero the time derivatives at $t=0$ would imply that the ideal equations would not be initially satisfied; later on we will analyse this case in detail.
Also, very importantly, note that if the initial data satisfies the ideal equations, then
the first order change of frame vanishes at $ t=0$. This means that we can study the effect of using different frames along the evolution without having to worry about the effect of changing frame in the initial data: we leave this as a another exercise to explore in Section \ref{sec:change_frame_ID} of the paper.
We choose $\overline{\mathcal{E}}=1$ and spatial domain of size $L \, \overline{T} \simeq 34.2$ with periodic boundary conditions. 

\begin{figure}[t]
	{\includegraphics[width=0.495\textwidth]{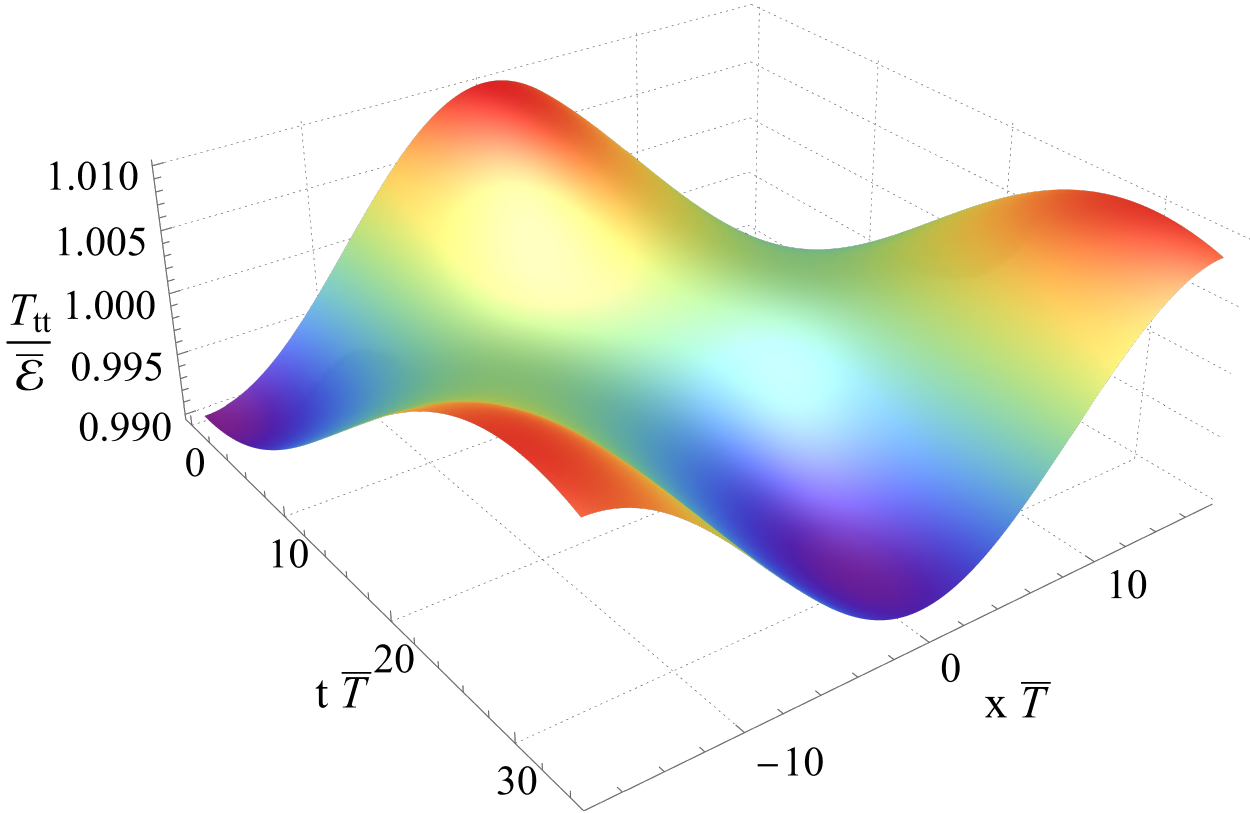}} 
	{\includegraphics[width=0.495\textwidth]{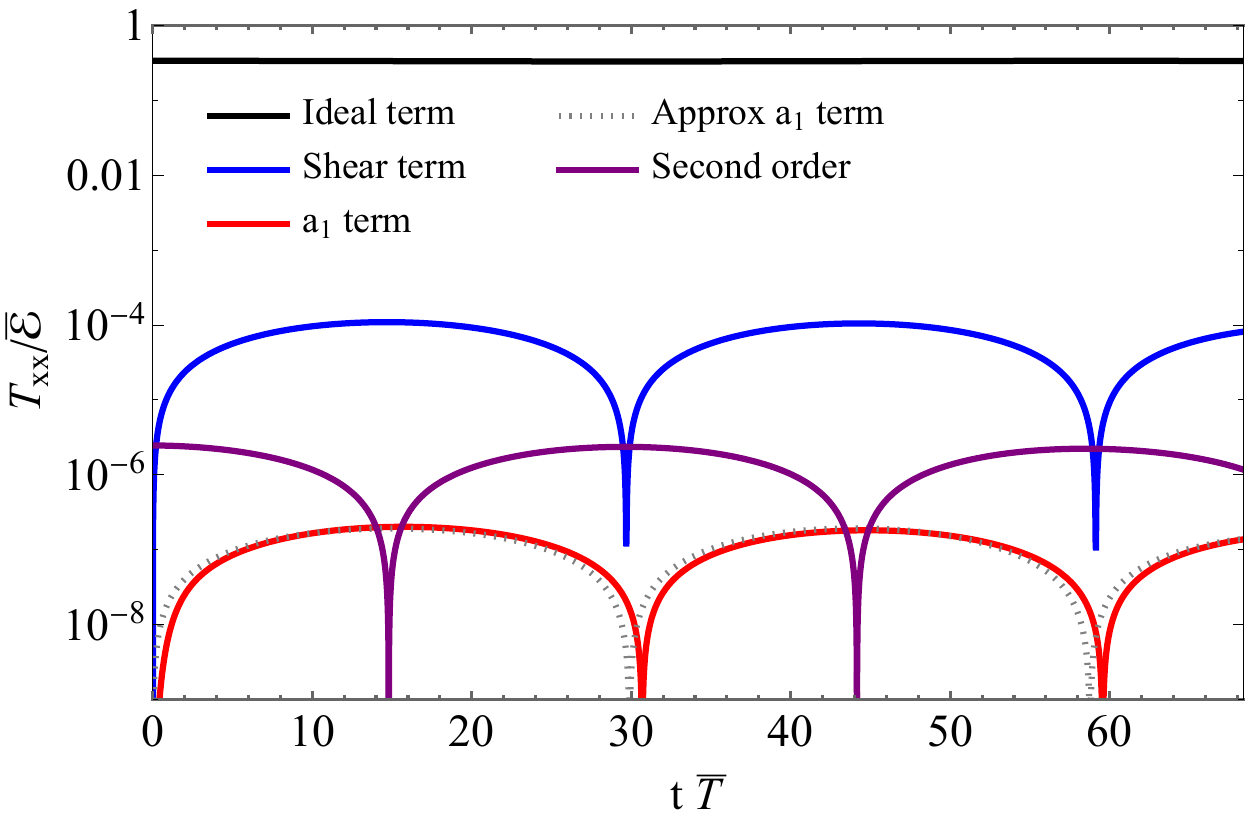}} 
	\caption{(Left): Spacetime evolution of the $T_{tt}$ component of the stress tensor of a sinusoidal perturbation of a homogeneous thermal state with initial data (\ref{Initial_data_sinusoidal}), momentum $k/\overline{T} \simeq 0.184$, box of size $L \, \overline{T} \simeq 34.2$ and energy $\overline{\mathcal{E}}=1$, evolved using the equations of first-order viscous hydrodynamics with frame $\{a_1,a_2\}=\{5,5\}$ and $N_x=2048$ points. (Right): For the evolution on the left, we show the log plot of the absolute value of the different contributions to the $T_{xx}$ component of the stress tensor \eqref{eq:tmunu11} at $x=0$ as a function of time: ideal term in solid black, shear term in solid blue and $a_1$ term in solid red ($a_2$ term is vanishing at $x=0$ by symmetry).
	We also include the result of the approximate expression (\ref{Residual0a}) in dotted grey, which allows to understand  the amplitude of the $a_1$ term. Additionally, we include the result of evaluating the second order terms \eqref{constitutive0sheartensor0} in solid purple.
	From this plot we can confirm that the system is well within the regime of hydrodynamics, as the shear term is much smaller than the ideal term and the second order term is much smaller than the shear term; more specifically, Criterion A is satisfied. The most important conclusion from this plot
	is that $a_1$, $a_2$ terms are much smaller than the shear term, as expected as they are of higher order on shell, and thus negligible for the physics up to first order; so, Criterion B is also satisfied. 
    }
	\label{sinusoidal_evolution_to_show_size_a1a2}
\end{figure}

We perform the real-time evolution of the non-linear first-order viscous hydrodynamics equations \eqref{conservationideal} with \eqref{eq:tmunu11}, initial conditions \eqref{Initial_data_sinusoidal} and hydrodynamic frame $\{a_1,a_2\}=\{5,5\}$. The resulting solution is shown in Fig. \ref{sinusoidal_evolution_to_show_size_a1a2} (left) for the $T_{tt}$ component of the stress tensor. This evolution captures the well known physics of the sound mode, with frequency and decay rate well approximated by \cite{Kovtun:2019hdm}: 
	\begin{align}
	\omega(k)&= c_s k - i \frac{\Gamma}{2(\epsilon+p)} k^2 +O(k^3)\, \,, 
	\label{Dispersion_relation_BDNK_k2}
\end{align}
with $c_s=1/\sqrt{3}$, $p=\epsilon/3$ and $\Gamma=4\eta/3$.

\subsubsection{The criteria}
We start by examining if the solution is within the regime of hydrodynamics, and for this we explain in detail our definition of Criterion A. We use the numerical solution for $\{\epsilon,u_x\}$ to evaluate pointwise in spacetime the different terms of the constitutive relations \eqref{eq:tmunu11}; in Fig. \ref{sinusoidal_evolution_to_show_size_a1a2} (right) we show the absolute value of the different terms in the $T_{xx}$ component of the stress tensor \eqref{eq:tmunu11},\footnote{The reason why we analyse the relative sizes of the terms of the $T_{xx}$ component of the stress tensor \eqref{eq:tmunu11} and not $T_{tx}$ or $T_{tt}$ is that in the shear tensor $\sigma_{\mu\nu}$, because of being transverse to the velocity, the component $\sigma_{tx}$ is suppressed by the velocity and  the component $\sigma_{tt}$ by the square of the velocity, and so these components are much smaller than $\sigma_{xx}$ for this solution, in which velocities are small. }
in a log scale at $x =0$ as functions of time.
Comparing the ideal term (solid black) with the shear term (solid blue), we conclude that first order terms are much smaller than ideal terms, indicating that the system is within the regime of hydrodynamics. To be precise,  we should compare the ideal term $T_{xx}^{(0)}$ with the full first order  $T_{xx}^{(1)}$ (shear term + $a_1$ term + $a_2$ term), and not only the shear term; however, as we will describe in detail below,  the $a_1$, $a_2$ terms are much smaller than the shear term and so $T_{xx}^{(1)} \simeq -\eta\, \sigma_{xx}$ is a very good approximation. 

As explained in the previous section, even if in the theory under consideration the stress tensor does not include second order terms, we will measure the size of such terms in our solutions since nothing prevent us from computing the required higher order derivatives.
We evaluate the second line of the constitutive relations \eqref{constitutive0sheartensor0} using our numerical solution for $\{\epsilon,u_x\}$, obtaining the solid purple line in Fig. \ref{sinusoidal_evolution_to_show_size_a1a2} (right). We use the transport coefficients (\ref{N=4SYMcoeffs}).
Note that these second order expressions in the second line of \eqref{constitutive0sheartensor0} are in the Landau frame, and our $\{\epsilon,u_x\}$ are not in the Landau frame; however, as we will verify below in detail, the difference between frames is of higher order, and hence negligible in our solution.  So, it is justified to use expression \eqref{constitutive0sheartensor0} as a very good approximation to the second order terms.

By comparing ideal, first order and second order terms, we find that the amplitude of each term in the gradient expansion is much smaller than the previous one. 
Ideally we would like to provide a quantitative criterion that is useful in a generic solution and  that does not rely on the amplitude of each term. 
For example, one possible criterion could be that
the system is in the regime of hydrodynamics if, pointwise in spacetime and for all components of the stress tensor, each term in the hydrodynamic expansion is not larger than a certain threshold value, e.g., 10\%, of any of the lower order terms.
However, a local, pointwise criterion has some disadvantages. For instance, at some specific points it might fail even if one considers a solution that is clearly in the regime of hydrodynamics: in our previous example, the definition above always fails when one of the relevant quantities crosses zero, and they do because they are oscillating. We would like to capture the fact that the system is well described by hydrodynamics in spite of these regions. 

One possible idea to define a criterion that avoids this problem of having regions where the relevant quantities cross zero is to depart from the pointwise analysis and consider some kind of integral version, i.e., a suitable norm that captures the global aspects of the solution. The idea would be to integrate in patches, possibly extended both in space and time. It would be reasonable to use patches whose size is similar to the characteristic scale of the problem: if they are too small we go back to the same problems as in the pointwise analysis, and if they are too large, we wash out the details of the system.
It would be interesting to perform an exploration of different such definitions, and we leave this systematic exploration for future work. In this paper we will use the following definition: we compute the $L_1$ norm (i.e., integral of the absolute value)
 of the relevant quantities over the the full spatial domain and also in time, with a time integration domain of the order of the characteristic timescale of the system.\footnote{The choice of the $L_1$ norm is only for convenience; other norms could be equally useful. Also, the choice of $10\%$ as a threshold value for acceptance is completely arbitrary. This value may depend on the system under consideration and the choice of initial conditions.}  

\vspace{2mm}

{ \it Criterion} A

\vspace{1mm}

The system is within the effective field theory regime at a given time $t$ if the $L_1$ norm over the spatial domain and over times $\{t-\tilde{t},t+\tilde{t}\}$, where $\tilde{t}$ is a characteristic time scale of the problem, of the shear term is smaller than $10\%$ of the $L_1$ norm  of the ideal term, and also the $L_1$ of the second order terms is smaller than $10\%$ the $L_1$ norm of the shear term, for all components of the stress tensor.

\vspace{2mm}

In the case of a sinusoidal perturbation, we consider $\tilde{t}$ as the period of oscillation. Relevant data for this criterion is ploted in Fig. \ref{sinusoidal_evolution_to_show_size_a1a2} (right) at $x=0$ and in Fig. \ref{Explicit_check_criteria_sinusoidal} (top, left) in the whole spacetime domain, where we
plot the $10\%$ of the shear term (in blue) and the second order term (in purple) in such a way that we can visually inspect their relative sizes. We do not include a plot of the ideal term because the shear term is orders of magnitude smaller, so it clearly satisfies the criterion. 
By computing the corresponding norms
we verified that Criterion A is well satisfied in our solution at all times, and thus we conclude that the system is in the effective field theory regime. The ratio of the $L_1$ norm of second order and first order terms is $2.2\%$, and of first order and ideal terms is $3.4\%$. We can then define the hydrodynamization time as the time that it takes for a system that is initially away from equilibrium, violating Criterion A, to relax to a regime satisfying Criterion A, and hence in the hydrodynamic regime.
\newline

We now proceed with a detailed definition of our second criterion, denoted by Criterion B.
We examine the size of the field redefinition dependent $a_1$, $a_2$  terms in \eqref{eq:tmunu11} and compare them to the shear viscosity term in \eqref{eq:tmunu11}, which is field redefinition independent. According to the discussion in Section \ref{First_order_hydrodynamics_and_effective_field_theory}, by using the equations of motion, the terms $a_1$, $a_2$ can be re-expressed, on shell, as second order terms, and thus in our solution they should be much smaller than the shear viscosity term; 
if they are much smaller, then the choice of frame will not affect the physics to first order.

\vspace{2mm}

{ \it Criterion} B

\vspace{1mm}

We will say that the physics to first order is independent of the chosen hydrodynamic frame at a time $t$ if the $L_1$ norm over the spatial domain and over times $\{t-\tilde{t},t+\tilde{t}\}$
of the spatial components of both the $a_1$ and $a_2$ terms in the stress energy tensor are smaller than $10\%$ of the $L_1$ norm of the spatial components of shear term (computed over the same spacetime domain).

\vspace{2mm}

Note that we only consider the spatial components of the stress tensor  in this definition. This is because the shear tensor is transverse to the velocity and the temporal components are suppressed when the velocity is small, as in our solution, while in the $a_1$, $a_2$ terms this suppression is not present. Therefore, for the temporal components of the stress tensor, the shear term may be smaller than the $a_1$, $a_2$ terms.

\begin{figure}[thbp]
	{\includegraphics[width=0.495\textwidth]{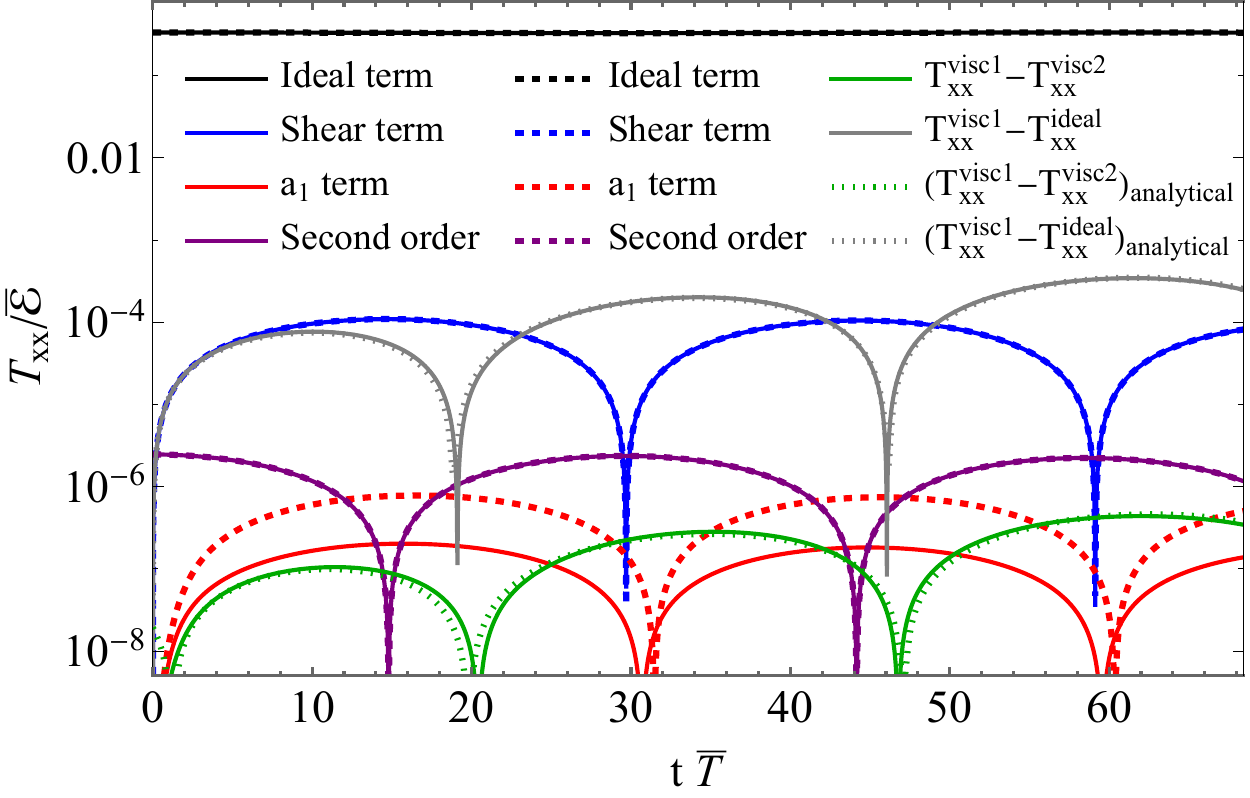}} 
	{\includegraphics[width=0.495\textwidth]{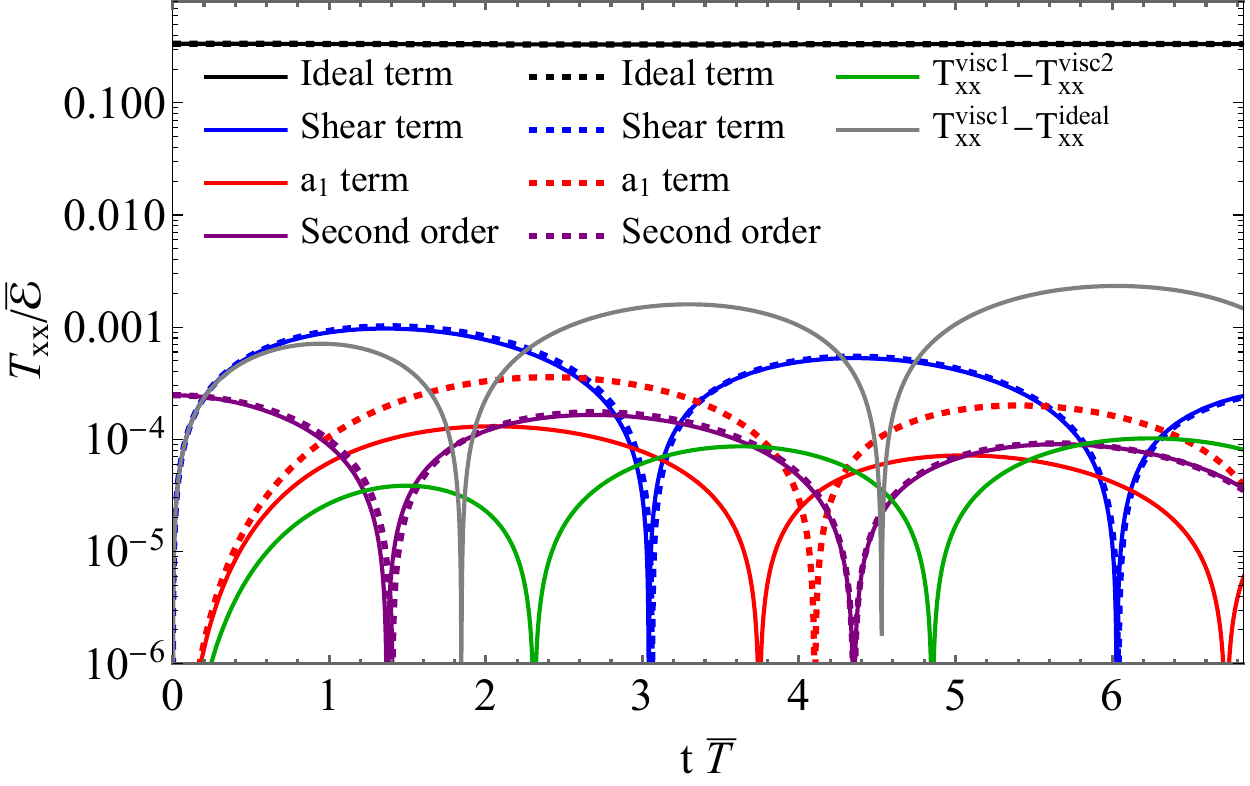}} 
	\caption{
 Left: We plot the same information as in Fig. \ref{sinusoidal_evolution_to_show_size_a1a2} (right) which was performed in frame $\{a_1,a_2\}=\{5,5\}$ in continuous lines, and now we add the data of an evolution with similar initial data \eqref{Initial_data_sinusoidal} and momentum $k/\overline{T}\simeq 0.184$, but now using frame $\{a_1,a_2\}=\{10,10\}$, in dashed lines and same color coding. The red solid and dashed lines correspond to the $a_1$ terms in each case, with an amplitude differing by a factor of around 4 as the approximate expression \eqref{Residual0a} indicates. 
We include the difference $T^{\text{visc1}}_{xx}-T^{\text{visc2}}_{xx}$ at $x=0$ of the two solutions in different frames  $\{a_1,a_2\}=\{5,5\}$ and  $\{a_1,a_2\}=\{10,10\}$, in solid green. Moreover, we also perform a numerical evolution of the ideal equations with initial data \eqref{Initial_data_sinusoidala}--\eqref{Initial_data_sinusoidalb} and obtain the difference $T^{\text{visc1}}_{xx}-T^{\text{ideal}}_{xx}$ of one viscous evolution and the ideal evolution, in solid grey. 
We include the analytical results for the differences $\left(T^{\text{visc1}}_{xx}-T^{\text{visc2}}_{xx}\right)_{\text{analytical}}$ and $\left(T^{\text{visc1}}_{xx}-T^{\text{ideal}}_{xx}\right)_{\text{analytical}}$  in dotted green and dotted grey respectively, finding good agreement. 
Right: Similar as (left) but with momentum ten times larger, $k/\overline{T}\simeq 1.84$, to analyse a situation which is marginally in the regime of hydrodynamics. 
See Fig. \ref{Explicit_check_criteria_sinusoidal} for plots in the spacetime domain and not just at $x=0$ of the relevant quantities to evaluate Criteria A, B and C in these runs. }
	\label{Diference_Ttt_of_two_runs_a1a2_5_a1a2_10}
\end{figure}

By computing the norms, we find that Criterion B is satisfied in our solution. In Fig. \ref{sinusoidal_evolution_to_show_size_a1a2} (right) we can observe that at $x=0$ the $a_1$ term is smaller than the shear term well below a $10\%$.
In Fig. \ref{Explicit_check_criteria_sinusoidal} (middle, left) we present the $a_1$ term (in red) and $a_1$ term (in orange) in the spacetime domain, together with $10\%$ the shear term (in blue). 
As additional information, we note that the ratio of amplitudes of the $a_1$ term and shear term is $0.19\%$ and of the $a_2$ term shear term is $0.06\%$. Thus, we can conclude that the physics up to first order in the hydrodynamic gradient expansion is independent of the arbitrarily chosen frame used to perform the actual numerical evolution, see Fig. \ref{Diference_Ttt_of_two_runs_a1a2_5_a1a2_10}. %

After this conclusion one could think to proceed in practical terms as follows.
If the system under consideration is well within the regime of hydrodynamics, one could think of the terms $a_1$, $a_2$ in the first-order hydrodynamic equations as mere regulators that render the equations well-posed since, on shell, they are of second order, and thus negligible to first order.
Then we can work as if we were in the Landau frame, as changing frame is a second order effect.
However, this way of thinking of the equations might be appropriate only when the system is well within the regime of hydrodynamics, and one must be careful if it is not clear that the system is  in that regime, such as in systems that are marginally in the regime of hydrodynamics, as the ones we study below.
\newline

Now we would like to find an easy way to understand the value of the amplitude of the $a_1$, $a_2$ terms in this solution. A natural approach would be to linearize the $a_1$, $a_2$ terms, but under linearization the detailed cancellations between the addends are spoiled. Recall that the $a_1$, $a_2$ terms are proportional to the equations of motion, which are nearly satisfied if the solution is in the effective field theory regime, explaining why there are detailed cancellations. 
Another option would be to use the equations of motion to replace the ideal hydrodynamic equations in $a_1$, $a_2$ by higher order terms. However, by linearising these higher order terms also cancellations among them are spoiled.

A way forward is as follows. In the Landau frame, i.e., $\{a_1,a_2\}=\{0,0\}$, the only contribution to the second order derivatives in the equations of motion comes from the shear term, and this contribution can be easily linearized. With this in mind, we can perform a field redefinition \eqref{changeframeconformal} from the Landau frame to a generic frame $\{a_1,a_2\}$ in which we replace the ideal hydrodynamic equations by the linearization of the contribution of the shear term to the terms with second order derivatives in the equations of motion.
With these expressions in a general frame, we can now perform an expansion in amplitude of the relevant part of the $a_1$, $a_2$ terms (without spoiling any detailed cancellation) obtaining 
\begin{subequations}
	\begin{align}
		\frac{3}{4}\frac{\dot{\epsilon}}{\epsilon}+\nabla \cdot u  &\simeq \frac{\eta}{\epsilon} (\partial_x u_x)^2 + \frac{3}{4}\, a_2\, \frac{\eta^2}{\epsilon^2} \partial_x^3 u_x   + ...\,\,, \label{Residual0a}\\
		\dot{u}^{\mu}+\frac{1}{4} \frac{\nabla_{\perp}^{\mu} \epsilon}{\epsilon} & \simeq \frac{\eta }{\epsilon} \partial_x^2 u_x+\frac{1}{4}\, a_1\,\frac{\eta^2}{\epsilon^2} \partial_t \partial_x^2 u_x+ ...\,\,. \label{Residual0b}
	\end{align}
	\label{Residual0}
\end{subequations}
The first term on the right hand side comes from the leading term in amplitude of the shear part and the second term from the leading term in amplitude of the $a_1$, $a_2$ contribution. 
Notice that in \eqref{Residual0a} the term linear in the shear is quadratic in the velocity, whilst the term proportional in $a_2$ part is linear in the velocity. Thus the latter dominates in our solution. In \eqref{Residual0b} both the shear and $a_1$ part are linear in the velocity, but the shear part is of lower order in derivatives, so it dominates.

Expressions (\ref{Residual0}) are very useful because they allow to estimate the size of the $a_1$, $a_2$ terms in the constitutive relations \eqref{eq:tmunu11} in the particular case of the sinusoidal perturbation of a thermal state. We can have analytical control over these terms, with explicit dependence on the values of the constants $a_1$ and $a_2$, amplitude, momentum, etc. 
Thus, we can perform an analysis of the size of these terms without the need to perform numerical simulations. 
In particular, it is interesting to analyze the limiting situation where $k/\overline{T}$ is not small and the system is starting to exit the regime of hydrodynamics and check to what extent the  $a_1$, $a_2$ terms are small or not.

\begin{figure}[thbp]
	\centering
	{\includegraphics[width=0.55\textwidth]{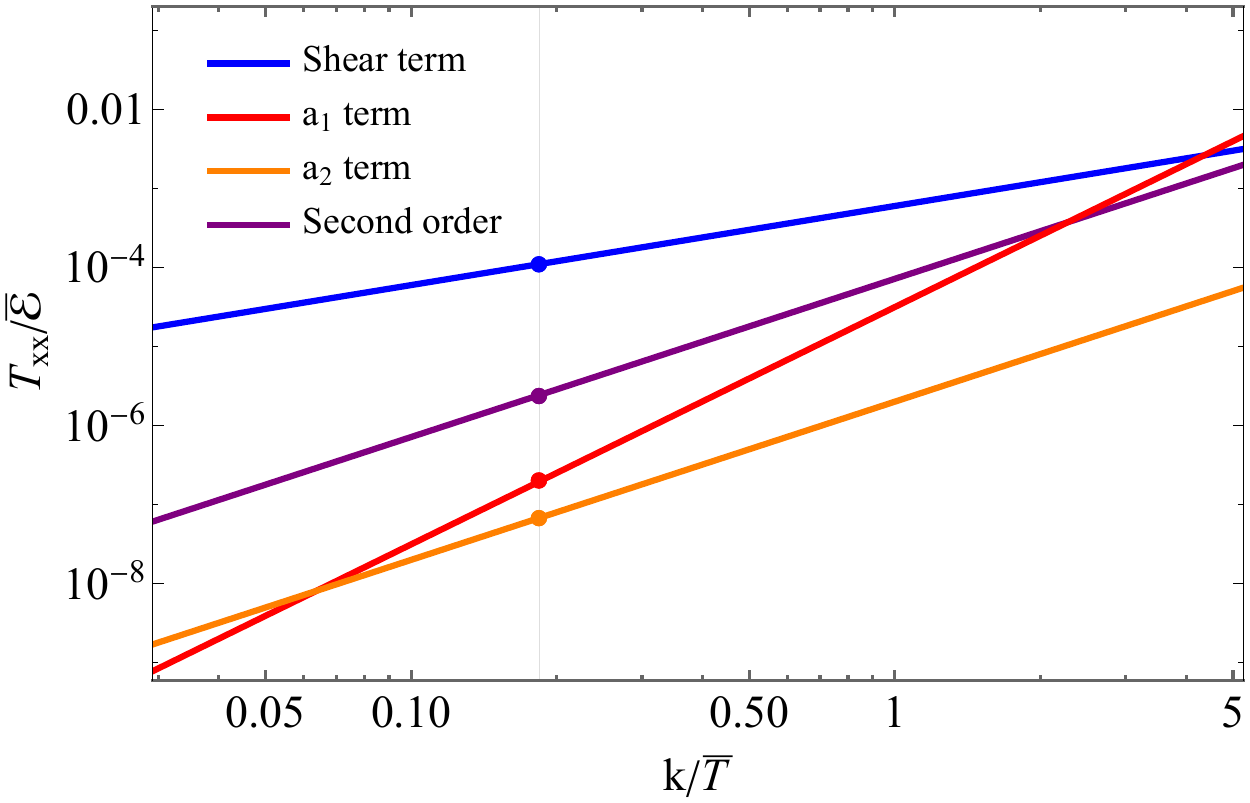}} 
	\caption{Log-log plot of the amplitude of the different terms of the $T_{xx}$ component of the stress tensor \eqref{eq:tmunu11} for initial data \eqref{Initial_data_sinusoidal} and frame $\{a_1,a_2\}=\{5,5\}$ as a function of momentum. To produce this plot we did not do any numerical evolutions; we simply computed the amplitude of each term in the initial data. For the shear term (solid blue) we linearize the expression in \eqref{eq:tmunu11} and for the $a_1$ and $a_2$ terms (solid red and solid orange respectively) we use the approximate expressions \eqref{Residual0}. We also include the amplitude of second order terms obtained by linearizing the second line in \eqref{constitutive0sheartensor0}, solid purple. 
	We observe that the $a_1$, $a_2$ terms are smaller than the shear term up to values $k/\overline{T}\simeq1$ or even beyond.  
 The vertical line shows the case studied in Fig. \ref{sinusoidal_evolution_to_show_size_a1a2}, with $k/\overline{T}\simeq 0.184$ and the dots correspond to the amplitudes obtained from that real time evolution.}
		\label{Sinusoidal_approximation_size_of_terms_vs_k}
\end{figure}

In Fig. \ref{Sinusoidal_approximation_size_of_terms_vs_k} we use the approximate expressions \eqref{Residual0} to obtain the amplitude of the $a_1$, $a_2$ terms of the $T_{xx}$ component of the stress tensor \eqref{eq:tmunu11} for a sinusoidal perturbation of a thermal state \eqref{Initial_data_sinusoidal}, as a function of momentum: $a_1$ term in red, $a_2$ term in orange. We also include the amplitude of the shear term in \eqref{eq:tmunu11} and the second order terms in \eqref{constitutive0sheartensor0} that we obtain by linearizing these expressions. 
From Fig. \ref{Sinusoidal_approximation_size_of_terms_vs_k} we can obtain the following conclusions. 	We observe that the $a_1$, $a_2$ terms are smaller than the shear term up to values $k/\overline{T}\simeq1$ or even beyond. Thus,  different frames provide the same physical description to first order. If we consider the criterion that the $a_1$, $a_2$ terms are smaller than the shear term by at least $10\%$, that is, Criterion A, in the case of Fig. \ref{Sinusoidal_approximation_size_of_terms_vs_k} this happens at $k/\overline{T}\simeq 1.37$;  however, this number will depend on the values of $a_1$, $a_2$ and on the amplitude of the perturbation (and of course on the choice of threshold). For larger values of $a_1$, $a_2$ this number will be smaller. This suggests that the larger $a_1$, $a_2$ are, the smaller the value of $k/\overline{T}$ where the ratio equals $10\%$ is. Thus,  if we are close to the boundary of the regime of applicability of hydrodynamics, it is preferable to work with smaller values of $a_1$, $a_2$ as this imposes a lower limit for the frame independence of the physics up to first order according to Criterion B.

We now consider a third criterion, which we denote by Criterion C. The motivation is the following. Above we have proposed Criterion B to determine when the physics up to first order is independent of the choice of frame. We can complement it with another criterion:  
perform two numerical evolutions in different causal frames with similar initial data and compare the two solutions after a certain time $t$.  It could happen that even if Criterion B indicates that the $a_1$, $a_2$ terms are small at all times, their effects accumulate over time, resulting large differences at late times. Another possibility is that even if the $a_1$, $a_2$ terms are not small, the solutions obtained in different frames remain close to each other. 

To asses this criterion,  in addition to the solution  in Fig. \ref{sinusoidal_evolution_to_show_size_a1a2} in frame $\{a_1,a_2\}=\{5,5\}$, we now perform another evolution with similar initial data \eqref{Initial_data_sinusoidal}, $k/\overline{T} \simeq 0.184$, but in the frame $\{a_1,a_2\}=\{10,10\}$;\footnote{One may wonder if the initial data should be changed to the corresponding working frame, but for this specific initial data the change of frame at $t=0$ is exactly vanishing; this is because it satisfies the ideal equations	\eqref{conservationidealexplicit} at $t=0$. Below we will address the question of changing frame in the initial data in cases where the change of frame is non trivial.}
we denote the two solutions by $T^{\text{visc1}}_{\mu\nu}$ and $T^{\text{visc2}}_{\mu\nu}$ respectively, shown in Fig. \ref{Diference_Ttt_of_two_runs_a1a2_5_a1a2_10} (left) in solid and dashed lines respectively. 
According to the discussion in Section \ref{First_order_hydrodynamics_and_effective_field_theory}, we choose two frames that are well separated, in this case by a factor of 2.
In Fig. \ref{Diference_Ttt_of_two_runs_a1a2_5_a1a2_10} (left) we present the difference of the stress tensor component $T_{xx}$ of the two solutions, $T^{\text{visc1}}_{xx}-T^{\text{visc2}}_{xx}$, at $x=0$ as a function of time in solid green.

\definecolor{ao(english)}{rgb}{0.0, 0.6, 0.0}

\begin{figure}[thbp]
	\centering
	{\includegraphics[width=0.45\textwidth]{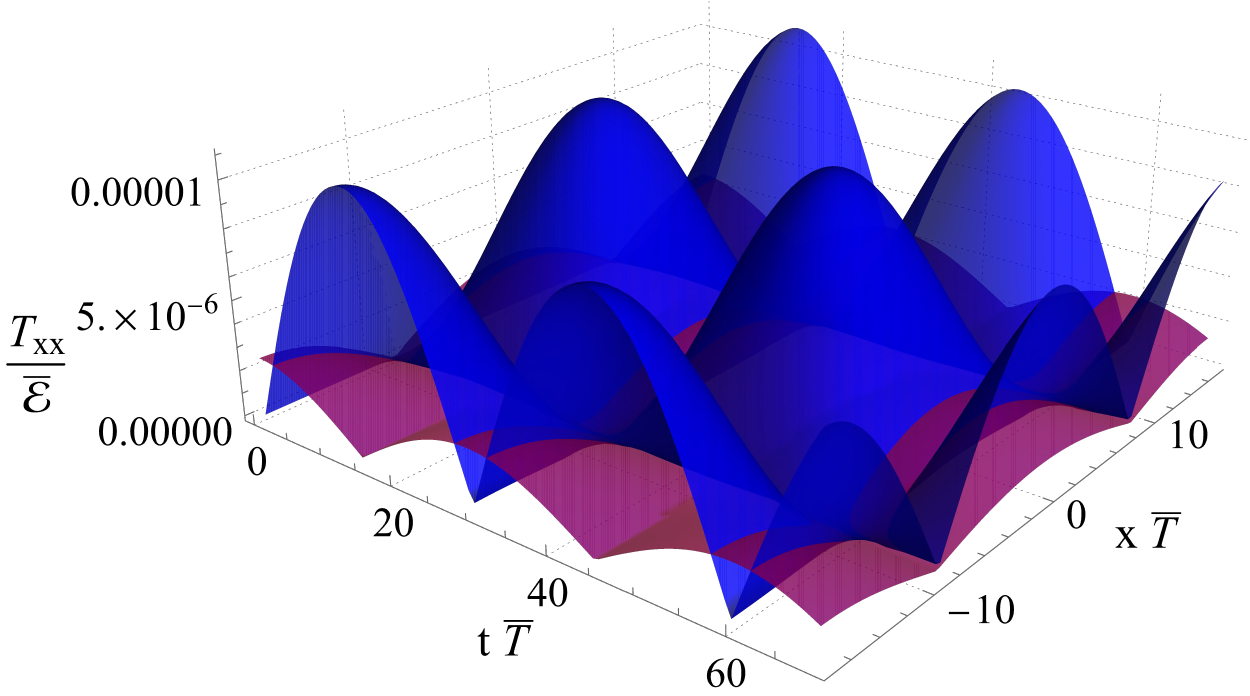}}
	\put (-29,109) {\large $\displaystyle Criterion$ $\displaystyle A$}
	\put (-24,100) {$\colorbox{blue}{\rule{0pt}{1pt}\rule{1pt}{0pt}}$}
	\put (-14,97.5) {\footnotesize $\textbf{10\%}$ \footnotesize Shear}
	\put (-24,90) {$\colorbox{violet}{\rule{0pt}{1pt}\rule{1pt}{0pt}}$}
	\put (-14,88.5) {\footnotesize Second order}
	{\includegraphics[width=0.45\textwidth]{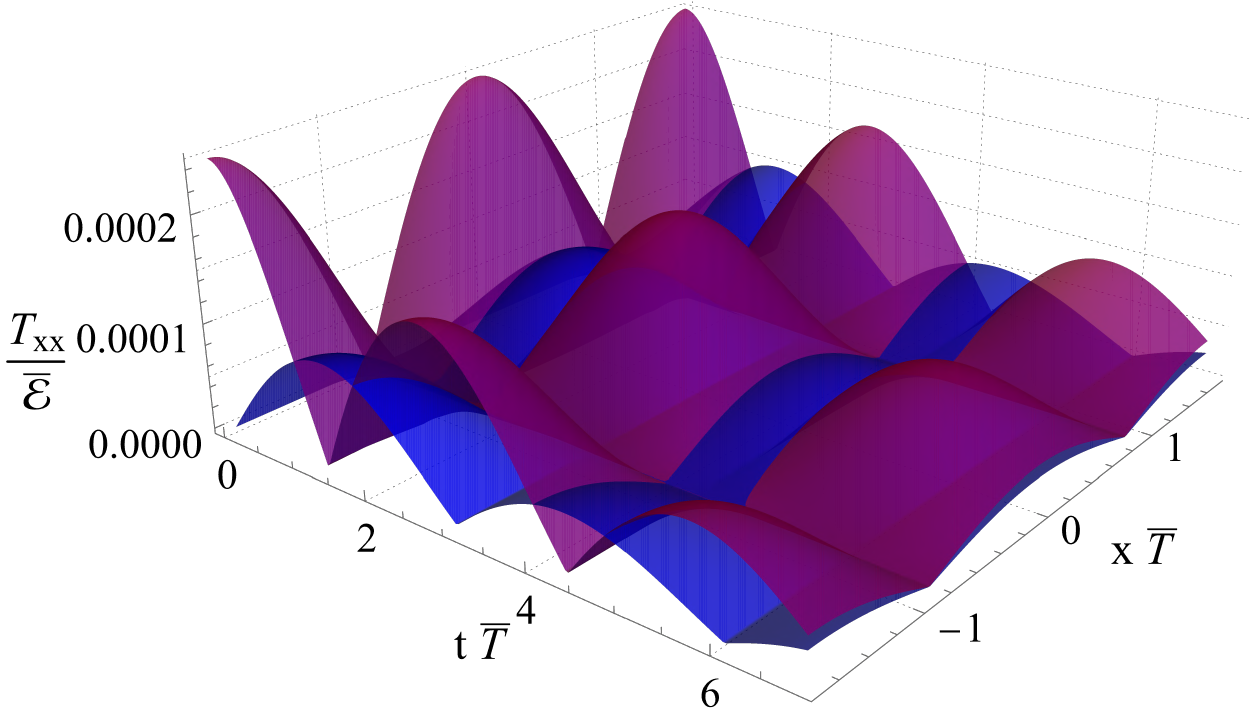}}
	{\includegraphics[width=0.45\textwidth]{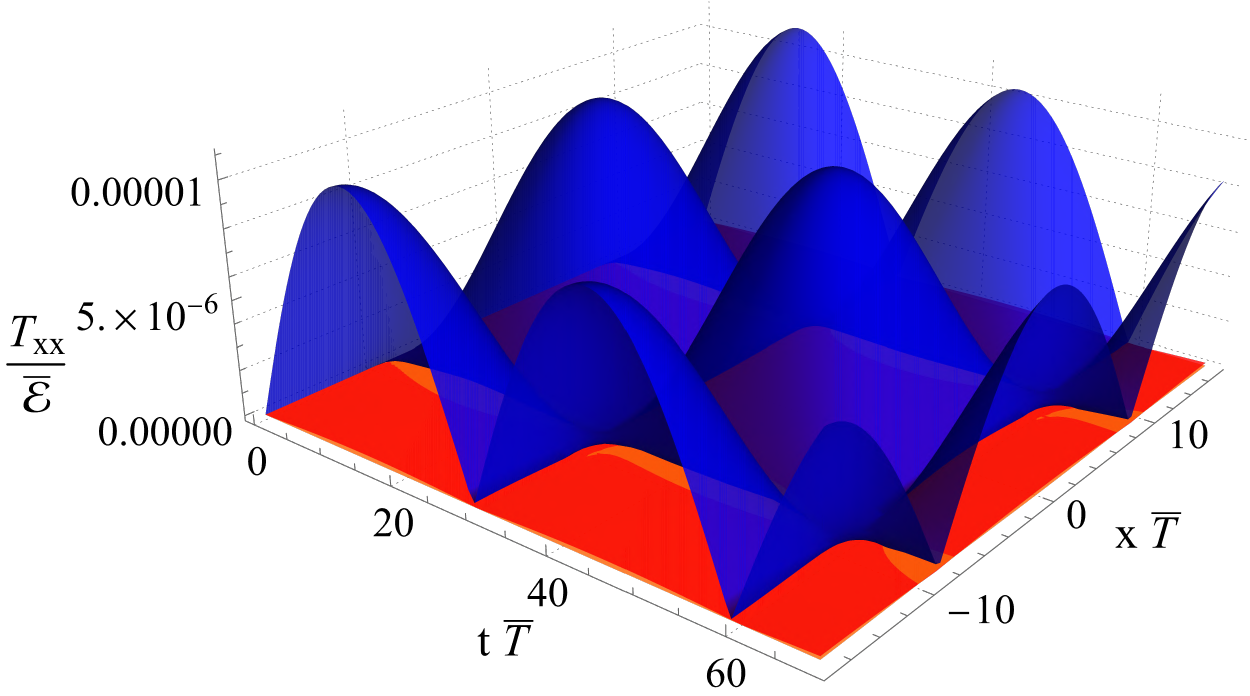}}
	\put (-25,110) {\large $\displaystyle Criterion$ $\displaystyle B$}
	\put (-18,101) {$\colorbox{blue}{\rule{0pt}{1pt}\rule{1pt}{0pt}}$}
	\put (-8,98.5) {\footnotesize $\textbf{10\%}$ \footnotesize Shear}
	\put (-18,91) {$\colorbox{red}{\rule{0pt}{1pt}\rule{1pt}{0pt}}$}
	\put (-8,88.5) {\footnotesize $a_1$ term}
	\put (-18,81) {$\colorbox{orange}{\rule{0pt}{1pt}\rule{1pt}{0pt}}$}
	\put (-8,79.5) {\footnotesize $a_2$ term}
	{\includegraphics[width=0.45\textwidth]{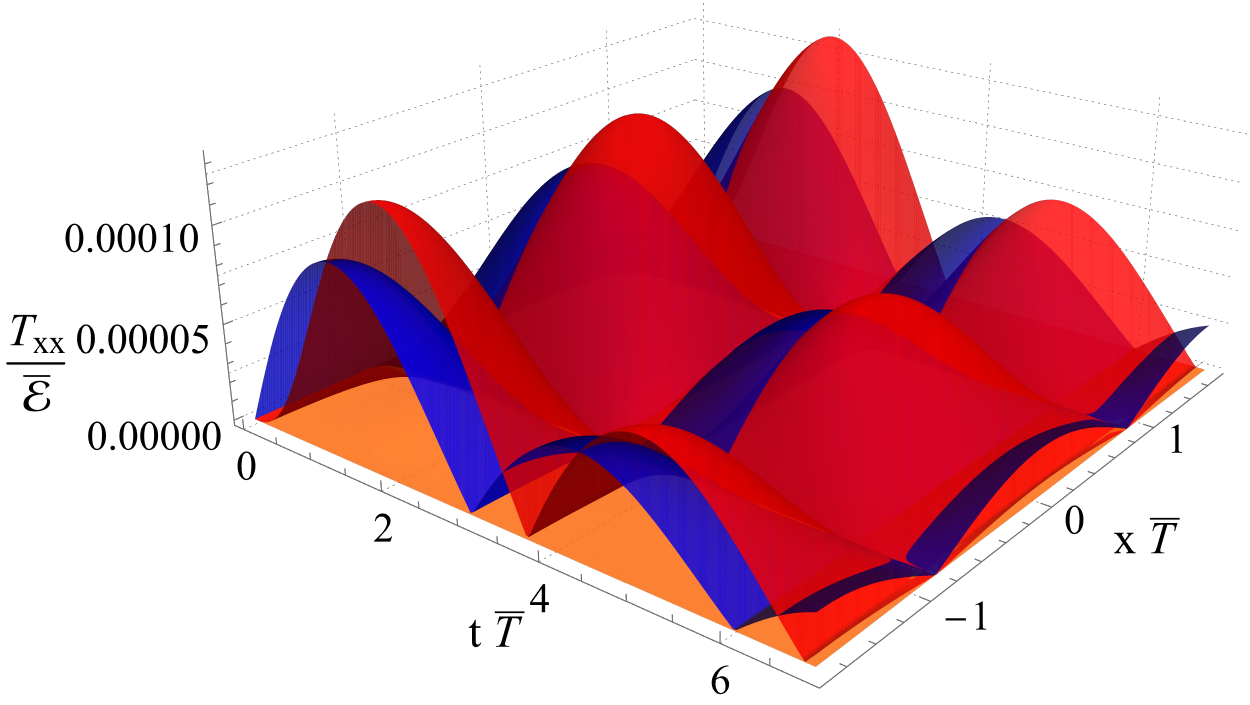}}
	{\includegraphics[width=0.45\textwidth]{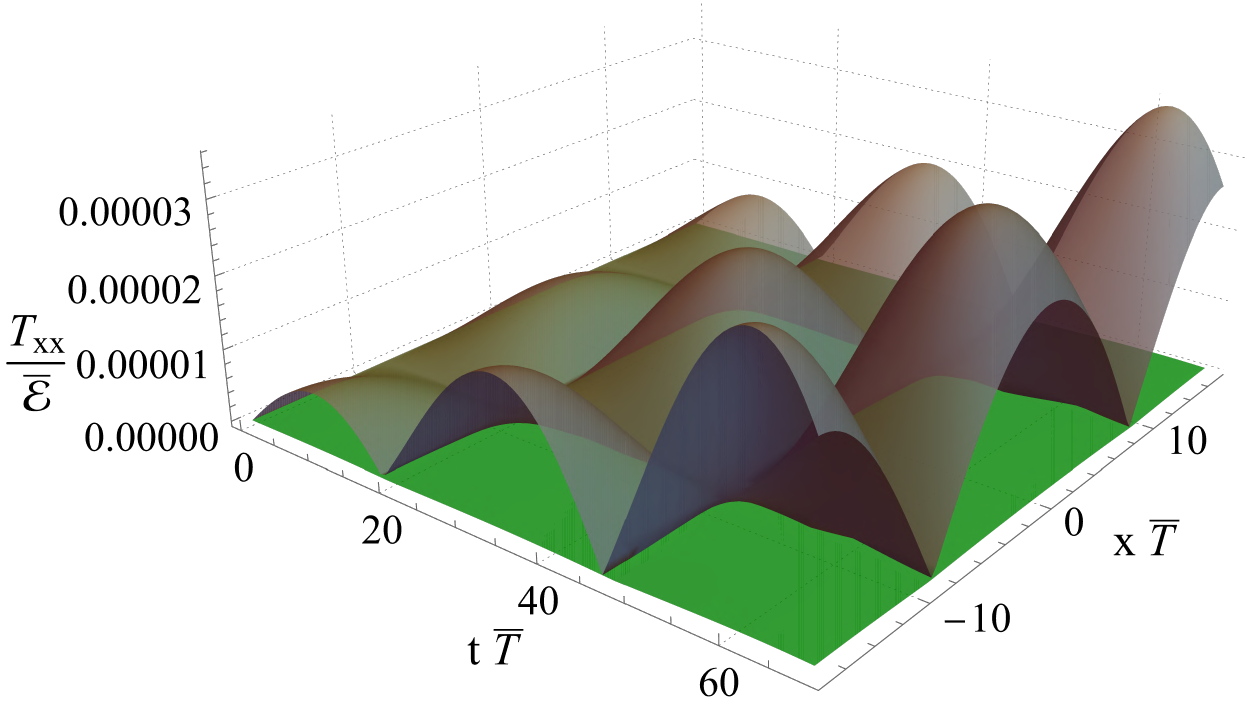}} 
	\put (-24,116) {\large $\displaystyle Criterion$ $\displaystyle C$}
	\put (-32,107) {$\colorbox{gray}{\rule{0pt}{1pt}\rule{1pt}{0pt}}$}
	\put (-22,104.5) {\footnotesize $\textbf{10\%}$ $\left(T^{\text{visc1}}_{xx}-T^{\text{ideal}}_{xx}\right)$}
	\put (-32,96) {$\colorbox{ao(english)}{\rule{0pt}{1pt}\rule{1pt}{0pt}}$}
	\put (-22,94.5) {\footnotesize  $T^{\text{visc1}}_{xx}-T^{\text{visc2}}_{xx}$}
	{\includegraphics[width=0.45\textwidth]{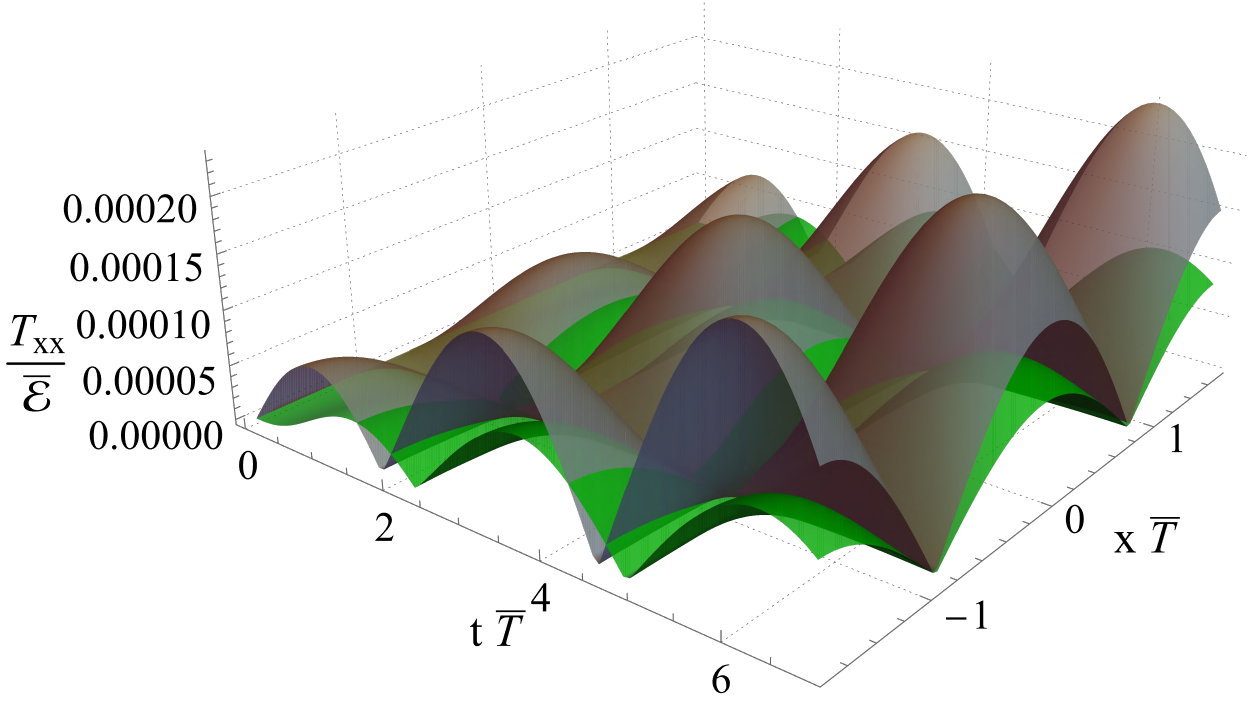}} 
	\caption{   For the evolutions presented in Fig. \ref{Diference_Ttt_of_two_runs_a1a2_5_a1a2_10} we show here relevant quantities that allow for an easy visual inspection of the different criteria A, B and C. 
		For this, in the top panels we plot a $10\%$ of the shear term (blue) and second order terms (purple). In middle panels we plot a $10\%$ of the ideal term (blue) and $a_1$, $a_2$ terms (red, orange). In bottom panels we plot a $10\%$ of $T^{\text{visc1}}_{xx}-T^{\text{ideal}}_{xx}$ in solid grey and $T^{\text{visc1}}_{xx}-T^{\text{visc2}}_{xx}$ in solid green. 
		Left plots correspond to the case $k/\overline{T} \simeq 0.184$, and right plots to $k/\overline{T} \simeq 1.84$.
		All quantities are in absolute values. Top an middle panels correspond to evolutions in frame $\{a_1,a_2\}=\{5,5\}$. 
	   }
	\label{Explicit_check_criteria_sinusoidal}
\end{figure}

Having computed the difference $T^{\text{visc1}}_{xx}-T^{\text{visc2}}_{xx}$ , we would like to find a relevant quantity to compare it with so that we can establish a criterion to decide when the system is invariant under field redefinitions to first order in the hydrodynamic gradient expansion.  
One possibility could be to compare this difference with $T_{xx}$ itself; however, this might not be sensible in our solution because the perturbation is small compared to the average value and it does not tell us about the first order physics.
Another possibility would be to compare this difference to the shear term; however, the former will typically increases with time while the shear term will decrease in such a way that the difference might eventually become larger than the shear term. 

We propose the following approach.  We consider a third solution, namely one obtained by solving the ideal hydrodynamics evolution equations using the initial data \eqref{Initial_data_sinusoidala}--\eqref{Initial_data_sinusoidalb},\footnote{Recall that ideal hydrodynamics equations are of first order in derivatives, thus we do not need to specify the time derivatives of the evolved quantities at the initial time. The latter were chosen so that the equations of ideal hydrodynamics are satisfied initially, so using this initial data for the ideal case is consistent. }
$k/\overline{T} \simeq 0.184$. 
We denote the stress energy tensor obtained from this third solution by $T^{\text{ideal}}_{\mu\nu}$.
We can now compute the difference of the stress tensors  of one viscous solution and  the ideal  solution, that is, $T^{\text{visc1}}_{\mu\nu}-T^{\text{ideal}}_{\mu\nu}$, 
and compare it with the difference of the stress tensors of the two viscous solutions,  $T^{\text{visc1}}_{\mu\nu}-T^{\text{visc2}}_{\mu\nu}$. 
In this way we are comparing the effect of changing frames with the effect of including viscous terms in the hydrodynamics description, which is precisely what we want. 

\vspace{2mm}

{ \it Criterion} C

\vspace{1mm}

 We will say that the physics to up first order in the derivative expansion is invariant under a  change of frame at time $t$ when the $L_1$ norm of the difference
 $T^{\text{visc1}}_{\mu\nu}-T^{\text{visc2}}_{\mu\nu}$ computed over the spatial domain and time interval $\{t-\tilde{t},t+\tilde{t}\}$
 is less  than $10\%$ of the $L_1$ norm of the difference $T^{\text{visc1}}_{\mu\nu}-T^{\text{ideal}}_{\mu\nu}$ (and $T^{\text{visc2}}_{\mu\nu}-T^{\text{ideal}}_{\mu\nu}$), for all components of the stress tensor.\footnote{Let us emphasize that both quantities  $T^{\text{visc1}}_{\mu\nu}-T^{\text{visc2}}_{\mu\nu}$ and $T^{\text{visc1}}_{\mu\nu}-T^{\text{ideal}}_{\mu\nu}$ are expected to grow in linearly in time (e.g., \cite{Reall:2021ebq}), but their ratio should have a weaker time dependence and hence this criterion could still be useful at much later times.}

\vspace{2mm}

In Fig. \ref{Diference_Ttt_of_two_runs_a1a2_5_a1a2_10} (left) we show the result of the difference $T^{\text{visc1}}_{xx}-T^{\text{visc2}}_{xx}$ in solid green and $T^{\text{visc1}}_{xx}-T^{\text{ideal}}_{xx}$  in solid grey at $x=0$. 
In Fig. \ref{Explicit_check_criteria_sinusoidal} (left, bottom) we show a plot in all spacetime of the quantities $T^{\text{visc1}}_{xx}-T^{\text{visc2}}_{xx}$ and $10\%$ of $T^{\text{visc1}}_{xx}-T^{\text{ideal}}_{xx}$; this allows for a direct visual inspection of the relevant quantities for Criterion C.
Computing the $L_1$ norms, 
we find that the first one is smaller than $10\%$ of the second one at all times, and thus we conclude that Criterion C is satisfied. The actual ratio of the $L_1$ norm of terms $T^{\text{visc1}}_{xx}-T^{\text{visc2}}_{xx}$ and $T^{\text{visc1}}_{xx}-T^{\text{ideal}}_{xx}$ is $0.14\%$.

Thus, criterion C also confirms that the frame dependent terms have a higher order effect that is negligible for the physics up to first order.
Also, we can conclude that the accumulated effect of the terms $a_1$, $a_2$ is small in this solution, and having locally small $a_1$, $a_2$ terms also implies having a small difference in the evolutions in different frames. 
\newline

Now we would like  to address the following question: would it be possible to have control on the difference $T^{\text{visc1}}_{xx}-T^{\text{visc2}}_{xx}$? 
We will obtain an analytical estimate of the difference $T^{\text{visc1}}_{xx}-T^{\text{visc2}}_{xx}$, which will allow to understand the size of this quantity. To this end, we use that the solution to the linear problem is known and given by the dispersion relation of the sound mode \cite{Kovtun:2019hdm}:
\begin{align}
		\omega(k)&= c_s k - i \frac{\Gamma}{2(\epsilon+p)} k^2 +  \frac{\Gamma^2}{8 c_s(\epsilon+p)^2} k^3-i a_2 \eta\frac{\Gamma^2}{2 (\epsilon+p)^3} k^4 +O(k^5)\, \,, 
			\label{Dispersion_relation_BDNK_k4}
\end{align}
 up to $O(k^4)$, with $c_s=1/\sqrt{3}$, $p=\epsilon/3$ and $\Gamma=4\eta/3$. Recall that this is the dispersion relation of the first-order hydrodynamics equations, that is, the truncated theory; beyond $O(k^2)$, the terms would be different if we had included second and/or higher order terms in the expansion of the stress energy tensor. 
Recalling that our initial data is $x\leftrightarrow-x$ symmetric, the general solution to the linear problem is a sum of functions 
\begin{align}
\left(  \tilde{A} e^{\omega_{Im} t} \cos(\omega_{Re} t) + \tilde{B} e^{\omega_{Im} t} \sin(\omega_{Re}t) \right) \cos(k x)\, \,, 
	\label{Linear_solution_dispersion_relation_BDNK_k4}
\end{align}
in harmonics of $k$, even if here we will restrict to the lowest harmonic as the other terms are very small in our solution. Here $\{\omega_{Re},\omega_{Im}\}$ are the values  obtained from evaluating 	\eqref{Dispersion_relation_BDNK_k4}.
Now, for the two solutions in Fig. \ref{Diference_Ttt_of_two_runs_a1a2_5_a1a2_10} (left) 
we use the analytical expression 	\eqref{Linear_solution_dispersion_relation_BDNK_k4}  to fit $T_{xx}$ and obtain the values of the constants $\tilde{A}$ and $\tilde{B}$ in each case. We now consider the difference of the two analytical expressions $\left(T^{\text{visc1}}_{\mu\nu}-T^{\text{visc2}}_{\mu\nu}\right)_{\text{analytical}}$ and include it in Fig. \ref{Diference_Ttt_of_two_runs_a1a2_5_a1a2_10} (left), in dotted green. 
We can conclude that we obtain a very good estimate of the difference $T^{\text{visc1}}_{\mu\nu}-T^{\text{visc2}}_{\mu\nu}$. Moreover, we can also perform a fit to the solution of ideal hydrodynamics using the ideal dispersion relation, and obtain the analytical difference $\left(T^{\text{visc1}}_{\mu\nu}-T^{\text{ideal}}_{\mu\nu}\right)_{\text{analytical}}$ that we plot in dotted grey, obtaining also a good estimate.

To be precise, we perform the aforementioned fits using data from some time $t>0$ with $t\overline{T}\simeq10$, and not from $t=0$. The reason is that we observe that the fit is better if we avoid the first instants of the evolution since there is a very small quasinormal mode that relaxes quickly and is not captured by \eqref{Linear_solution_dispersion_relation_BDNK_k4}. It would be interesting to study in detail this quasinormal mode, but we leave it for future work.
In principle $\tilde{A}$ and $\tilde{B}$ can be obtained from the initial data analytically, as the initial data is also analytical. However, the fit that we  obtain in doing so is not so good precisely because of the presence of the aforementioned quasinormal mode.

Using the analytical solution we can now provide an answer to the question: even if the $a_1$, $a_2$ terms are not small compared to the shear term, is it possible that solutions obtained in different frames capture the same physics up to first order? In other words: can Criterion B be violated while Criterion C be satisfied? The answer is yes, and we have an example in which this is under control: In the dispersion relation \eqref{Dispersion_relation_BDNK_k4} the constant $a_2$ enters at $O(k^4)$ (and $a_1$ enters at  $O(k^6)$); thus, if we consider a frame with large value of $a_2$, the corresponding $a_2$ term in the stress tensor becomes large compared to the shear term but the analytical solution is barely modified because it only receives corrections at  $O(k^4)$ and hence the evolution of the stress tensor in different frames will be very similar. 

 Even if we could consider these solutions as counterexamples to the statement that Criterion A implies Criterion B, they are obtained by choosing unreasonably large values of $a_1$ and $a_2$. If we restrict  $\{a_1,a_2\}$  to values that are close to saturating the inequalities \eqref{hyperbolicity_conditions_BDNK}, then we can exclude these counterexamples. Below we provide further arguments in favor of this choice. 

\subsubsection{A solution marginally in the effective field theory regime}
Up to this point we have considered solutions that are in the regime of hydrodynamics by construction.
In the experimental description of the QGP created in heavy-ion collisions, when the hydrodynamics codes are initialized, gradients might be large so it is unclear whether the QGP should be well-described by (viscous) hydrodynamics. Motivated by this picture, we now explore a system that is on the verge of the hydrodynamics regime. We consider another pair of simulations, also in the frames $\{a_1,a_2\}=\{5,5\}$ and $\{a_1,a_2\}=\{10,10\}$, and initial data \eqref{Initial_data_sinusoidal} but now with momentum a factor of ten larger, $k/\overline{T} \simeq 1.84$.
In Fig. \ref{Diference_Ttt_of_two_runs_a1a2_5_a1a2_10} (right) we show the relevant quantities of these two evolutions at $x=0$, and in 
 Fig. \ref{Explicit_check_criteria_sinusoidal} (right)
 we depict the relevant quantities to evaluate the different criteria A, B and C in the whole spacetime.

We observe that Criterion A is violated, that is, the system is slightly outside the regime of hydrodynamics since the ratio of the $L_1$ norms of the second order and first order terms is $18\%$. Furthermore, the ratio between the $L_1$ norm of the $a_1$ term and the shear term is $15\%$, so Criterion B is also violated. However, Criterion C is satisfied, that is, the two solutions obtained in different frames are close to each other since the ratio between the $L_1$ norms of $T^{\text{visc1}}_{xx}-T^{\text{visc2}}_{xx}$ and $T^{\text{visc1}}_{xx}-T^{\text{ideal}}_{xx}$ is $5.3\%$. This indicates that, according to Criterion C, the invariance of the physics up to first order under frame redefinitions is robust, even if the system is marginally in the effective field theory regime.

The fact that Criterion B is satisfied or not in a situation in where the system is close to the boundary of  the regime of hydrodynamics may depend on the specific values of $\{a_1,a_2\}$. If for a pair of values $\{a_1,a_2\}$ the system is about to violate Criterion B, then choosing slightly larger values $\{a_1,a_2\}$ will trivially lead to a violation of this criterion (assuming that the threshold value for acceptance is kept the same). Thus, the fact that Criterion B is satisfied or not in these limiting cases depends on the values of $\{a_1,a_2\}$ chosen. Therefore, to avoid ambiguities, it is preferable to work with values $\{a_1,a_2\}$ that are close to saturate the bound 	\eqref{hyperbolicity_conditions_BDNK}. 
On the other hand, Criterion C is still robust in these circumstances.

One would also like to understand how this system hydrodynamizes. That is, how long it takes for the system to become well described by hydrodynamics according to Criterion A. 
The answer is that it seems that this system does not hydrodynamize in times comparable to $\overline{T}$ or even much larger. 
Linear physics indicates that 
the solution will remain away from the effective field theory regime, as second order terms decay at the same rate as first order terms, with a ratio of amplitudes that is constant in time and thus does not decrease. It could be that non-linear effects change the picture at much later times.

\subsubsection{The choice of frame}

We now comment on the choice of  frame  in practical applications.
Above we have discussed in detailed examples that show that physics up to first order in the gradient expansion is independent of the chosen hydrodynamic frames $\{ a_1, a_2\}$ as long as the system is within the regime of hydrodynamics. So, in principle, one could use any values of $\{ a_1, a_2\}$ that obey the hyperbolicity  bounds \eqref{hyperbolicity_conditions_BDNK} and obtain equivalent physical descriptions up to first order. 
However, working  with large values of $\{ a_1, a_2\}$ is not practical. 
The first reason is that the characteristic velocities of the PDEs depend on the values of $\{ a_1, a_2\}$, and for larger values of $\{ a_1, a_2\}$ these velocities are smaller. Then, if there are velocities in the system that are larger than the characteristic velocities the code will crash; we provide further comments in the study of shockwaves below.
The second reason is that larger values of  $\{ a_1, a_2\}$ will set more restrictive bounds on the size of corresponding terms, which may result in rather artificial violations of Criterion B. %
For example, in the case studied in Fig. 	\ref{Changing_initial_data} (right), if we used  $\{ a_1, a_2\}=\{ 100, 100\}$, instead of $\{ a_1, a_2\}=\{ 10, 10\}$, we would obtain that the difference is already $O(1)$ between the two simulations. If we have done the same in (Left) we would still be well in the regime of hydrodynamics. Thus, this example illustrates the fact that using larger values of $\{ a_1, a_2\}$ may set a more restrictive bounds for the validity of Criterion B.  

Recall that $\{ a_1, a_2\}$ have to satisfy the hyperbolicity bounds  \eqref{hyperbolicity_conditions_BDNK}. Thus, to avoid artificially large effects from changing frames it is convenient to work values of $\{ a_1, a_2\}$ that are close to saturating \eqref{hyperbolicity_conditions_BDNK}. In fact, the equality region in \eqref{hyperbolicity_conditions_BDNK}  was coined `sharply causal region', and it is interesting because the characteristics have exactly the speed of light and thus one  can in principle evolve shocks of arbitrarily large amplitudes \cite{Freistuhler:2021lla,Pandya:2021ief}. %
Hence, for these reasons it is convenient to work in the sharply causal frames; it is a sweet spot satisfying several interesting conditions. One such example would be: $\{ a_1, a_2\}=\{ \frac{25}{4}, \frac{25}{7}\}$. A typical value used in this paper is $\{ a_1, a_2\}=\{5,5\}$ which is not far from this sharply causal region, and while it is quite arbitrary, it is equally valid.

\subsubsection{Changing frame in the initial data}
\label{sec:change_frame_ID}

In the previous analyses we performed simulations in two different frames $\{a_1,a_2\}=\{5,5\}$ and $\{a_1,a_2\}=\{10,10\}$, using the same initial data \eqref{Initial_data_sinusoidal} in both cases. This was consistent because for that particular choice of initial data the effect of changing frames \eqref{changeframeconformal} is exactly vanishing. In more realistic applications,  e.g.,  to describe experimental data in the context of heavy-ion collisions, one will be given initial data for the stress tensor, and not directly initial data for the evolution variables $\{\epsilon,u_x\}$ in the chosen causal frame. 
Thus, in practice, one has to obtain the initial data for the evolution variables in the working causal frame from a given stress tensor. Assuming that the initial stress tensor in the regime of validity of effective field theory, one possible way to proceed is the following: start from initial data for the stress tensor and diagonalize it to obtain $\{\epsilon,u_x\}$ in the Landau frame and then change frame using \eqref{changeframeconformal}.

In generic circumstances the initial data will not satisfy exactly the ideal hydrodynamics equations and the change of frame expressions will be non-trivial at that initial time. However, if there are reasons to believe that the system is well described by hydrodynamics, then the ideal hydrodynamics equations should be nearly satisfied and corrections should appear in a hierarchy of ever decreasing terms. %
For this reason, considering initial data so that the time derivatives $\{\partial_t \epsilon,\partial_t u_x\}$ are chosen to satisfy the ideal equations seems to be a good approximation. Moreover, this choice allowed us to analyse the effect of changing frames in the evolutions without having to address the question of changing frame in the initial data. 
Other choices, such as vanishing initial time derivatives of the evolution variables , tend to result in configurations for which the ideal hydrodynamics equations are far from being satisfied at $t=0$, which could result in significant differences in the initial data in different causal frames. In this section we address the effect of changing frame in the initial data.
\newline

Let us be more precise about our procedure to  change frames in the initial data. To change frames one uses the field redefinitions \eqref{changeframeconformal}, where the constants $\{a_1,a_2\}$ in this expression specify the new causal frame. Note that this expression assumes that the system is in the regime of applicability of hydrodynamics since  it ignores second (and higher) order derivatives. Therefore, to change  from frame $\alpha$ to frame $\beta$, the constants in \eqref{changeframeconformal} are $\{a_1^{\alpha}-a_1^{\beta},a_2^{\alpha}-a_2^{\beta}\}$, with the quantities on the right hand side  in frame $\alpha$ and the ones on the left hand side in frame $\beta$. In particular, the change of frame from the Landau frame to the causal frame $\{a_1,a_2\}$ is given by expression \eqref{changeframeconformal} with the constants given by $\{-a_1,-a_2\}$ and the quantities on the right hand side in the Landau frame. One could consider other procedures to change frames at the level of the initial data and they would differ from the one described here by second and higher order gradients. Of course, if the system is not in the hydrodynamic regime, then different procedures will result in significantly different initial conditions.

Recall that to solve the first order viscous hydrodynamics equations we need to provide initial data for the time derivatives of the fundamental variables $\{\partial_t \epsilon,\partial_t u_x\}$. We now comment on possible ways to consistently compute these time derivatives when changing the frame of the initial data. First, note that the time derivatives are of first order in gradients and hence they should be the same in all frames, up to higher order terms that should be ignored. In practice, if we happen to know the thermodynamic variables over a time interval\footnote{This would be the case if the experimental data is given as a time series, or we have computed the solution using some other scheme such as MIS.} then we can change frame in the entire interval and compute the time derivatives in the new frame variables. This is what we will do in the examples below. Alternatively, if one only has data at $t=0$, one can compute the time derivative of expression \eqref{changeframeconformal} and either ignore the second time derivatives or impose the equations of motion at $t=0$ in the original frame to compute them.
\newline

\begin{figure}[thbp]
	{\includegraphics[width=0.495\textwidth]{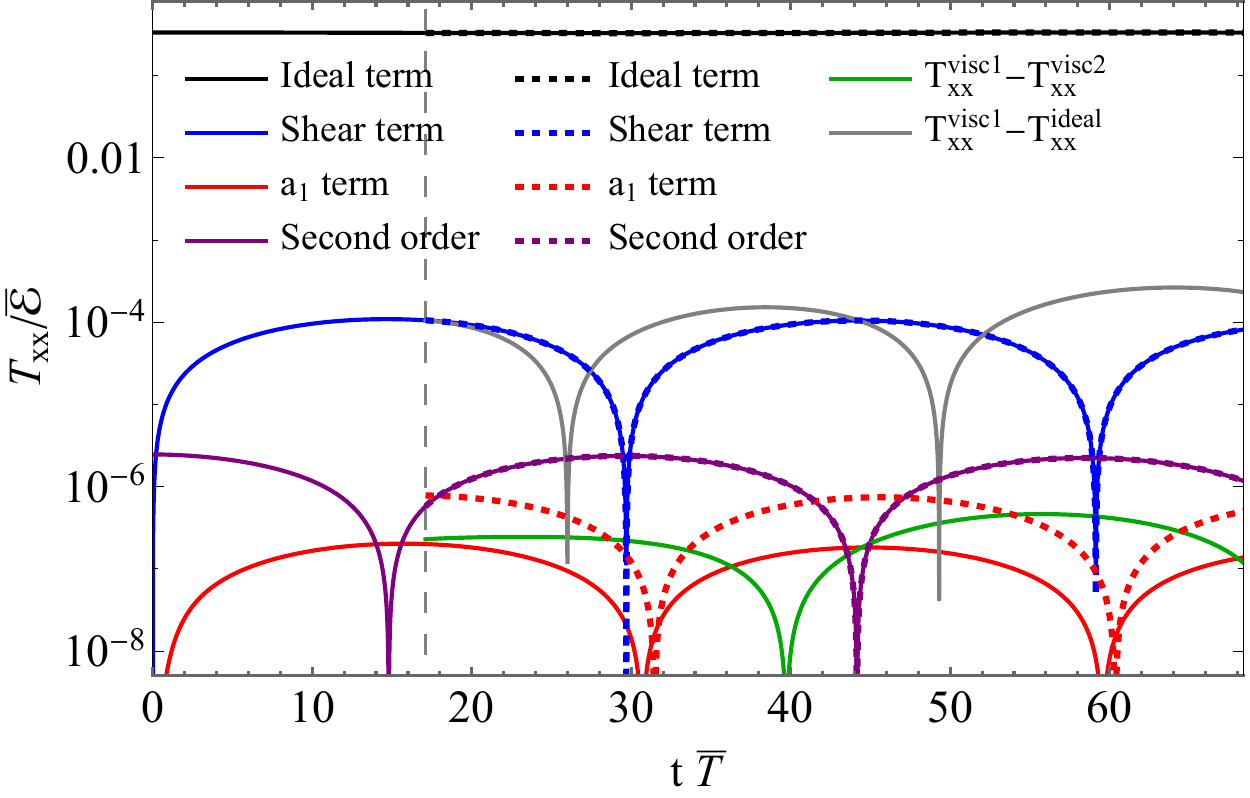}} 	{\includegraphics[width=0.495\textwidth]{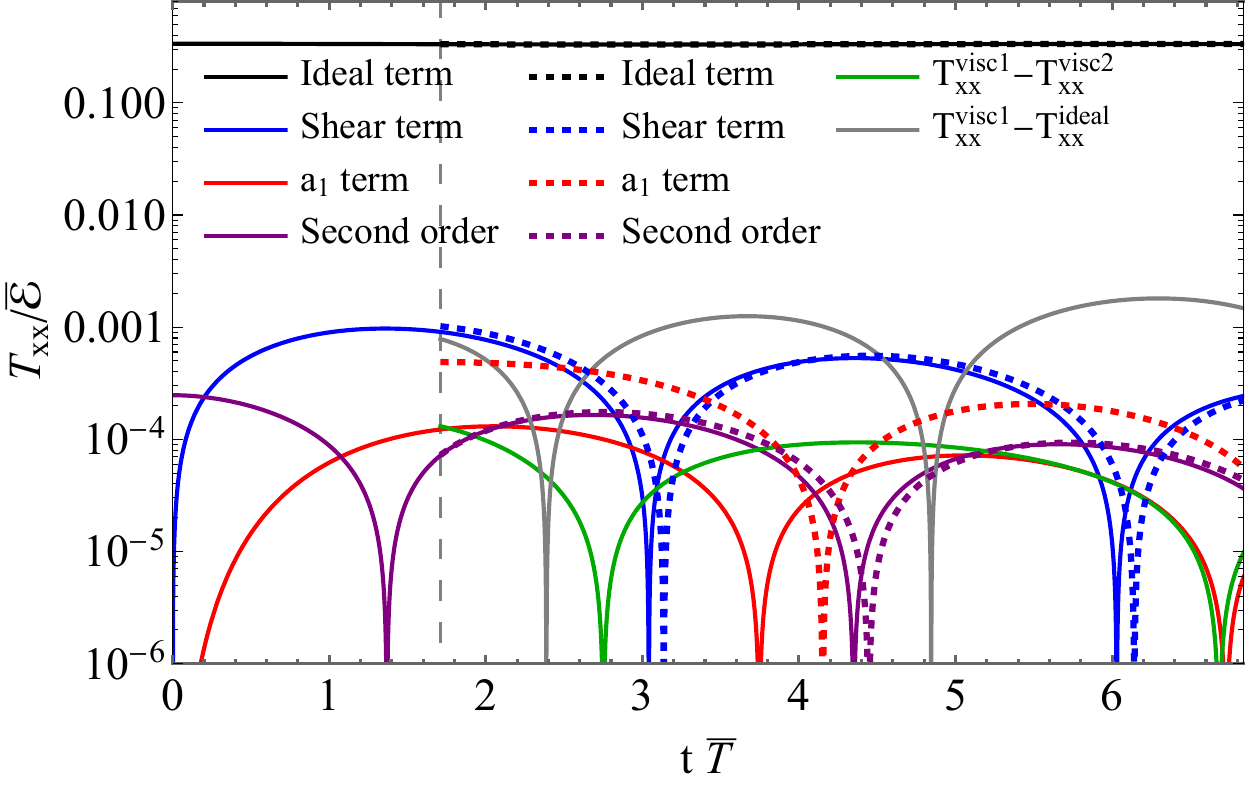}}
	\caption{ (Left) Solid lines correspond to the same evolution as in Fig. \ref{Diference_Ttt_of_two_runs_a1a2_5_a1a2_10} (left) in frame $\{a_1,a_2\}=\{5,5\}$. We include in dashed lines and same color coding the result of using the data at $t \overline{T} \simeq 17.1$ (time indicated by a vertical dashed line) to perform a change of frame to $\{a_1,a_2\}=\{10,10\}$ and use it as initial data for an evolution in frame $\{a_1,a_2\}=\{10,10\}$. We show in solid green the difference $T^{\text{visc1}}_{xx}-T^{\text{visc2}}_{xx}$. Moreover, we perform an evolution of the ideal hydrodynamics equations and we show the difference $T^{\text{visc1}}_{xx}-T^{\text{ideal}}_{xx}$ in solid grey.
    (Right) Similar exercise as in (left) for the evolution with momentum ten times larger $k/\overline{T}\simeq 1.84$, to analyse a situation which is marginally in the regime of hydrodynamics. The main conclusion from these plots is that changing frame in the initial data according to the prescription of effective field theory introduces a small difference comparable to the difference in the evolutions in different frames analysed in Fig. 	\ref{Diference_Ttt_of_two_runs_a1a2_5_a1a2_10}.}
	\label{Comparison_ideal_evolution}
\end{figure}

In order to explore the effect of changing frame in the initial data we consider two different exercises. 
First, we consider a numerical solution in a given frame (see Fig. \ref{Diference_Ttt_of_two_runs_a1a2_5_a1a2_10}) at $t>0$, change frames and thus obtain initial data to evolve the equations in the new frame.
For concreteness consider the solution in  Fig. \ref{sinusoidal_evolution_to_show_size_a1a2}, which is in frame $\{a_1,a_2\}=\{5,5\}$, at time $t \overline{T}\simeq 17.1$. We change to frame $\{a_1,a_2\}=\{10,10\}$ using \eqref{changeframeconformal} and use this as initial data for an evolution in the new frame. In Fig. \ref{Comparison_ideal_evolution} (left) we show the original evolution in frame $\{a_1,a_2\}=\{5,5\}$ at $x=0$, in solid lines, and the new evolution in frame $\{a_1,a_2\}=\{10,10\}$, in dashed lines; the difference $T^{\text{visc1}}_{xx}-T^{\text{visc2}}_{xx}$ is depicted in solid green. 
To carry out these simulations we computed the time derivatives for the initial data in the new causal frame $\{a_1,a_2\}=\{10,10\}$; using instead the time derivatives computed in the original frame does not make a noticeable difference, as one might have expected.
For comparison, we have also performed an evolution of the ideal hydrodynamics equations using initial data at $t \overline{T}\simeq 17.1$ from the solution in frame $\{a_1,a_2\}=\{5,5\}$ (using data in the frame $\{a_1,a_2\}=\{10,10\}$ does not make a difference).  Notice that although  the ideal hydrodynamics equations are not exactly satisfied at $t \overline{T}\simeq 17.1$, the error terms are small,
as they should be since the system is well in the regime of hydrodynamics by construction. The difference $T^{\text{visc1}}_{xx}-T^{\text{ideal}}_{xx}$ in is shown in solid grey. 

From these results we conclude that 
the difference $T^{\text{visc1}}_{xx}-T^{\text{visc2}}_{xx}$ in the two evolutions in different frames due to the change of frame in the initial data, see Fig. \ref{Comparison_ideal_evolution} (left), is  comparable to the difference introduced solely upon time evolution in different frames studied in Fig. \ref{Diference_Ttt_of_two_runs_a1a2_5_a1a2_10} (left) where, recall, the change of frame was vanishing at $t=0$.
Moreover, we conclude that criteria A, B and C are satisfied at a similar level in this situation. 
We repeated this exercise for initial data at some other times $t \overline{T}\simeq  8.5, 25.6$, obtaining similar conclusions.
\newline

Motivated by the physics of the QGP,
we now study a case which is marginally in the regime of hydrodynamics by performing a 
similar exercise as the previous one but with momentum ten times larger, $k/\overline{T}\simeq 1.84$. In Fig. \ref{Comparison_ideal_evolution} (right) we show the result of using initial data at time $t \overline{T}\simeq 1.71$.
Again, we conclude that
the change of frame in the initial data introduces a difference $T^{\text{visc1}}_{xx}-T^{\text{visc2}}_{xx}$ which is comparable to the difference generated solely upon time evolution studied in the example in Fig. \ref{Diference_Ttt_of_two_runs_a1a2_5_a1a2_10} (right). 
We find that Criteria A and B are violated at a similar level as in the situation in Fig. \ref{Diference_Ttt_of_two_runs_a1a2_5_a1a2_10} (right), and Criterion C is still satisfied at a similar level. 
We repeated the same exercise at times $t \overline{T}\simeq  0.85, 2.56$ obtaining similar conclusions. 

Thus, we can conclude that the change of frame performed in this way in the initial data does not change the picture regarding the applicability of Criteria A, B and C in these solutions: they are satisfied or violated at a similar level as in the situation of Fig. \ref{Diference_Ttt_of_two_runs_a1a2_5_a1a2_10}.
\newline

We present now our second example. Recall that the initial data \eqref{Initial_data_sinusoidal} was chosen so that the ideal hydrodynamics equations are identically satisfied (and hence effects of changing frames are identically zero). Here we want to study a situation where initially the equations of ideal hydrodynamics are far from being satisfied. To do so, we assume that initial data is in the Landau frame and modify \eqref{Initial_data_sinusoidal} by imposing $\partial_t u_x|_{t=0}=0$, that is: 
\begin{subequations}
	\begin{align}
		\epsilon |_{t=0}&= \overline{\mathcal{E}} \left( 1+ 0.01 \cos(k x) \right) \, \,, \label{Initial_data_sinusoidala2}\\
		\partial_t \epsilon|_{t=0}&= 0\,\,, \label{Initial_data_sinusoidalb2}\\
		u_x|_{t=0}&= 0 \, \,, \label{Initial_data_sinusoidalc2}\\
		\partial_t	u_x|_{t=0}&= 0\, \,. \label{Initial_data_sinusoidald2}
	\end{align}
	\label{Initial_data_sinusoidal2}
\end{subequations}
Changing frames to our working causal frame $\{a_1,a_2\}$ using the effective field theory prescription \eqref{changeframeconformal} gives the following initial data:
\begin{subequations}
	\begin{align}
		\epsilon |_{t=0}&= \overline{\mathcal{E}} \left( 1+ 0.01 \cos(k x) \right) \, \,, \label{Initial_data_sinusoidala_change_of_frame}\\
		\partial_t \epsilon|_{t=0}&= 0\,\,, \label{Initial_data_sinusoidalb_change_of_frame}\\
		u_x|_{t=0}&=\frac{3}{16} a_2 \eta \frac{0.01 k \sin(kx)}{\overline{\mathcal{E}}\left(1+0.01 \cos(kx)\right)^2} \, \,, \label{Initial_data_sinusoidalc_change_of_frame}\\
		\partial_t	u_x|_{t=0}&= 0\, \,. \label{Initial_data_sinusoidald_change_of_frame}
	\end{align}
	\label{Initial_data_sinusoidal2_chage_frame}
\end{subequations}

One can consider the time derivatives in \eqref{Initial_data_sinusoidalb_change_of_frame} and \eqref{Initial_data_sinusoidald_change_of_frame} in the Landau frame or in the causal frame, as the difference is of higher order; here  we have chosen to keep them in the Landau frame. %
The fact that the time derivatives at $t=0$ in \eqref{Initial_data_sinusoidal2_chage_frame} (or in \eqref{Initial_data_sinusoidal2}) vanish, necessarily imply that  the ideal equations are not satisfied and the error terms are not small.
This might not correspond to a physically relevant situation, where the system should be in (or close to) the regime of applicability of hydrodynamics, but it allows to assess the effect of the  ideal equations not being satisfied in the initial data in a situation in which this effect is large.

\begin{figure}[thbp]
	{\includegraphics[width=0.495\textwidth]{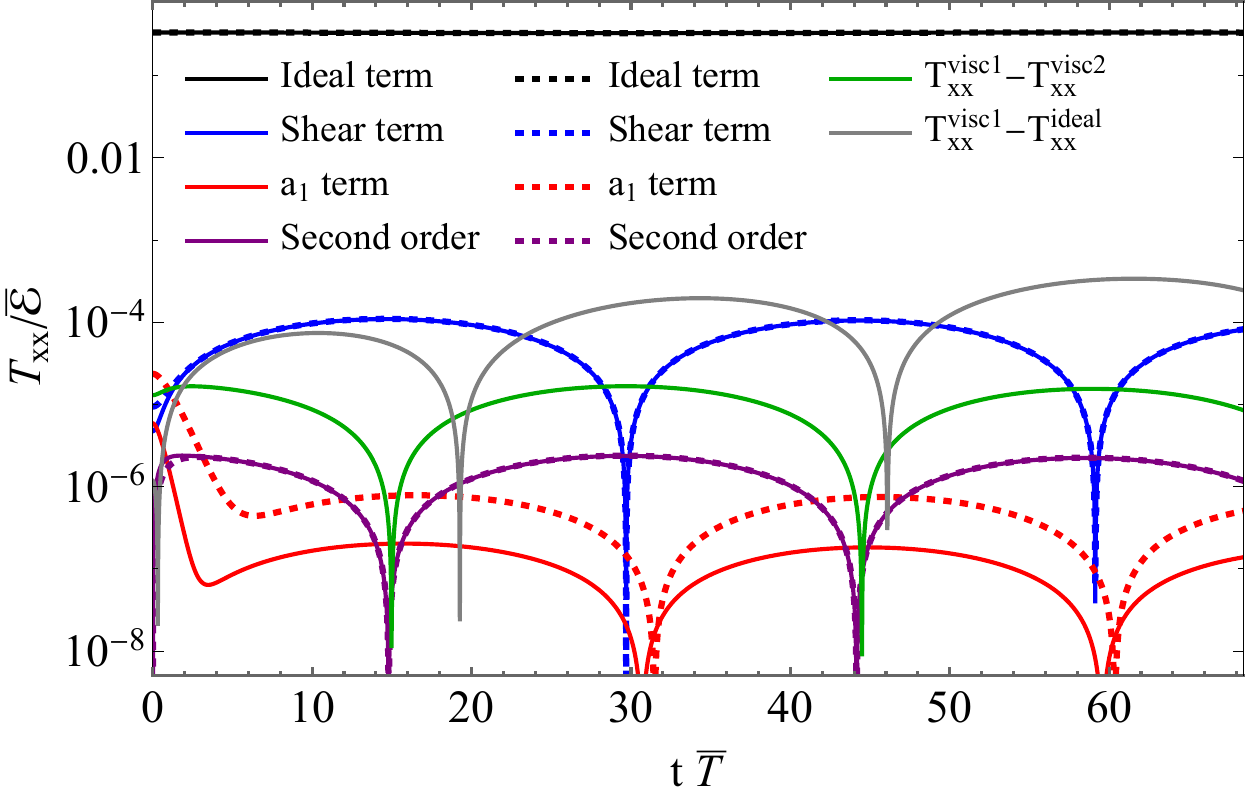}} 
	{\includegraphics[width=0.495\textwidth]{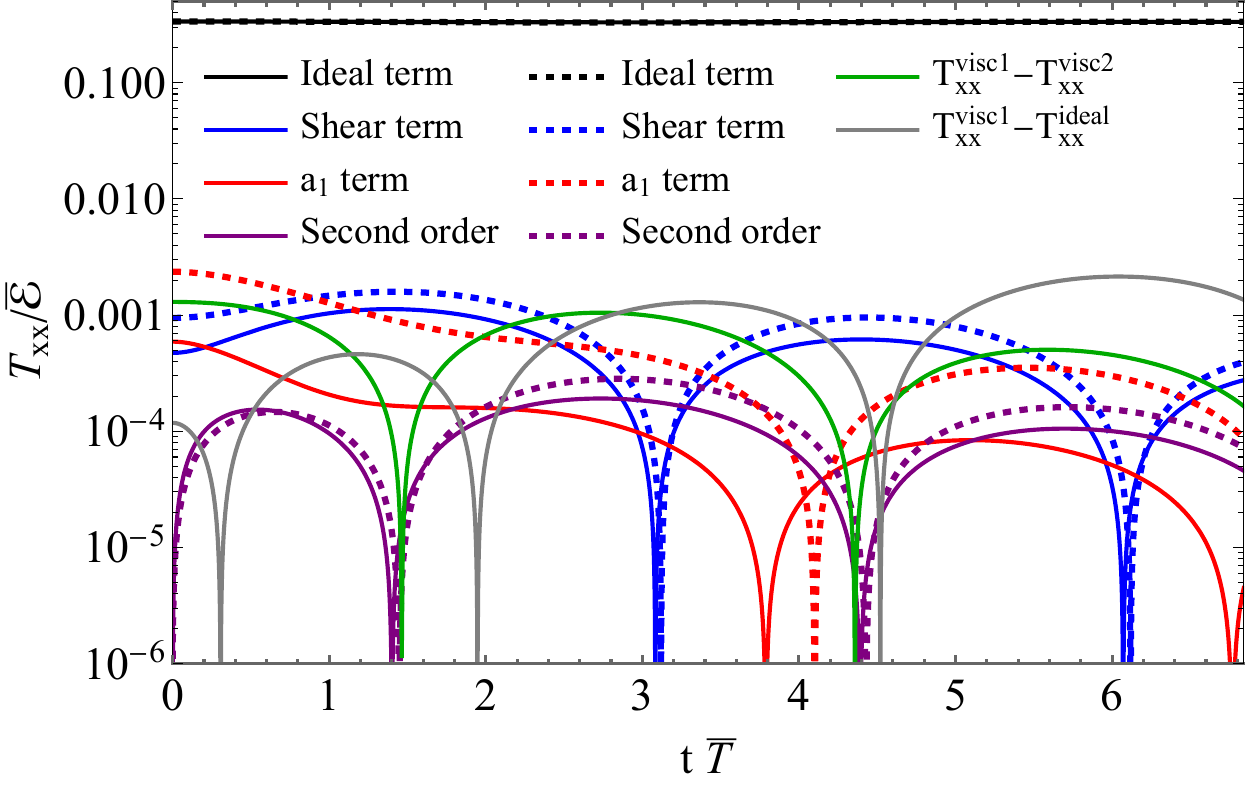}} 
	\caption{(Left) Evolutions with initial data \eqref{Initial_data_sinusoidal2_chage_frame} and momentum $k/\overline{T} \simeq 0.184$ in frames $\{a_1,a_2\}=\{5,5\}$, continuous lines, and  $\{a_1,a_2\}=\{10,10\}$, dashed lines, at $x=0$. We show the difference $T^{\text{visc1}}_{xx}-T^{\text{visc2}}_{xx}$ in solid green. We evolve the ideal equations with initial data \eqref{Initial_data_sinusoidala2},\eqref{Initial_data_sinusoidalc2} and show the result $T^{\text{visc1}}_{xx}-T^{\text{ideal}}_{xx}$ in solid grey.
		(Right) Same exercise as in (left) but with momentum ten times larger $k/\overline{T} \simeq 1.84$. 
		The conclusion is that changing frame in the initial data introduces a difference $T^{\text{visc1}}_{xx}-T^{\text{visc2}}_{xx}$ that is sustained along the evolution.
	}
	\label{Changing_initial_data}
\end{figure}

Let us consider initial data \eqref{Initial_data_sinusoidal2_chage_frame} in frame $\{a_1,a_2\}=\{5,5\}$; the corresponding numerical evolution is shown in solid lines in Fig. \ref{Changing_initial_data} (left). The same exercise in frame $\{a_1,a_2\}=\{10,10\}$ is shown in dashed lines, and the difference $T^{\text{visc1}}_{xx}-T^{\text{visc2}}_{xx}$ is depicted in solid green. In addition we compute the solution of the ideal hydrodynamics equations with initial data \eqref{Initial_data_sinusoidala2},\eqref{Initial_data_sinusoidalc2} and the corresponding difference $T^{\text{visc1}}_{xx}-T^{\text{ideal}}_{xx}$ is shown in solid grey.\footnote{For the initial data of the ideal equations we have three natural posibilities: use the data in the Landau frame \eqref{Initial_data_sinusoidala2},\eqref{Initial_data_sinusoidalc2} or in any of the two causal frames \eqref{Initial_data_sinusoidala_change_of_frame},\eqref{Initial_data_sinusoidalc_change_of_frame}. We computed the three cases to check that the difference does not change the conclusions, even if it is not very small. In Fig. \ref{Changing_initial_data} we plot only the first case.}

The difference $T^{\text{visc1}}_{xx}-T^{\text{visc2}}_{xx}$ introduced by the initial data \eqref{Initial_data_sinusoidalc_change_of_frame} at $x=0$ can be seen in Fig. \ref{Changing_initial_data} (left), where the green line intersects the vertical axis. 
Upon time evolution the amplitude of the difference $T^{\text{visc1}}_{xx}-T^{\text{visc2}}_{xx}$ does not vary considerably. 
If we compare this example to the one in Fig. \ref{Diference_Ttt_of_two_runs_a1a2_5_a1a2_10} (left) we find that the initial data for which the ideal hydrodynamics equations are far from being satisfied initially, introduces a difference which is much larger than the difference $T^{\text{visc1}}_{xx}-T^{\text{visc2}}_{xx}$ generated solely via time evolution (with vanishing initial change of frame) by more than one order of magnitude. 
On the other hand, in the example in Fig. \ref{Comparison_ideal_evolution} (left), in which the initial is such that the ideal hydrodynamics equations are nearly satisfied, the difference in the initial data is  of the same order as the difference generated upon time evolution, see Fig. \ref{Diference_Ttt_of_two_runs_a1a2_5_a1a2_10} (left).

As we are using initial data for which the ideal hydrodynamics equations are far from being satisfied one may suspect that at initial times Criterion A is violated.  However, Criterion A is satisfied at all times and in particular at $t=0$. This is not just the effect of integrating over the spatial domain and one period in time (which is the characteristic scale of the system), but even locally in time the amplitudes of ideal, first order and second order terms are each smaller than $10\%$ the previous one, even at initial times.
We find curious that in spite of the ideal equations are largely violated at $t=0$, even initially there still is a hierarchy in the gradient expansion.

Criterion B is satisfied at all times. In this case we observe that at initial times, and for a timescale much smaller than the period, the amplitude of the $a_1$ term is even larger than the shear term, but it quickly decays. In spite of this initial transient, integrating over a period in time the conditions of Criterion B are satisfied.\footnote{This example suggest that it could be interesting to extend the definition of our criteria to consider integration in different scales. For example, if we integrated in scales ten times smaller than the period, criterion B would be violated at initial times.}  
The fact that the $a_1$ term is large at $t=0$ is a consequence of the choice of the initial data.
Criterion C is satisfied at all times. However, if we compare locally in time the amplitudes of the difference  $T^{\text{visc1}}_{xx}-T^{\text{visc2}}_{xx}$ 
with $T^{\text{visc1}}_{xx}-T^{\text{ideal}}_{xx}$, 
we find that initially it is not smaller than a $10\%$ and this is restored at times $t\overline{T}\simeq 4$. 
\newline

In Fig. \ref{Changing_initial_data} (right) we repeat the previous exercise using similar initial data \eqref{Initial_data_sinusoidal2_chage_frame} but with momentum ten times larger, $k/\overline{T} \simeq 1.84$, to consider a situation that is marginally in the regime of hydrodynamics.
We find that the difference $T^{\text{visc1}}_{xx}-T^{\text{visc2}}_{xx}$ along the evolution due to change of frame in the initial data is considerably larger than the difference introduced solely upon time evolution (that is, with vanishing change of frame at $t=0$) studied in Fig.	\ref{Diference_Ttt_of_two_runs_a1a2_5_a1a2_10} (right).
Due to this larger difference, in this case all criteria A, B and C are now violated for the times shown in the plot.

\subsection{Large  amplitude Gaussian perturbation}
\label{Gaussian_section}
In the previous subsection we studied solutions that were in the linear regime by construction,  even if we solved the full non-linear evolution equations. We now extend these studies to solutions that are well in the non-linear regime. 
For this purpose we consider a thermal homogeneous state with a large Gaussian perturbation of the energy density as initial data:
\begin{subequations}
	\begin{align}
		\epsilon |_{t=0}&= \overline{\mathcal{E}} \left( 1+ e^{-\frac{x^2}{2\sigma^2}} \right) \, \,, \label{Initial_data_gaussiana0}\\
		\partial_t \epsilon|_{t=0}&= 0\,\,, \label{Initial_data_gaussianb0}\\
		u_x|_{t=0}&= 0 \, \,, \label{Initial_data_gaussianc0}\\
		\partial_t	u_x|_{t=0}&=\frac{x e^{-\frac{x^2}{2\sigma^2}}}{4 \sigma^2 \, \left( 1+ e^{-\frac{x^2}{2\sigma^2}} \right)}  \, \,. \label{Initial_data_gaussiand0}
	\end{align}
	\label{Initial_data_gaussian}
\end{subequations}
By tuning the width  $\sigma$  of the Gaussian we can have a system that is well in the regime of hydrodynamics (large $\sigma$, i.e., small gradients), or a system that is far from the regime of hydrodynamics (small $\sigma$, i.e., large gradients).
As in \eqref{Initial_data_sinusoidal}, the time derivatives at $t=0$, eqs. \eqref{Initial_data_gaussianc0} and \eqref{Initial_data_gaussiand0}, are chosen such that the ideal hydrodynamics equations are initially satisfied.
As in previous subsection, $\overline{T}$ and $\overline{{\mathcal{E}}}$ are the temperature and energy density of the thermal state, related by $\overline{\mathcal{E}}=\frac{3}{4} \pi^4\,\overline{T}^{4} $.

\begin{figure}[thbp]
	{\includegraphics[width=0.495\textwidth]{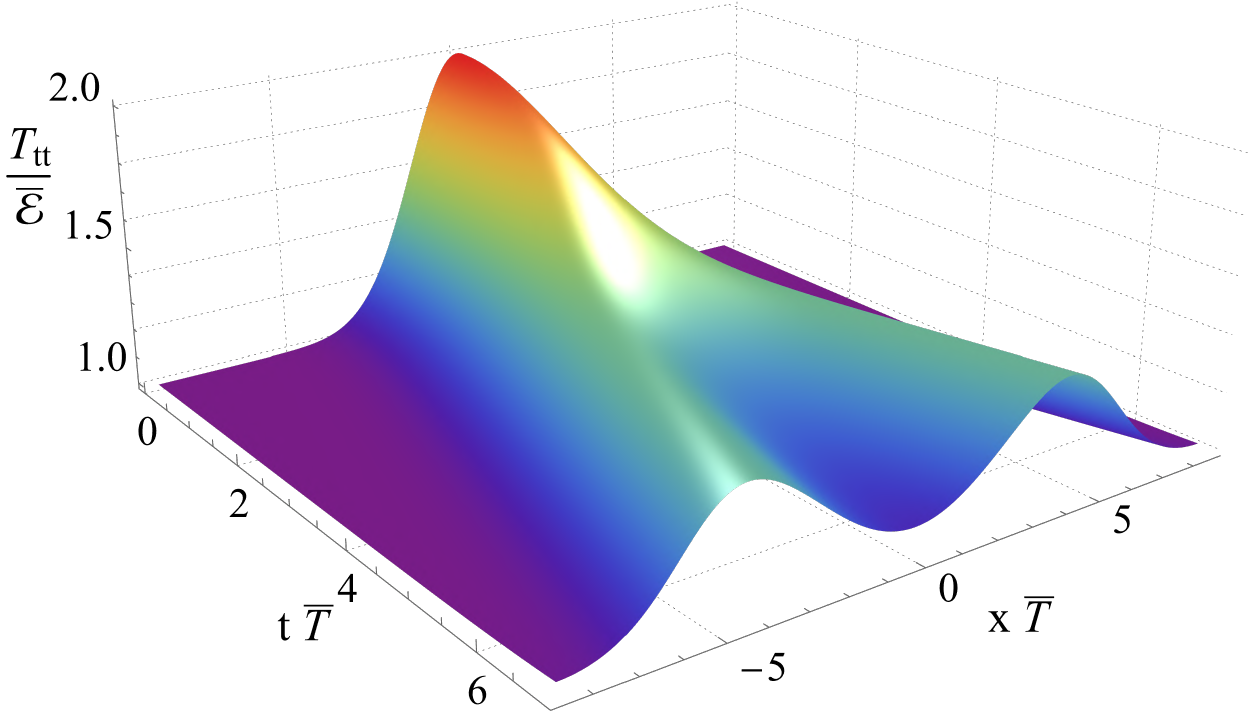}} 
	{\includegraphics[width=0.495\textwidth]{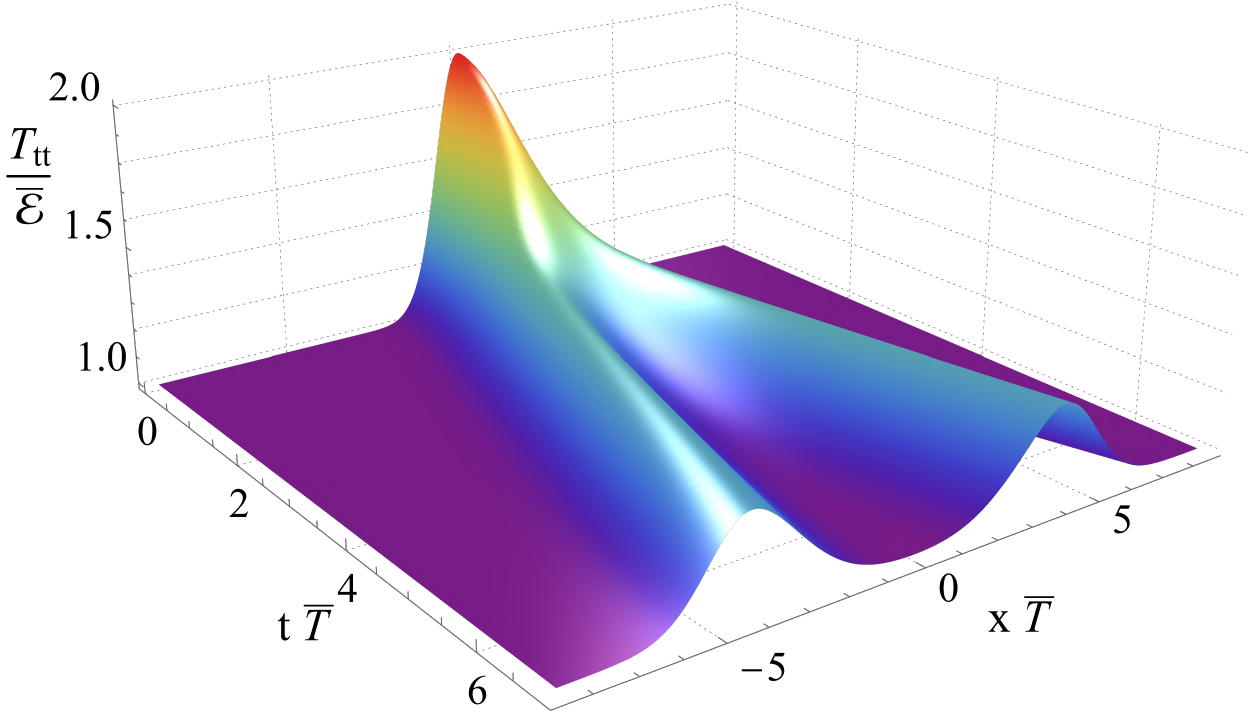}} 
	{\includegraphics[width=0.495\textwidth]{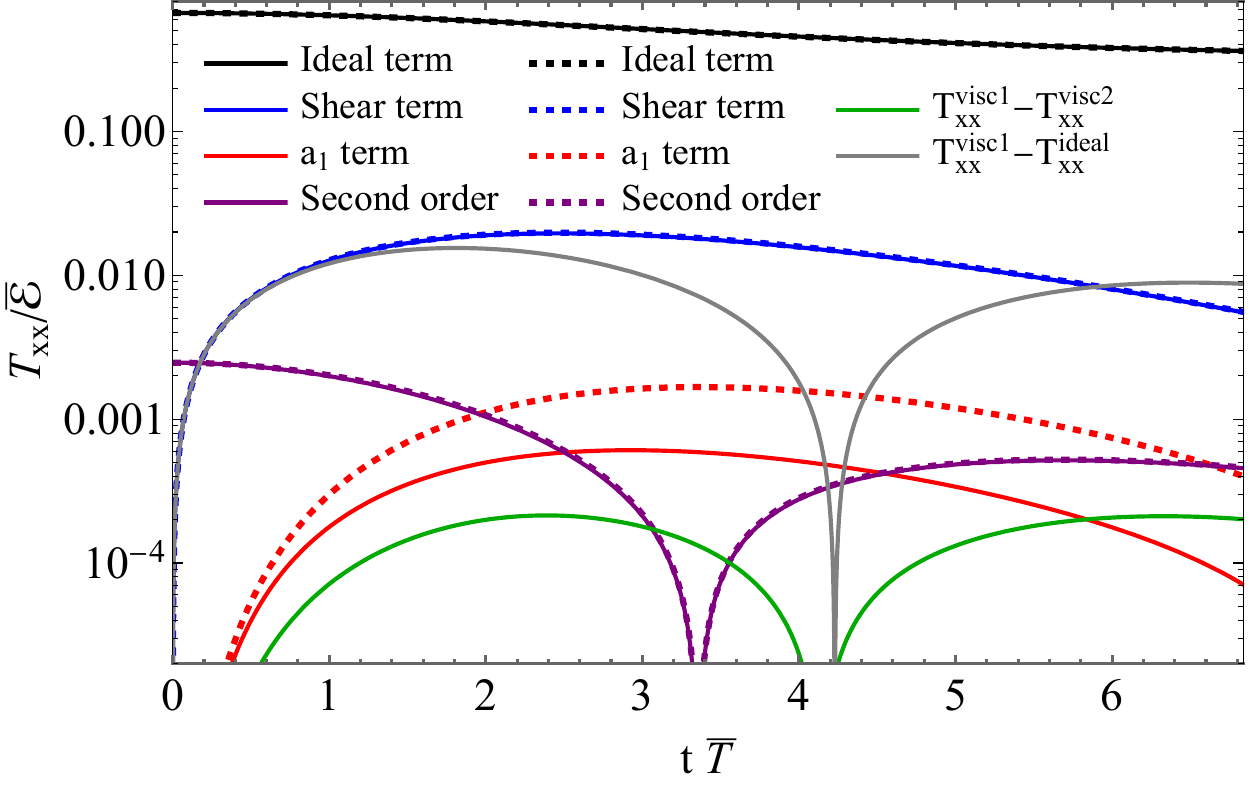}} 
	{\includegraphics[width=0.495\textwidth]{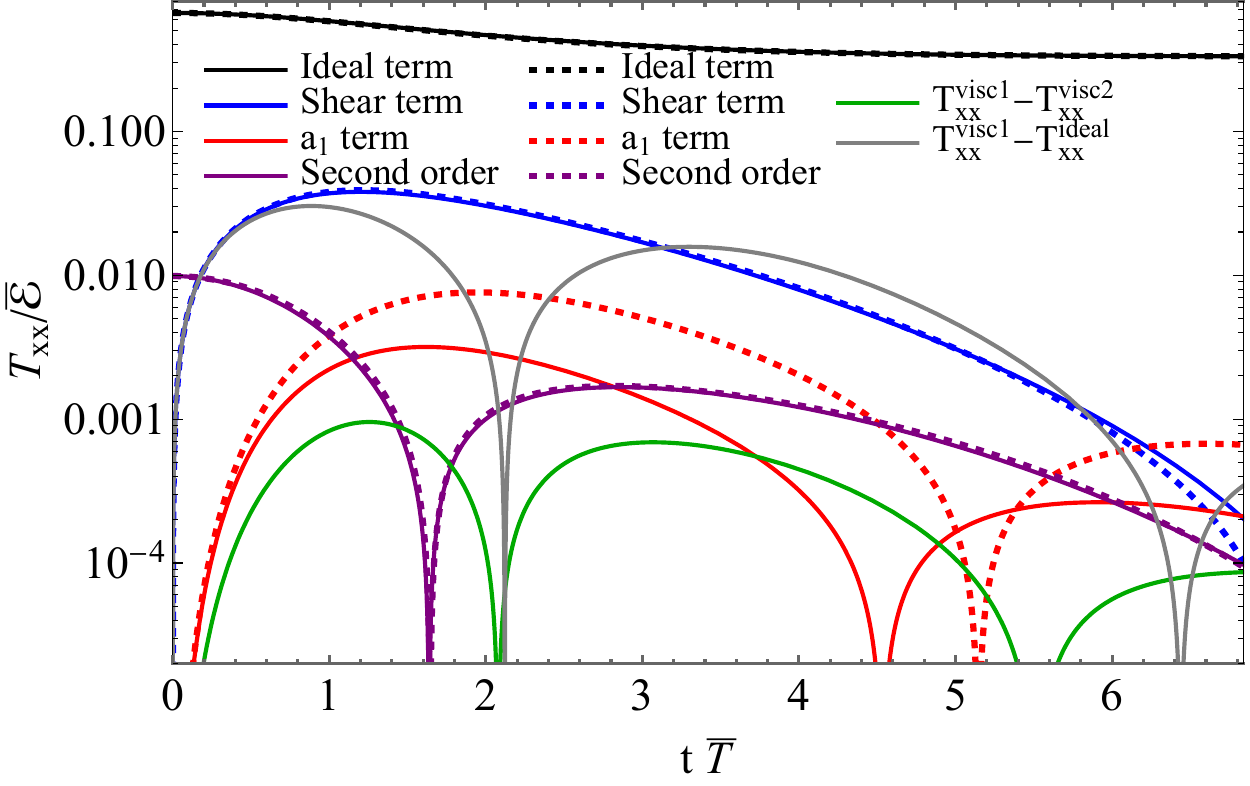}} 
	\caption{(top, left): Spacetime evolution of the $T_{tt}$ component of the stress tensor with initial data of a gaussian, large amplitude perturbation of a homogeneous thermal state \eqref{Initial_data_gaussian}, width $\sigma \overline{T}\simeq 1.45 $ and frame $\{a_1,a_2\}=\{5,5\}$. 
	In (bottom, left) we show the log plot of the absolute value of the different contributions to the $T_{xx}$ component of the stress tensor \eqref{eq:tmunu11} at $x=0$ as a function of time: ideal term in solid black, shear term in solid blue, $a_1$ term in solid red ($a_2$ term is vanishing at $x=0$ by symmetry).
We include the result of evaluating the second order terms \eqref{constitutive0sheartensor0} in solid purple. 
We also include the data for another run, with similar initial data, performed in a different frame $\{a_1,a_2\}=\{10,10\}$, in dashed lines.
We plot the difference $T^{\text{visc1}}_{xx}-T^{\text{visc2}}_{xx}$ in solid green and the difference $T^{\text{visc1}}_{xx}-T^{\text{ideal}}_{xx}$ in solid grey. See Fig. 	\ref{3Dplots_gaussian_criteria} for plots in all spacetime domain.
In the plots on the right we present the result of performing the same exercise but with smaller width $\sigma \overline{T}\simeq 0.73$ in the initial data, and it corresponds to a system marginally in the regime of hydrodynamics.}
	\label{gaussain_evolution_to_show_size_a1a2}
\end{figure}

We first consider the evolution of the initial data \eqref{Initial_data_gaussian} with width $\sigma \overline{T}\simeq1.45$  in the frame $\{a_1,a_2\}=\{5,5\}$. This width is of the order of the microscopic scale and hence we expect that the system is on the verge of the regime of applicability of hydrodynamics. In Fig. \ref{gaussain_evolution_to_show_size_a1a2} (top, left) we show the evolution of $T_{tt}$ component of the stress tensor.

\begin{figure}[thbp]
	{\includegraphics[width=0.495\textwidth]{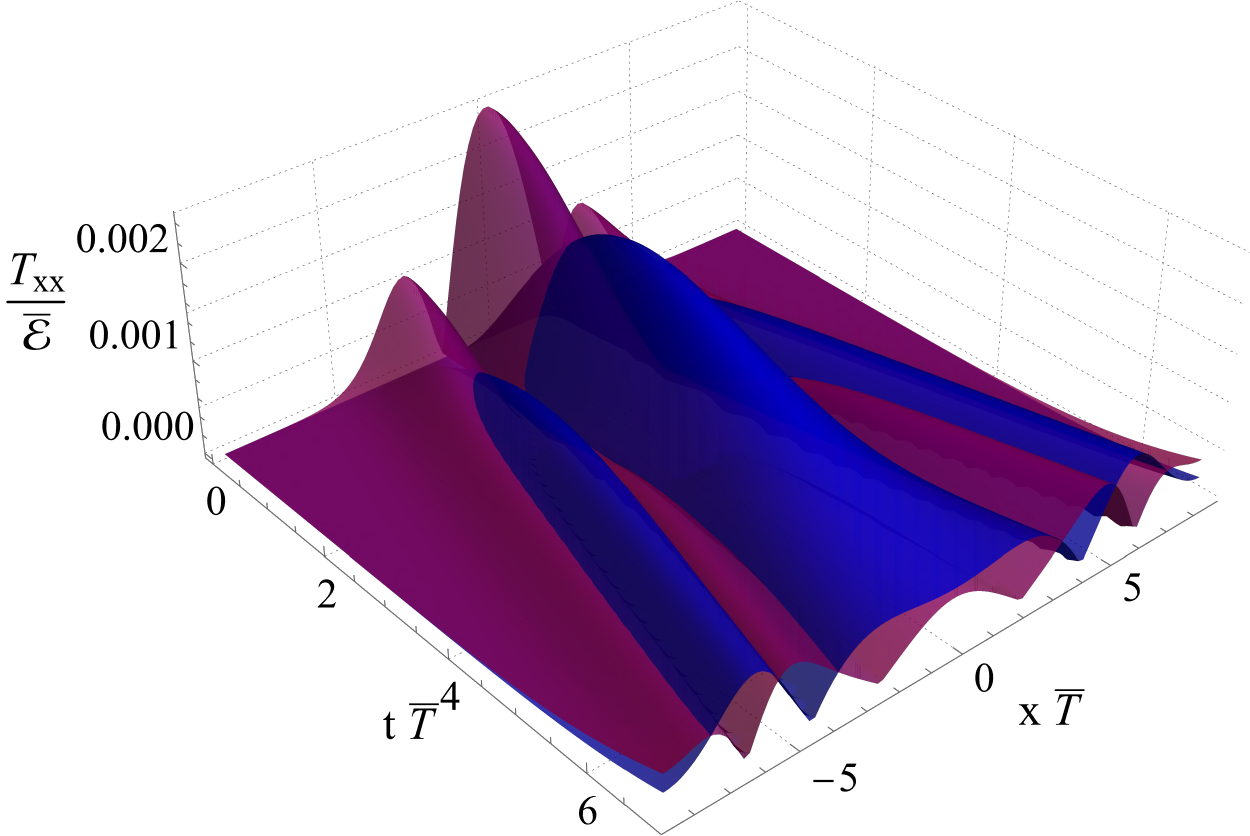}} 
	\put (-29,138) {\large $\displaystyle Criterion$ $\displaystyle A$}
	\put (-24,129) {$\colorbox{blue}{\rule{0pt}{1pt}\rule{1pt}{0pt}}$}
	\put (-14,126.5) {\footnotesize $\textbf{10\%}$ \footnotesize Shear}
	\put (-24,119) {$\colorbox{violet}{\rule{0pt}{1pt}\rule{1pt}{0pt}}$}
	\put (-14,118.5) {\footnotesize Second order}
	{\includegraphics[width=0.495\textwidth]{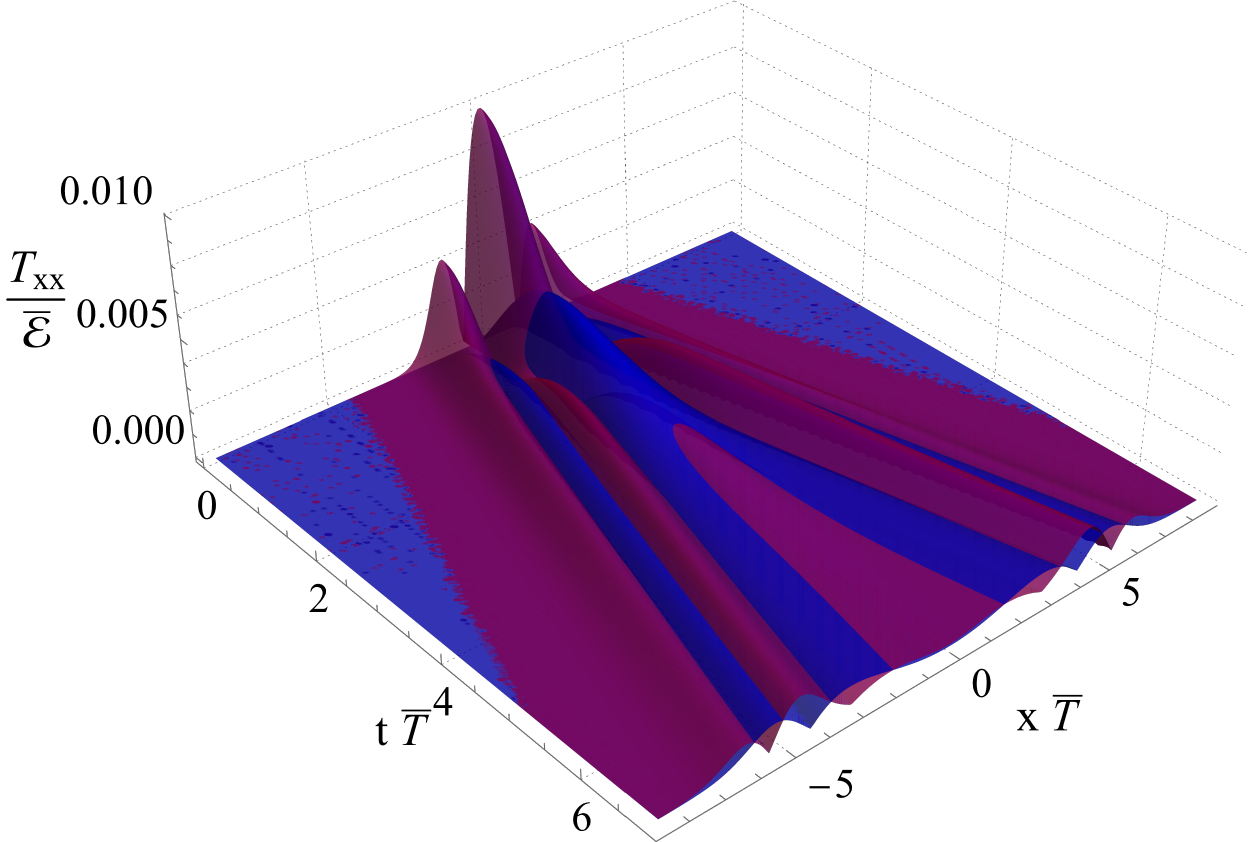}} 
	{\includegraphics[width=0.495\textwidth]{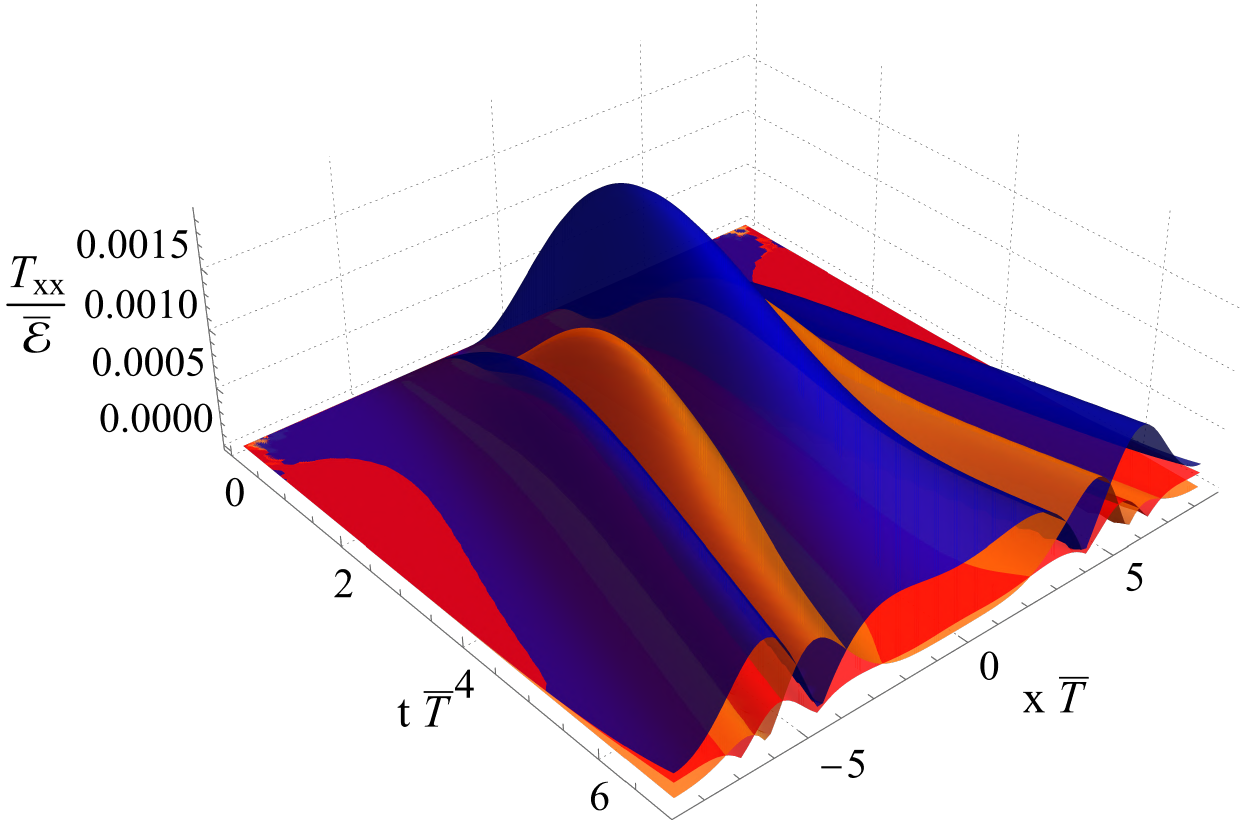}}
	\put (-25,139) {\large $\displaystyle Criterion$ $\displaystyle B$}
	\put (-18,130) {$\colorbox{blue}{\rule{0pt}{1pt}\rule{1pt}{0pt}}$}
	\put (-8,127.5) {\footnotesize $\textbf{10\%}$ \footnotesize Shear}
	\put (-18,120) {$\colorbox{red}{\rule{0pt}{1pt}\rule{1pt}{0pt}}$}
	\put (-8,118.5) {\footnotesize $a_1$ term}
	\put (-18,110) {$\colorbox{orange}{\rule{0pt}{1pt}\rule{1pt}{0pt}}$}
	\put (-8,108.5) {\footnotesize $a_2$ term}
	{\includegraphics[width=0.495\textwidth]{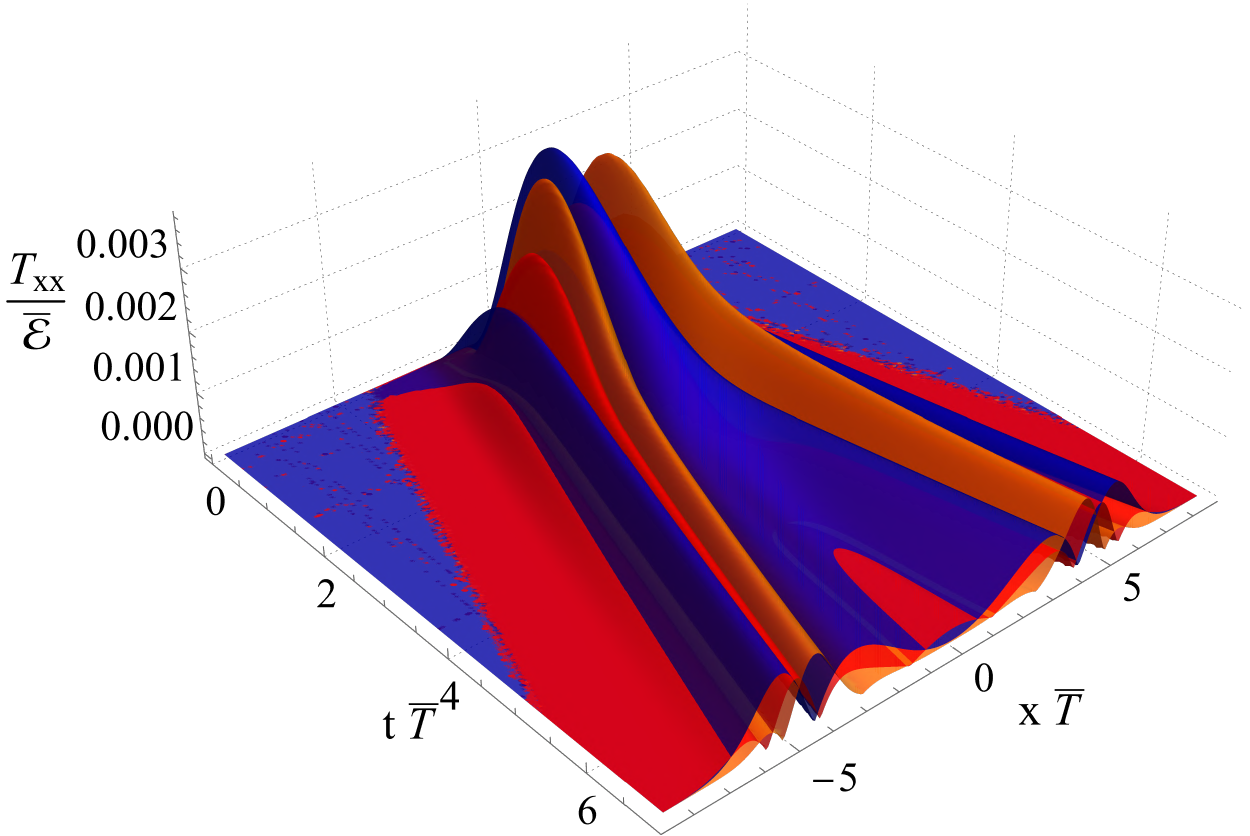}}
	{\includegraphics[width=0.495\textwidth]{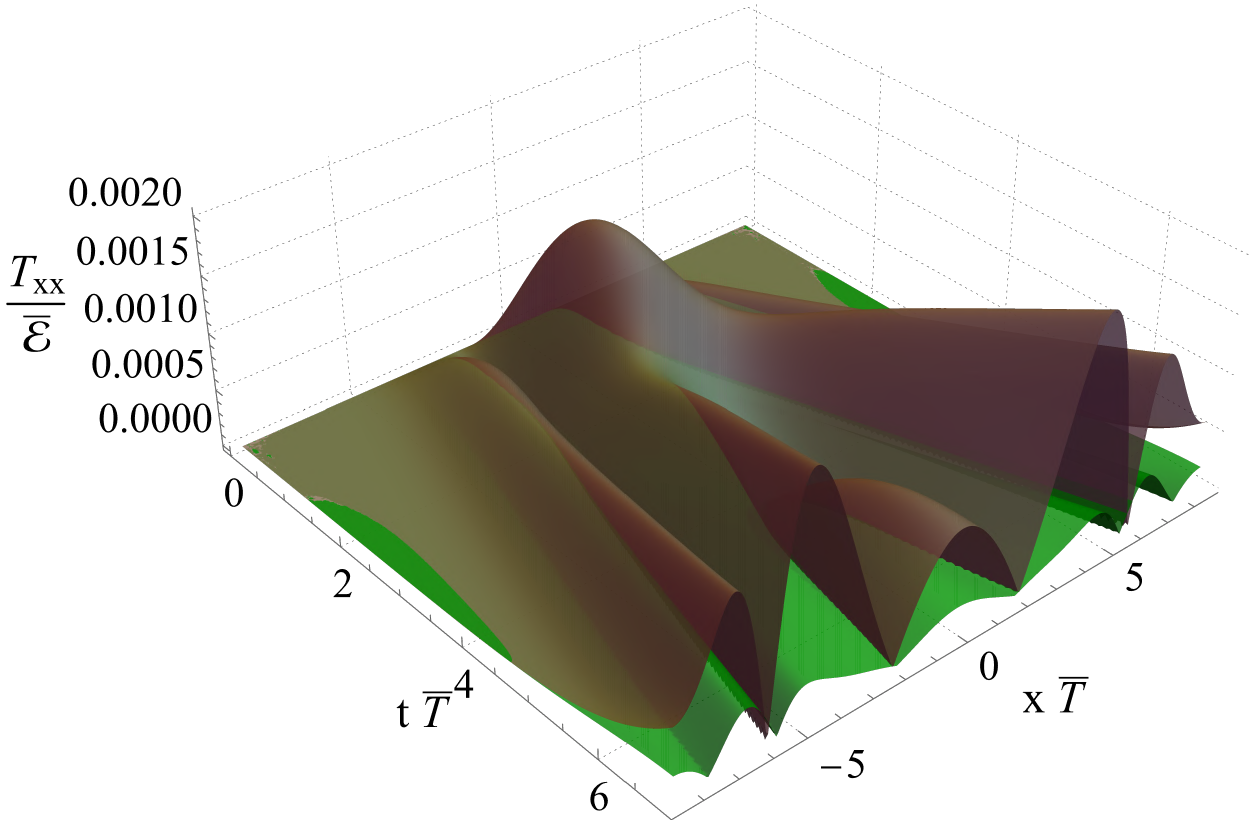}}
	\put (-24,140) {\large $\displaystyle Criterion$ $\displaystyle C$}
	\put (-32,131) {$\colorbox{gray}{\rule{0pt}{1pt}\rule{1pt}{0pt}}$}
	\put (-22,128.5) {\footnotesize $\textbf{10\%}$ $\left(T^{\text{visc1}}_{xx}-T^{\text{ideal}}_{xx}\right)$}
	\put (-32,120) {$\colorbox{ao(english)}{\rule{0pt}{1pt}\rule{1pt}{0pt}}$}
	\put (-22,118.5) {\footnotesize  $T^{\text{visc1}}_{xx}-T^{\text{visc2}}_{xx}$}
	{\includegraphics[width=0.495\textwidth]{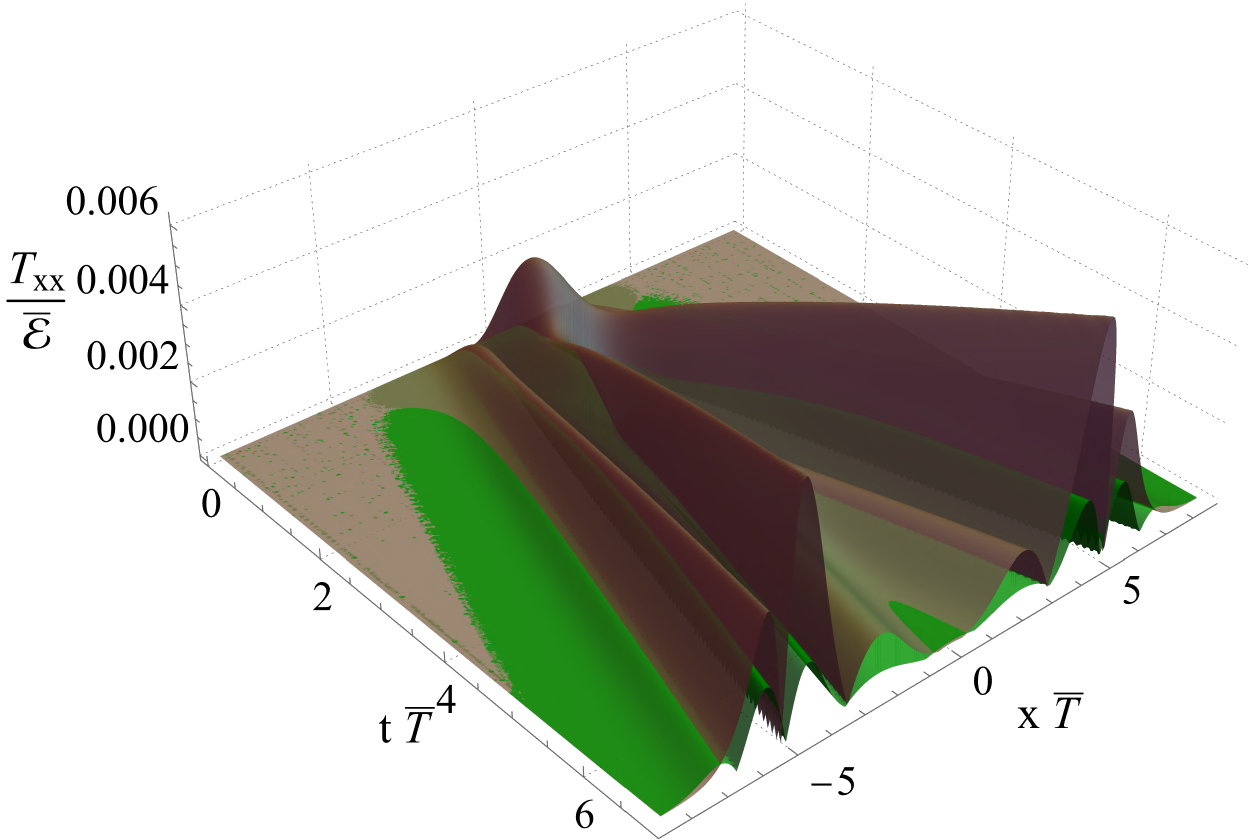}} 
	\caption{ For the evolutions presented in Fig. \ref{gaussain_evolution_to_show_size_a1a2} we show here relevant quantities for criteria A, B and C in the spacetime domain.
		In the top panels we plot a $10\%$ of the shear term in blue and second order terms in purple (ideal terms are much larger than shear, so we do not plot them). In middle panels we plot a $10\%$ of the shear term (blue) and $a_1$, $a_2$ terms (red, orange). In bottom panel we plot a $10\%$ of $T^{\text{visc1}}_{xx}-T^{\text{ideal}}_{xx}$ in solid grey and $T^{\text{visc1}}_{xx}-T^{\text{visc2}}_{xx}$ in solid green. 
		Left plots correspond to the case $\sigma \overline{T}\simeq 1.45$, and right plots to $\sigma \overline{T}\simeq 0.73$.
		All quantities are in absolute values. Top an middle panels correspond to evolutions in frame $\{a_1,a_2\}=\{5,5\}$. 
	}
	\label{3Dplots_gaussian_criteria}
\end{figure}

With the criteria defined in the previous subsection, we proceed with the analysis of the solution.
We start by considering Criterion A. We perform this comparison for the $T_{xx}$ component of the stress tensor \eqref{eq:tmunu11};  in Fig. \ref{gaussain_evolution_to_show_size_a1a2} (bottom, left) we depict this component at $x=0$,  and in Fig. \ref{3Dplots_gaussian_criteria} (top, left) we show the second order terms and $10\%$ of the shear term in $T_{xx}$ in the whole spacetime.
Note that since the initial data has vanishing velocity,  initially the shear term is zero but the presence of spatial gradients implies that the second order terms are non-vanishing. Upon time evolution the system quickly relaxes to the regime of hydrodynamics, with first order terms much smaller than ideal terms, and second order terms smaller than shear term, thus satisfying Criterion A. More precisely, we find that the $L_1$ norm of the second order terms is larger than $10\%$ of the $L_1$ norm of the shear term from $t \overline{T}=0$ to $t \overline{T}\simeq 1.57$. After this time, the $L_1$ norm of the second order terms is smaller than $10\%$ of the $L_1$ norm of the shear term. Therefore, we conclude that Criterion A is violated at initial times and restored at $t \overline{T}\simeq 1.57$; this may be considered as the hydrodynamization time. 

We may wonder why the system is initially away from the regime of hydrodynamics if the gradients are not very large. This is a consequence of the initial data: since the velocity is initially vanishing, the shear term is exactly zero while second order terms are non-zero. So, even for very large $\sigma$ and very small gradients, we have to wait for the system to evolve for some time until first order terms become larger than second order ones.

We continue by studying Criterion B. In 
Fig. \ref{gaussain_evolution_to_show_size_a1a2} (left, bottom) we plot the $a_1$ term and the shear term at $x=0$. Moreover, in
Fig. \ref{3Dplots_gaussian_criteria} (middle, left) we plot the $a_1$ and $a_2$ terms together with $10\%$ of the shear term in all spacetime. 
We find that the $a_1$, $a_2$ terms are smaller than $10\%$ the shear term in most of the domain,  except for small regions where the shear term vanishes.
However, computing the $L_1$ norms of the $a_1$ and  $a_2$ terms  and the shear term shows that Criterion B is satisfied at all times.

In order to employ criterion C we perform another real-time evolution with similar initial data but in frame $\{a_1,a_2\}=\{10,10\}$. In Fig. \ref{gaussain_evolution_to_show_size_a1a2} (bottom,  left) we show the result of this evolution in dashed lines. 
Recall that it is justified to use exactly the same initial data as the change of frame is exactly vanishing at $t=0$, as it satisfies the ideal hydrodynamics equations.
We also perform an evolution using the equations of ideal hydrodynamics with similar initial data, eqs. \eqref{Initial_data_gaussiana}, \eqref{Initial_data_gaussianb}. 
We plot the difference $T^{\text{visc1}}_{xx}-T^{\text{visc2}}_{xx}$ in absolute value in solid green in Fig. \ref{gaussain_evolution_to_show_size_a1a2} (bottom, left), and the difference $T^{\text{visc1}}_{xx}-T^{\text{ideal}}_{xx}$ in solid grey. We also plot these two quantities in the whole spacetime in Fig. \ref{3Dplots_gaussian_criteria} (bottom, left).
 We observe that the first difference is smaller than $10\%$ the second difference in almost all the domain, and by computing the corresponding $L_1$ norms we conclude that Criterion C is satisfied.

In the previous subsection we learned that the ratio between the $a_1$ terms in two evolutions in different frames $\{a_1,a_2\}=\{5,5\},\{10,10\}$ was about a factor of 4, according to the approximate expression \eqref{Residual0}. Even if the approximate expression is not applicable in these solutions, as they are well in the non-linear regime, we still find a difference which is roughly  a factor of 4.

Let us also comment that in Fig. \ref{3Dplots_gaussian_criteria}, top and middle plots are for the case $\{a_1,a_2\}=\{5,5\}$. For the case $\{a_1,a_2\}=\{10,10\}$, the top plot would be basically similar with Criterion A satisfied or violated in the same regions, but the middle plot would be qualitatively different: the $a_1$, $a_2$ terms are much larger, roughly by a factor of 4 and 2 respectively. This makes that Criterion B is slightly violated in this case, but it is restored at times $t \overline{T}\simeq 4.55$. This is expected as larger values of $\{a_1,a_2\}$ make the $a_1$, $a_2$ terms trivially larger. Actually, the values $\{a_1,a_2\}=\{10,10\}$ could be considered large as they are far from saturating the inequalities \eqref{hyperbolicity_conditions_BDNK}; for this reason we do not consider this as a counterexample to our main conclusion.
\newline

We proceed now with another simulation with initial data \eqref{Initial_data_gaussian} and a smaller width, $\sigma \overline{T}\simeq0.73$, so that the system now has larger gradients and it is slightly outside the regime of hydrodynamics.
We plot the result in Fig. \ref{gaussain_evolution_to_show_size_a1a2} (top, right) for the $T_{tt}$ component of the stress tensor. In Fig. \ref{gaussain_evolution_to_show_size_a1a2} (bottom, right) we plot the different elements of the $T_{xx}$ component of the stress tensor in solid lines.
This case shares an important aspect with the QGP created in heavy-ion collisions, which is that  initially both systems are marginally in the regime of hydrodynamics. Therefore,  even though the two systems are very different, one may hope that the conclusions obtained here might be relevant for the implementation of the first order viscous hydrodynamic equations to describe the QGP in experiments.

The system violates Criterion A 
up to times $t \overline{T}\simeq 12.3$, when it is restored; this is the hydrodynamization time. The initial Gaussian explodes into two waves that travel in opposite directions, as we observe in Fig. \ref{gaussain_evolution_to_show_size_a1a2} (top, right). Initially, each of these waves has large gradients, 
which are dissipated as they travel, resulting in smoother profiles; at times $t \overline{T}\simeq 12.3$ they are smooth enough so that Criterion A is satisfied (in the plots we show times only up to $t \overline{T}\simeq 6.84$ but we have run the simulations for longer). Relevant plots are Fig.  \ref{gaussain_evolution_to_show_size_a1a2} (bottom, right) and Fig. \ref{3Dplots_gaussian_criteria} (top, right).

Criterion B is slightly violated at initial times, and it is restored around times $t \overline{T}\simeq 2.1$, and satisfied from this time on. Relevant plots are Fig.  \ref{gaussain_evolution_to_show_size_a1a2} (bottom, right) and Fig. \ref{3Dplots_gaussian_criteria}(middle, right).

In order to evaluate criterion C we perform another evolution in frame $\{a_1,a_2\}=\{10,10\}$ with similar initial data and also an evolution using the ideal hydrodynamics equations. We plot the difference $T^{\text{visc1}}_{xx}-T^{\text{visc2}}_{xx}$ in absolute value in solid green in Fig. \ref{gaussain_evolution_to_show_size_a1a2} (bottom, right), and the difference $T^{\text{visc1}}_{xx}-T^{\text{ideal}}_{xx}$ in solid grey. We also plot these quantities in the whole spacetime in Fig. \ref{3Dplots_gaussian_criteria} (bottom, right).
We find that criterion C is satisfied at all times.
Thus, this is another example of a system that is marginally in the regime of hydrodynamics (actually slightly away according to our criterion A) and yet Criterion C is satisfied, which indicates the robustness of the frame independence of the first order physics. We have checked that Criterion C continues to be satisfied for values of $\sigma$ all the way down to $\sigma \overline{T}\simeq0.24$.
\newline

The main conclusions obtained from the analysis above are the following.
We observe that if Criterion A is satisfied, then criteria B and C are satisfied. 
We find that even if Criterion A is slightly violated (and B is also violated), Criterion C is still satisfied. This suggests that the frame independence of the first order physics is robust and it applies slightly beyond the hydrodynamics regime. 

\subsubsection{Changing frame in the initial data}

\begin{figure}[thbp]
	{\includegraphics[width=0.495\textwidth]{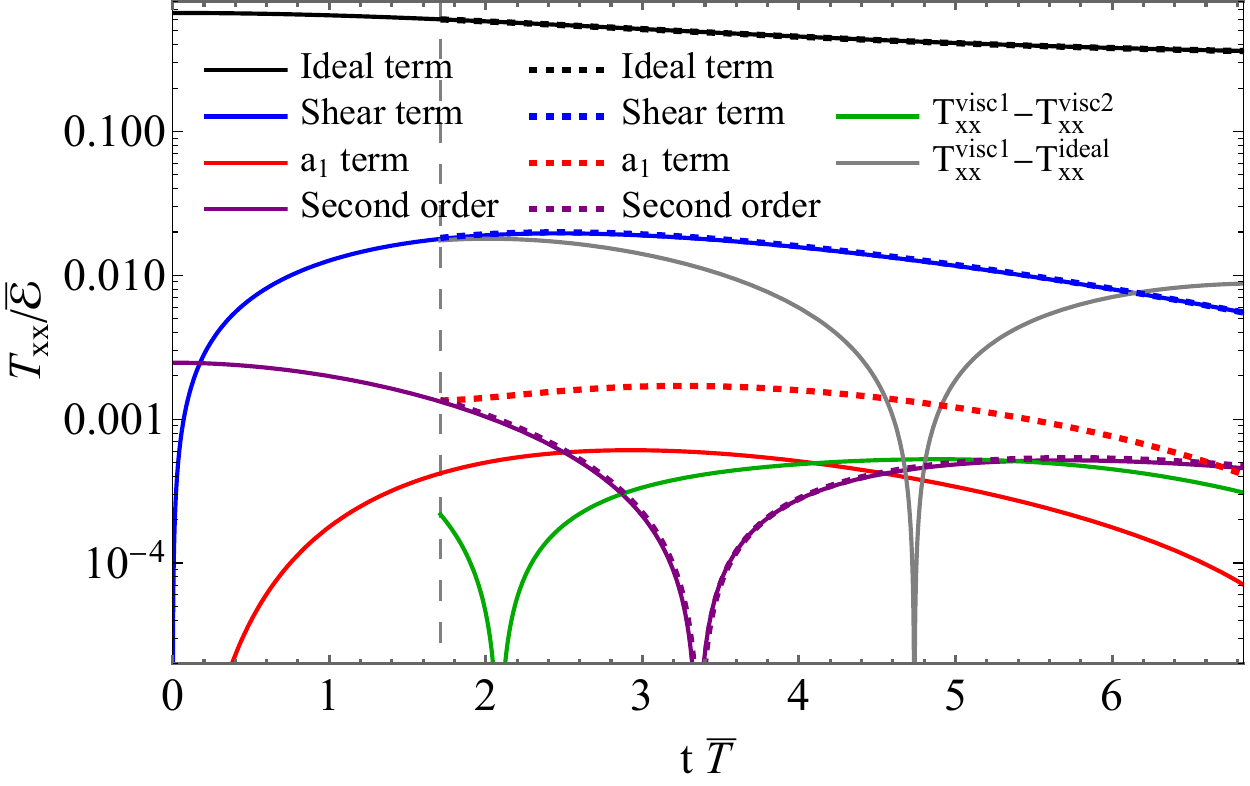}} 	{\includegraphics[width=0.495\textwidth]{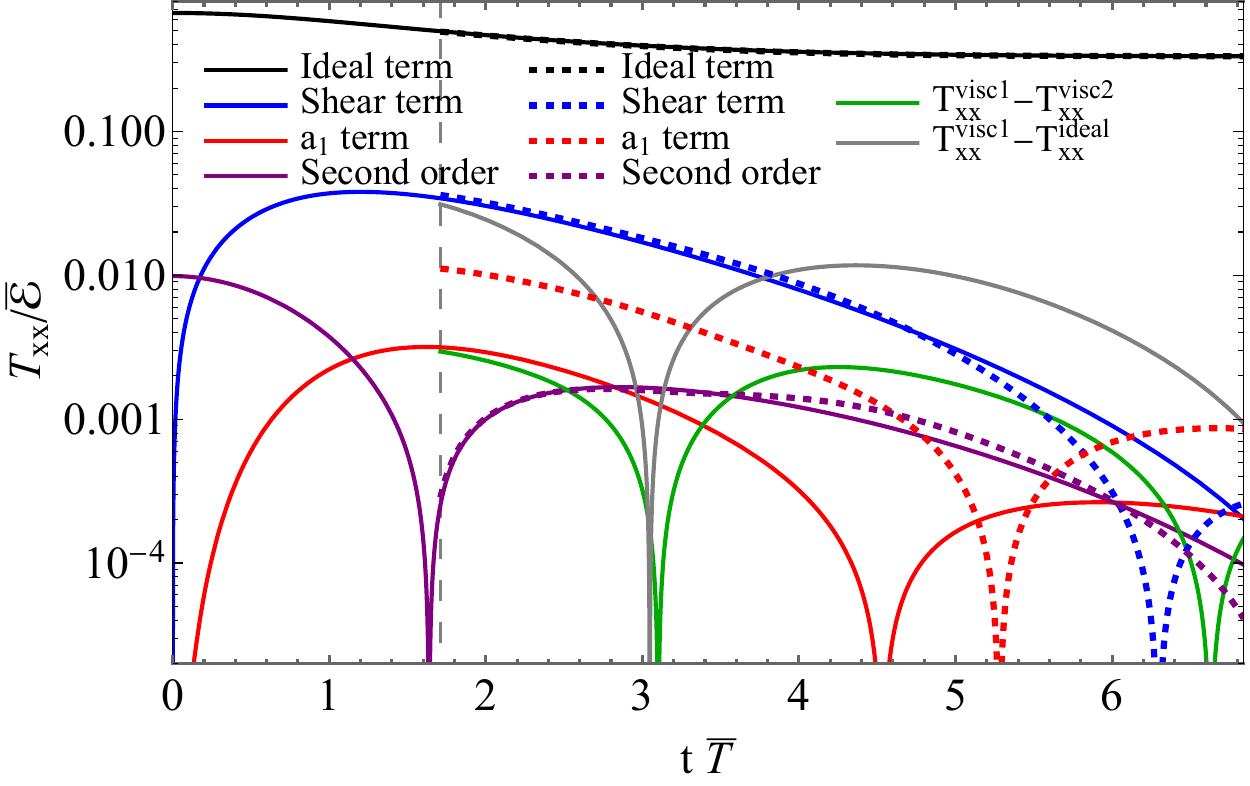}}
	\caption{ (Left) Solid lines correspond to the same evolution as in Fig. \ref{gaussain_evolution_to_show_size_a1a2} (left), which is in frame $\{a_1,a_2\}=\{5,5\}$. We include in dashed lines and same color coding the result of using data at $t \overline{T} \simeq 1.71$ (time indicated by a vertical dashed line) to perform a change of frame to $\{a_1,a_2\}=\{10,10\}$ and use it as initial data for an evolution in frame $\{a_1,a_2\}=\{10,10\}$. We show in solid green the difference $T^{\text{visc1}}_{xx}-T^{\text{visc2}}_{xx}$. Moreover, we perform an evolution of the ideal hydrodynamics equations and we show the difference $T^{\text{visc1}}_{xx}-T^{\text{ideal}}_{xx}$ in solid grey.     (Right): Similar exercise as in left for the case $\sigma \overline{T}\simeq 0.73$, to analyse a situation which is marginally in the regime of hydrodynamics. The main conclusion from these plots is that changing frame in the initial data according to the prescription of effective field theory introduces a small difference comparable to the difference due to perform evolutions in different frames analysed in Fig. 	\ref{gaussain_evolution_to_show_size_a1a2} .}
	\label{Gaussian_later_times_0}
\end{figure}

We now explore the effect of changing frame in the initial data. To do so,  we perform a similar exercise as in Fig. \ref{Comparison_ideal_evolution}; we start from the evolutions in the frame $\{a_1,a_2\}=\{5,5\}$  presented in Fig. \ref{gaussain_evolution_to_show_size_a1a2} and use the solution at a timeslice $t>0$ to obtain initial data for the frame $\{a_1,a_2\}=\{10,10\}$  via \eqref{changeframeconformal}. In Fig. \ref{Gaussian_later_times_0} (left, dashed lines) we present the same evolution as in Fig. \ref{gaussain_evolution_to_show_size_a1a2} (left) but now performed in frame $\{a_1,a_2\}=\{10,10\}$, initialized with data from the frame $\{a_1,a_2\}=\{5,5\}$  at $t\overline{T}\simeq 1.71$. 
In Fig. \ref{Gaussian_later_times_0} (right) we present the same exercise for the case $\sigma \overline{T}\simeq 0.73$.

The conclusions are similar to those in the previous exercise described in Fig.  \ref{Comparison_ideal_evolution}: The difference $T^{\text{visc1}}_{xx}-T^{\text{visc2}}_{xx}$  introduced by changing frame in the initial data, Fig.~\ref{Gaussian_later_times_0}, is comparable to the difference  introduced by the time evolution, Fig.~\ref{gaussain_evolution_to_show_size_a1a2}. Criteria A, B and C are satisfied at a similar level as for the evolutions with vanishing initial change of frame effects,  Fig \ref{gaussain_evolution_to_show_size_a1a2}, with similar times for when each criterion is violated or satisfied. Moreover, for completeness we repeated both exercise for times $t\overline{T}\simeq 3.42,5.13 $ obtaining similar conclusions.

\subsection{Shockwaves}

In this section we study the effects of using different causal frames in shockwave solutions. Shockwaves are discontinuous solutions of the ideal hydrodynamic equations, which are smoothed out once viscosity is included. 
The notion of shockwave may sound like a strong disturbance and even if viscous hydrodynamics provides a continuous profile, one could think that gradients are still large, suggesting that the solution is not in the effective field theory regime.
However, this intuition is not correct. 
Shockwave solutions of viscous hydrodynamics are in the regime of validity of effective field theory as long as their amplitude is small
 \cite{LandauBook}.\footnote{Small amplitude shockwaves admit analytical solutions \cite{LandauBook}, in which the width is inversely proportional to the amplitude of the shockwave, and thus if the amplitude is  small, the width is large and gradients are small. This picture is confirmed by numerical results, in which we can consider shockwaves of finite amplitude and check up to which amplitude Criterion A is still satisfied. In the examples below we confirm that they satisfy Criterion A up to velocities of the order of $v=0.66$.} 

We start by considering a smoothed out version of the Riemann problem with initial data given by 
\begin{subequations}
	\begin{align}
		\epsilon |_{t=0}&=\overline{\mathcal{E}}\left[\textstyle 1+\frac{\Sigma-1}{2}\left( \tanh(\frac{x-\Delta}{\tilde{\sigma}})-\tanh(\frac{x+\Delta}{\tilde{\sigma}}) \right) \right] \, \,, \label{Initial_data_gaussiana}\\
		\partial_t \epsilon|_{t=0}&= 0\,\,, \label{Initial_data_gaussianb}\\
		u_x|_{t=0}&= 0 \, \,, \label{Initial_data_gaussianc}\\
		\partial_t	u_x|_{t=0}&=0  \, \,. \label{Initial_data_gaussiand}
	\end{align}
	\label{Initial_data_Riemann}
\end{subequations}
Since we impose periodic boundary conditions, we consider initial data  that is $x\leftrightarrow-x$ symmetric. In the following plots we only show the region $x>0$.\footnote{We do not impose that the ideal equations are satisfied at $t=0$ because we will focus on the shockwave formed at later times, so for this purpose it is not relevant if the ideal equations are initially satisfied or not.}
We perform a numerical evolution of the initial data \eqref{Initial_data_Riemann} with parameters $\overline{\mathcal{E}}=1$, $\Sigma=3
$, $\Delta \overline{T} \simeq 20.5
$, $\tilde{\sigma} \overline{T}\simeq 0.684$, on a spatial domain $L\overline{T} \simeq 136.8
$ and frame $\{a_1,a_2\}=\{5,5\}$. In this setup the energy $\overline{\mathcal{E}}$ and temperature $\overline{T}$ are the ones of the low energy homogeneous region of the initial data.

\begin{figure}[thbp]
	\centering
	{\includegraphics[width=0.495\textwidth]{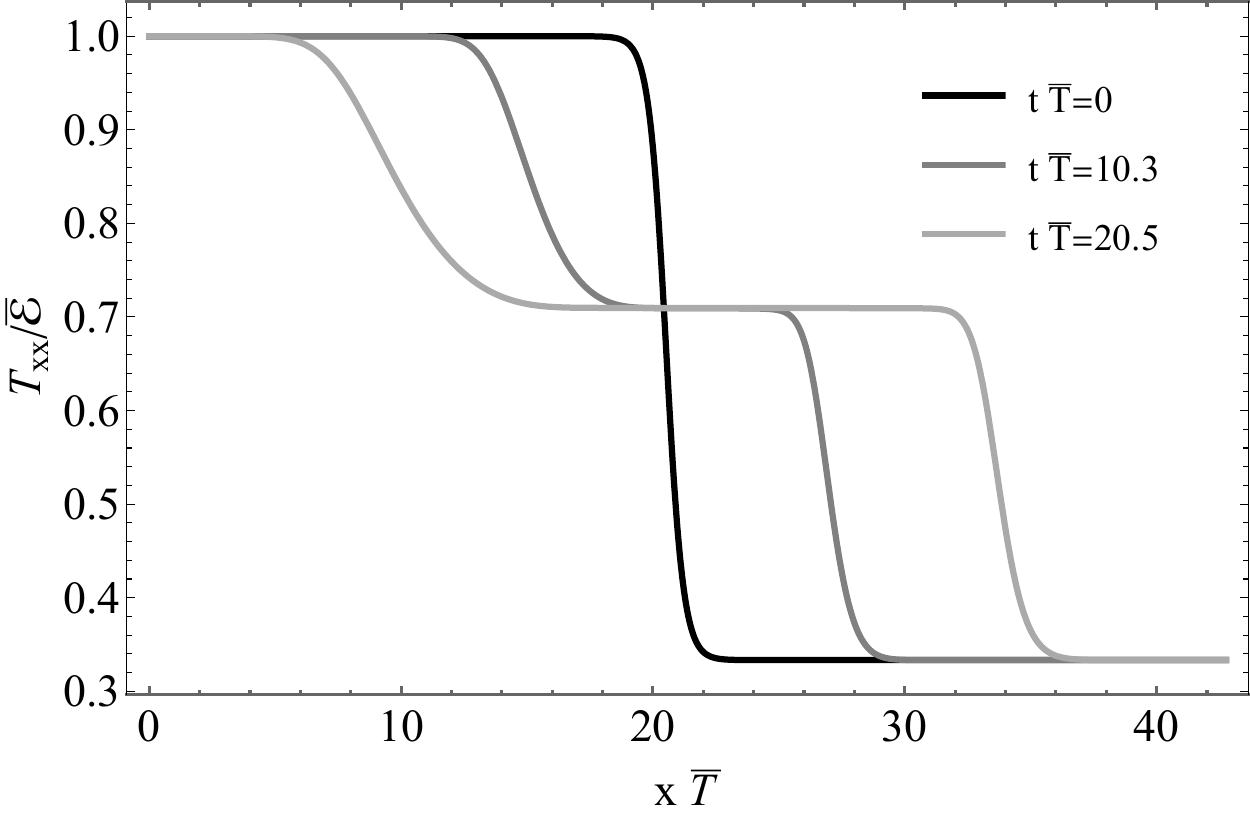}} 
	{\includegraphics[width=0.495\textwidth]{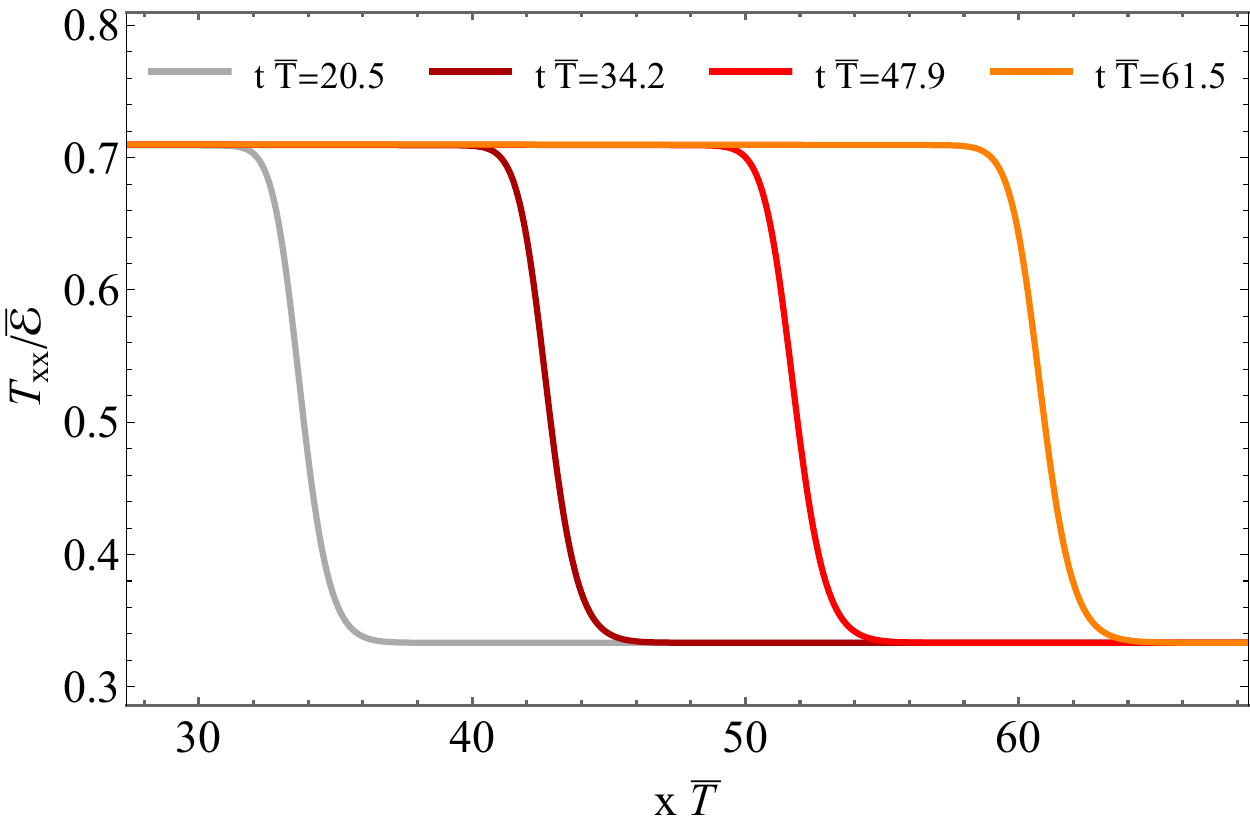}}
	{\includegraphics[width=0.495\textwidth]{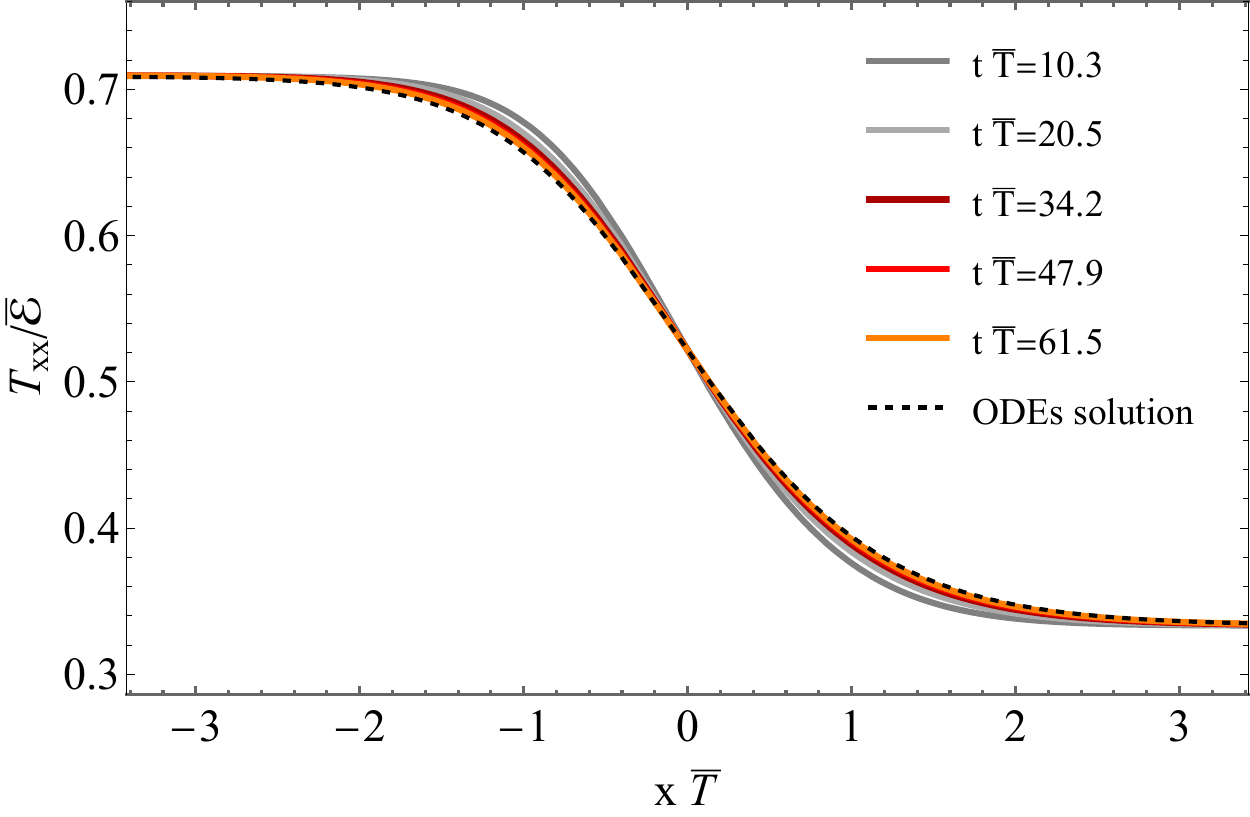}} 
	{\includegraphics[width=0.495\textwidth]{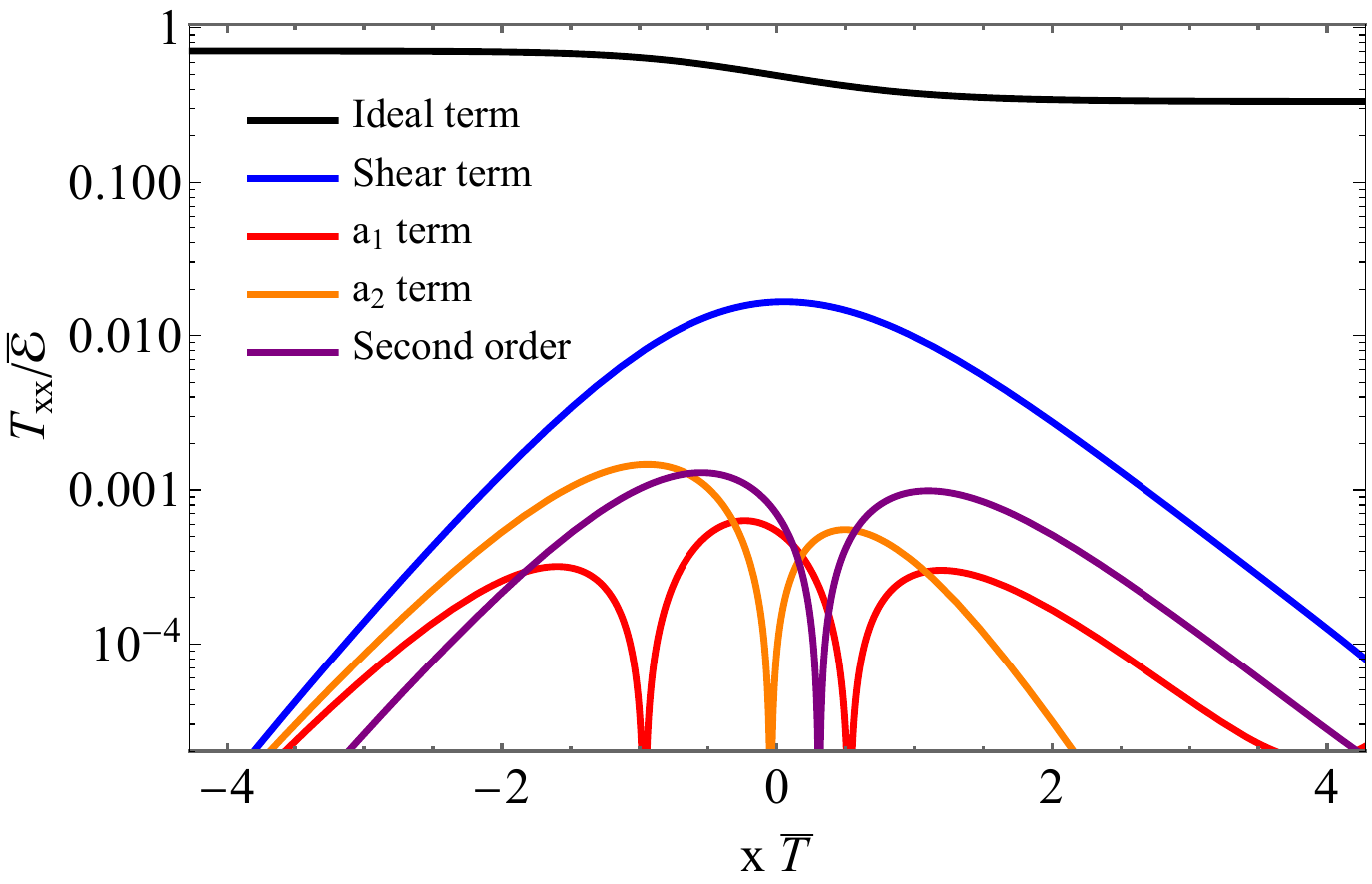}} 
	\caption{ (Top, left) Profile of the $T_{xx}$ component of the stress tensor at $t \overline{T}=0$ corresponding to the initial data \eqref{Initial_data_Riemann} and at later times when the system has evolved into a rarefaction wave propagating to the left, a shockwave propagating to the right and a homogeneous region in between. (Top, right) Profile of the $T_{xx}$ component of the stress tensor of the shockwave at times $t \overline{T}=\{20.5, 34.2, 47.9, 61.5\}$. (Bottom, left) Same profiles as in (top, right) shifted to have their midheight at $x=0$. We also include in dashed black the solution obtained by solving the ODEs. 
	(Bottom, right) Different contributions to the $T_{xx}$ component of the stress tensor of the shockwave at late times. We also include second order terms as in previous sections. 
	Criteria A and B are satisfied for this solution.}
	\label{Shockwave_profiles}
\end{figure}

The initial sharp profile evolves into a rarefaction wave that propagates to the left and a shockwave that propagates to the right, with a homogeneous region in between. See Fig. \ref{Shockwave_profiles} (top, left) for the profile of the $T_{xx}$ component of the stress tensor. This is the expected solution in relativistic hydrodynamics \cite{RezzollaBook}; studies of the Riemann problem in the case of first-order viscous hydrodynamics were done in 
\cite{Pandya:2021ief,Pandya:2022pif,Pandya:2022sff}.
In Fig. \ref{Shockwave_profiles} (top, right) we show the profile of the $T_{xx}$ component of the stress tensor in the shockwave solution at times $t \overline{T}=\{20.5, 34.2, 47.9, 61.5\}$. In Fig. \ref{Shockwave_profiles} (bottom, left) we show the same profiles shifted to have their mid height at $x=0$ so that they appear superimposed. This plot shows how the profile of the shockwave evolves in time and it asymptotes to a final configuration at late times. 
Also, the velocity of the shockwave asymptotes to a terminal velocity  $v_{\text{shockwave}}\simeq0.66$. 

We can compare the asymptotic solution that we obtain from our numerical evolutions with the solutions of the ODEs that follow from 
considering the system in the local rest frame of the shockwave and assuming stationarity  \cite{Pandya:2021ief}. Under these conditions, the conservation equation for the stress tensor $\partial_x T^{x\mu}=0$ can be easily integrated to obtain $T^{xt}=C_1$ and $T^{xx}=C_2$, where $C_1$ and $C_2$ are two real constants. Solving for $\{\epsilon',v'\}$ gives

\begin{subequations}
	\begin{align}
		 \epsilon' &={ \frac{-4\epsilon \sqrt{1-v^2} \left( (a_1-4)C_1 +  ((3a_2+4a_1-4)C_2-(4+a_2)\epsilon)v-3(a_1+2a_2)C_1v^2+3a_2(C_2+\epsilon)v^3 \right)}{ 9a_1 a_2 \eta (v-c_1)(v-c_2)(v-c_3)(v-c_4)} }\,\,, \\
		v' &=  {\frac{-(1-v^2)^{\frac{3}{2}} \left( a_2(\epsilon -3C_2)+3 (a_1+2a_2)C_1 v-3 ((4 a_1+a_2)C_2+a_2\epsilon)v^2 +9a_1 C_1 v^3 \right)}{9a_1 a_2 \eta (v-c_1)(v-c_2)(v-c_3)(v-c_4)} } \,\,, 
	\end{align}
	\label{ODEs_shockwaves}
\end{subequations}
where $c_i$, $i=1,2,3,4$ are the characteristic velocities,
\begin{equation}
		c_i=\pm \sqrt{\frac{a_1(2+a_2)\pm 2\sqrt{a_1(a_1+a_1a_2+a_2^2)}}{3 a_1 a_2}}\,.
	\label{characteristic_velocities}
\end{equation}
%
Boosting the solutions of these ODEs so that the fluid in front of the shock is at rest, we obtain a profile for the shockwave that can be compared with our asymptotic solutions,  dashed black curve in Fig. \ref{Shockwave_profiles} (bottom, left). 
This comparison confirms that indeed the late time asymptotic profile coincides with the solution of the ODEs.  The  latter are particularly interesting and useful because they are computationally cheaper (and more accurate) to obtain. For this reason, in the rest of this section we will use the solutions to the ODEs to study the effect of using different frames in shockwaves. 

The two asymptotic regions in a shockwave solution as well as its velocity are related to each other by the stress tensor conservation, which gives the Rankine-Hugoniot conditions. Viscous hydrodynamics provides a smooth transition between the two asymptotic states, but the latter and the velocity of the shockwave do not depend on using the viscous theory. 
Thus, the change of frames will only affect the specific shape of the profile but  not the asymptotic regions nor the velocity of the shock. 
In the following we study the profile of $T_{xx}$ in a shockwave solution in a Lorentz frame such that the fluid in front of the shock is at rest; in the local rest frame of the shockwave this analysis would be trivial as $T_{xx}$ is a constant.

In Fig. \ref{Shockwave_profiles} (bottom, right) we show the different contributions of the gradient expansion to the $T_{xx}$ component of the stress tensor.
By computing the norms, we find that Criterion A is only marginally satisfied since the ratio of the norms of the second order and the shear terms is $9\%$. Therefore, according to this criterion, the solution is marginally within the effective field theory regime.
%
In addition,  the ratio of the norms of the $a_1$ term ($a_2$ term) and the norm of the shear term  is $3.9\%$ ($7.3\%$). Hence, Criterion B is also satisfied.

\begin{figure}[thbp]
	\centering
	{\includegraphics[width=0.495\textwidth]{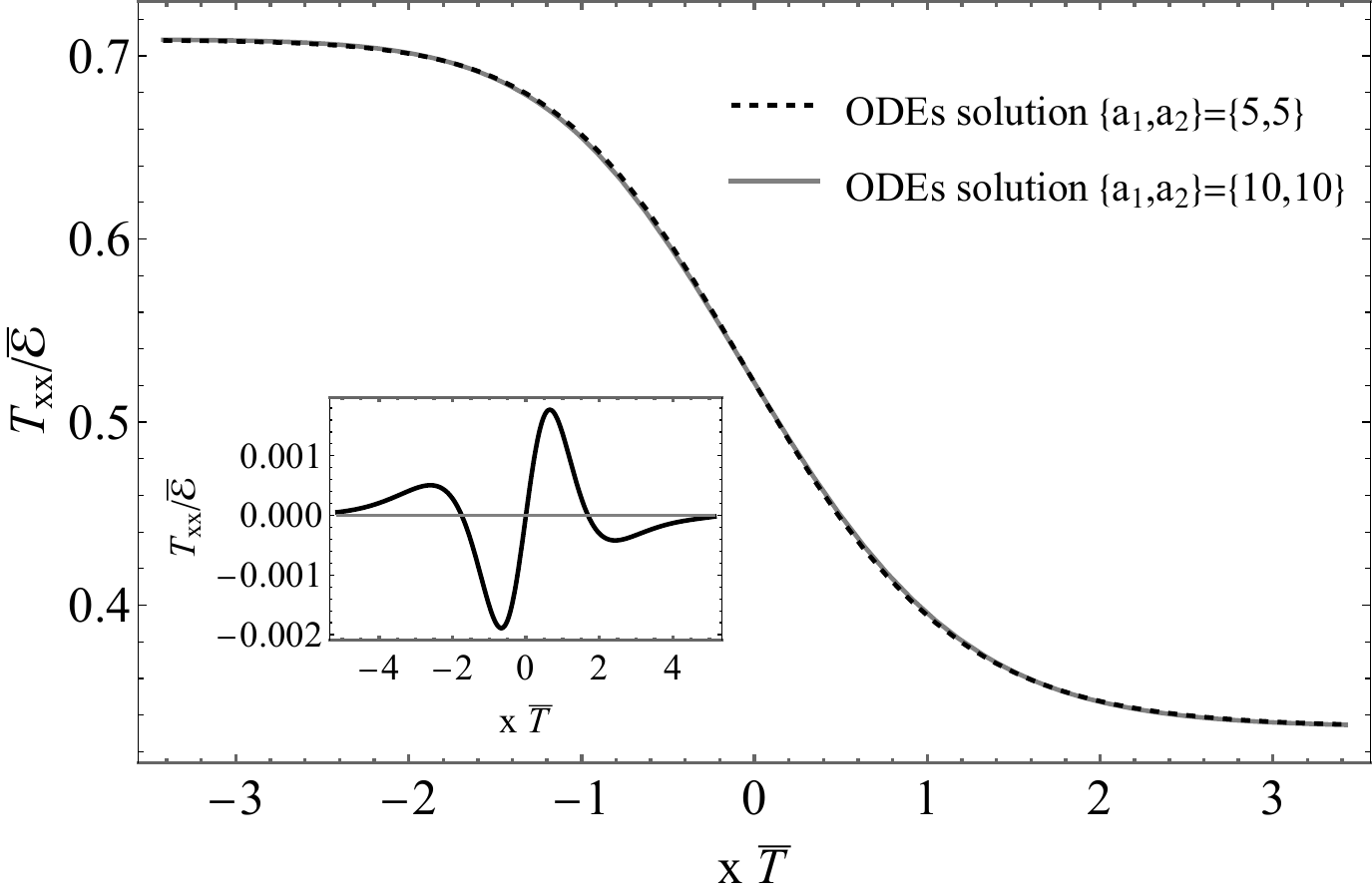}} 
	{\includegraphics[width=0.495\textwidth]{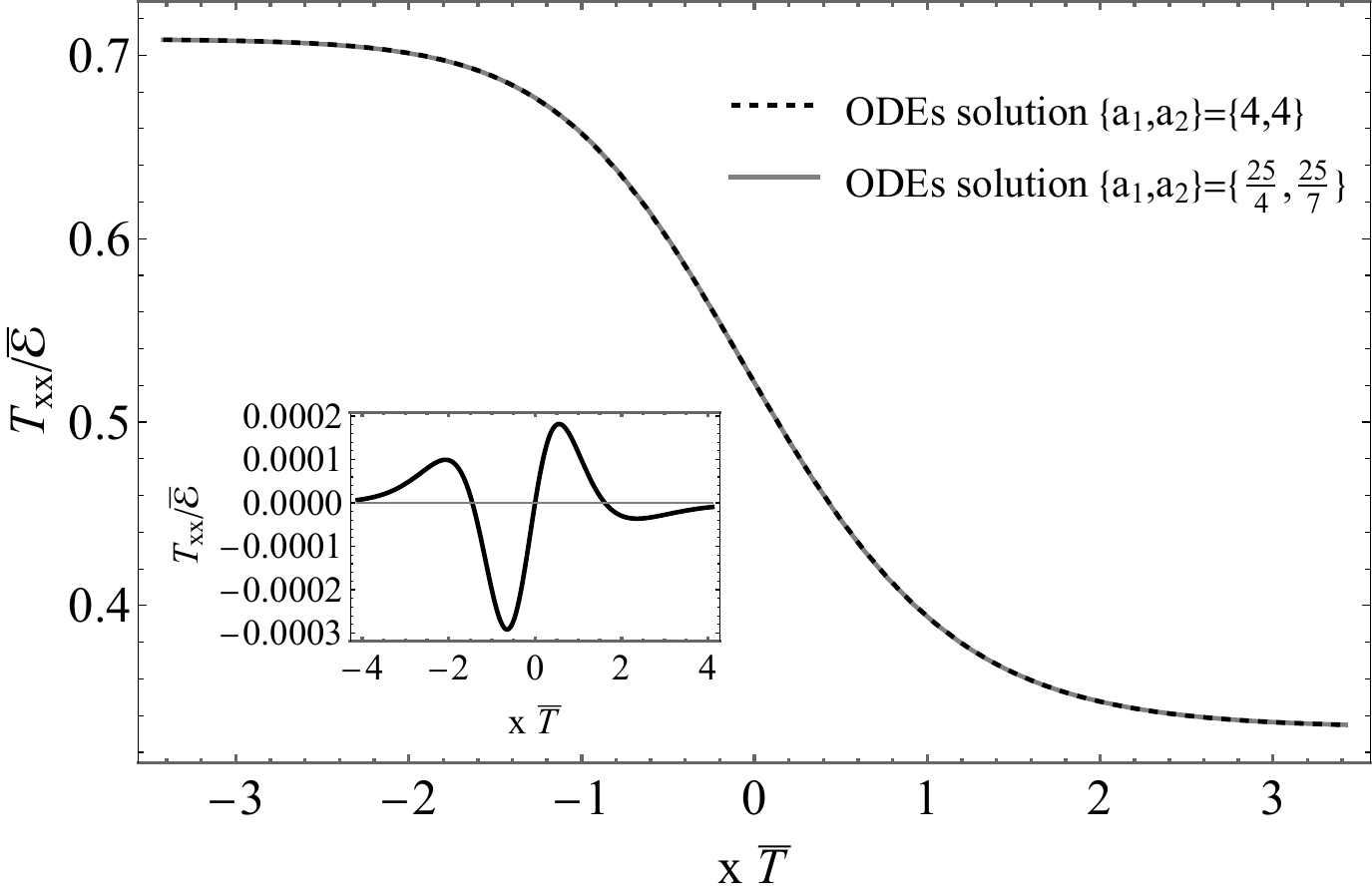}} 
	\caption{Comparison of the $T_{xx}$ component of the stress tensor of shockwave solutions with similar asymptotic regions and different frames. For these comparisons we use solutions obtained by solving the ODEs and boosting the solution such that the fluid in front of the shock is at rest. Solutions are shifted to have their midheight at $x=0$. The inset corresponds to the difference of the two solutions. (Left) The black dashed line corresponds to the same solution as in Fig. \ref{Shockwave_profiles} (bottom, left) which is in frame $\{a_1,a_2\}=\{5,5\}$, and we compare it with the solution with frame $\{a_1,a_2\}=\{10,10\}$, in solid grey. The ratio of the maximum difference to the amplitude of the shockwave is $0.48\%$.      
    (Right) Similar comparison as in (left) and with similar asymptotics, but in this case we use two sharply causal frames $\{a_1,a_2\}=\{4,4\}$ and $\{a_1,a_2\}=\{\frac{25}{4},\frac{25}{7}\}$. In this case the ratio is $0.08\%$.}
	\label{Shockwave_profiles_Txx_comparison}
\end{figure}

In order to study  Criterion C we construct a shockwave solution in the frame $\{a_1,a_2\}=\{10,10\}$ with the same asymptotics, see Fig. \ref{Shockwave_profiles_Txx_comparison} (left) in solid grey for the $T_{xx}$ component of the stress tensor, together with the solution in frame $\{a_1,a_2\}=\{5,5\}$ in dashed black.
We observe that both profiles are very similar to each other, see the inset for a detailed view of the difference. 
We note that Criterion C as previously defined is not applicable in shockwave solutions because in ideal hydrodynamics the shocks are discontinuous. In this case we use the following alternative criterion: we consider that the criterion is satisfied if the maximum difference between the two profiles divided by the difference between the asymptotic states is smaller than $5\%$.\footnote{As before, this particular choice of threshold is arbitrary.}
In our solution this ratio is $0.48\%$, and hence this criterion would be satisfied.
 
 References \cite{Freistuhler:2021lla,Pandya:2021ief,Pandya:2022pif,Pandya:2022sff} provide evidence that the first-order viscous hydrodynamics equations admit shockwave solutions as long as their velocity profile does not cross any characteristic velocity. We have verified that this is indeed the case for the chosen frames, $\{a_1,a_2\}=\{5,5\},\,\{10,10\}$, for which the characteristic velocities are always larger than the upstream velocity. We also verified that for frames with characteristic velocities that cross the velocity profile of the shockwave (in the local rest frame), the real-time evolutions crash and the ODEs do not seem to admit smooth shockwave solutions, in agreement with the results in \cite{Freistuhler:2021lla,Pandya:2021ief,Pandya:2022pif,Pandya:2022sff}.
 
Sharply causal frames, that is, frames saturating the inequalities \eqref{hyperbolicity_conditions_BDNK}, are specially interesting as they allow to capture arbitrarily large shocks while respecting causality. 
In Fig. \ref{Shockwave_profiles_Txx_comparison} (right) we repeat the same exercise as above but in this case with two frames in the sharply causal region, namely $\{a_1,a_2\}=\{4,4\}$ and $\{a_1,a_2\}=\{\frac{25}{4},\frac{25}{7}\}$. We find that for these solutions Criteria A, B and C are satisfied. In this case the ratio of the maximum difference and the differences of the asymptotic states is $0.08\%$; this ratio is smaller than above as the two frames here are  `close' to each other.  It would be interesting to have a precise notion of the separation between two frames in the space of frames (with, for example, some metric), which could be used in the definition of Criterion C. Otherwise  two solutions can trivially be very close to each other just because the parameters defining the corresponding frames are very close to each other,  even if such solutions are not in the effective field theory regime. 
We leave this for future work.

\begin{figure}[thbp]
	\centering
	{\includegraphics[width=0.49\textwidth]{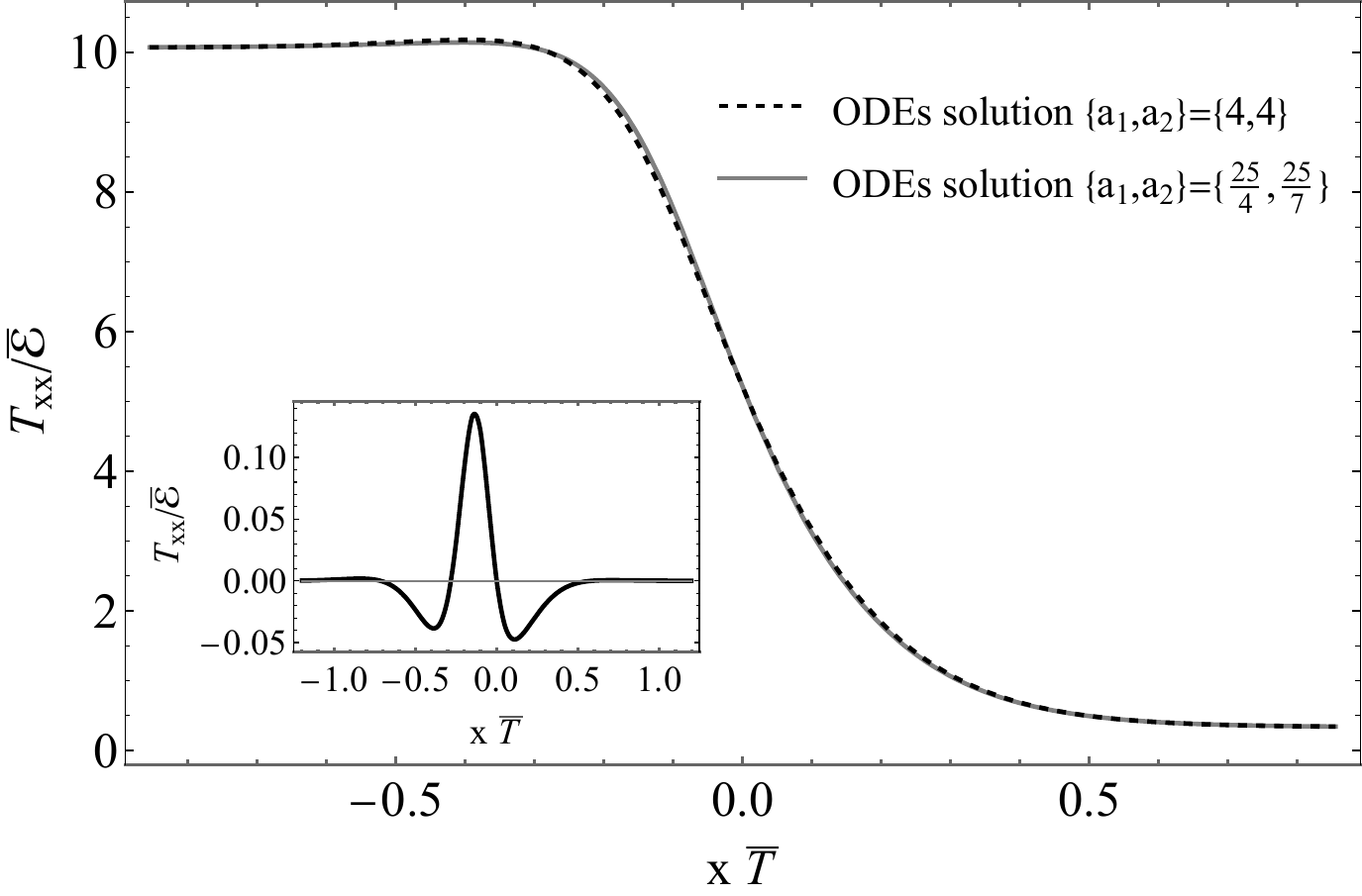}} 
	{\includegraphics[width=0.495\textwidth]{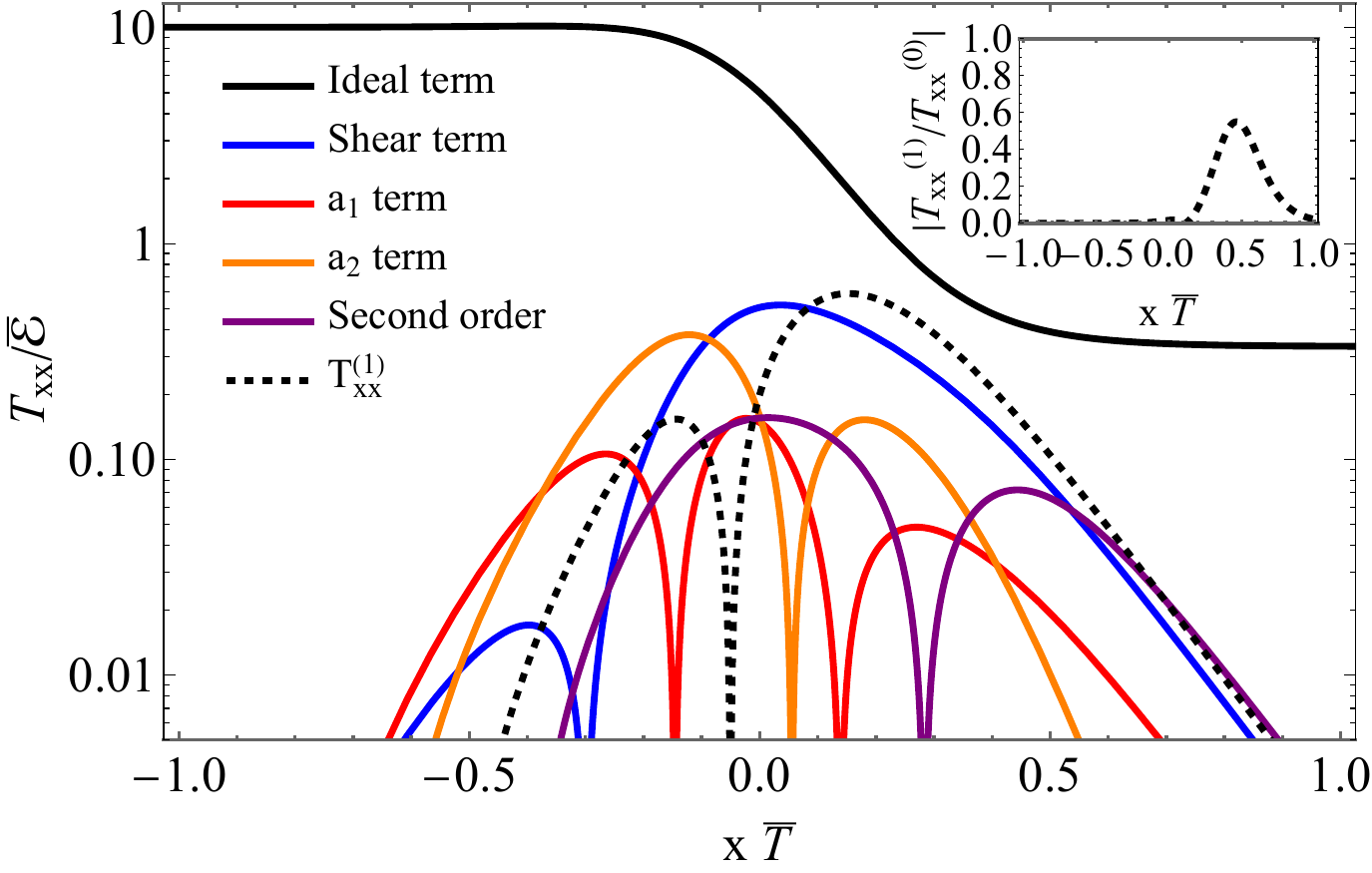}} 
	\caption{(Left) Shockwave solutions with velocity $v_{shockwave}=0.85$ in frames $\{a_1,a_2\}=\{4,4\}$, dashed black, and $\{a_1,a_2\}=\{\frac{25}{4},\frac{25}{7}\}$, solid grey. Solutions are shifted to have their midheight at $x=0$. The inset shows the difference of the two profiles.
	(Right) 
	Different contributions to the $T_{xx}$ component of the stress tensor for the solution in frame $\{a_1,a_2\}=\{4,4\}$. We include the value of the full first order term $T_{xx}^{(1)}$ in black dashed, as in this case the $a_1$, $a_2$ terms are not much smaller than the shear term.
	Criteria A and B are violated and criterion C satisfied. The ratio of the maximum deviation to the amplitude of the shockwave is $1.4\%$. 
	}
	\label{Strong_shocks}
\end{figure}

We now consider stronger shocks. We use the ODEs
to obtain shockwave solutions with velocity $v_{shockwave}=0.85$ in the frames $\{a_1,a_2\}=\{4,4\}$ and $\{a_1,a_2\}=\{\frac{25}{4},\frac{25}{7}\}$. In Fig. \ref{Strong_shocks} (left) we display  the corresponding $T_{xx}$ component of the stress tensor, after performing a boost to a Lorentz frame in which the fluid in front of the shock is at rest and has energy $\overline{\mathcal{E}}=1$. We find that criteria A and B are violated whilst Criterion C is satisfied. This indicates that the system is not in the effective field theory regime and yet the two profiles are still very close to each other, with a ratio of the maximum deviation to the amplitude of the shockwave of $1.4\%$.
We note that the $a_1$, $a_2$ terms are not much smaller than the shear term, so in Fig. \ref{Strong_shocks} (right, dashed black) we plot the full  first order term $T_{xx}^{(1)}$ in the stress tensor.

We conclude that also in shockwave solutions criterion C is robust even if the system is exiting the effective field theory regime.  
On the other hand, for sufficiently large velocities Criterion C is also violated. For example, for solutions with $v_{shockwave}=0.9999$ in frames $\{a_1,a_2\}=\{4,4\}$ and $\{a_1,a_2\}=\{\frac{25}{4},\frac{25}{7}\}$ we find that 
 the ratio of the maximum difference to the amplitude of the shockwave is $41\%$.
%

\section{Discussion}
\label{sec:Discussion}

In this paper we studied real-time numerical evolutions of the relativistic first-order viscous hydrodynamics equations for a conformal fluid. Our main goal was to explore the consequences of using different hydrodynamic frames (or field redefinitions) within the causal set of frames \eqref{hyperbolicity_conditions_BDNK} in practical situations. 
By performing evolutions with different sets of initial data and by considering specific quantitative criteria, we have made precise and provided evidence for the statement that 
the physics up to first order remains invariant under field redefinitions as long as the system is within the effective field theory regime.

Our studies are motivated by the future implementation of the relativistic first-order viscous hydrodynamics equations to model physical systems of interest such as the QGP created in heavy-ion collision experiments or astrophysical systems like neutron star mergers.
An important step towards this  goal is to understand to which extent the physics is independent of the choice of hydrodynamic frame. Providing specific quantitative criteria for this assessment is the main motivation for this paper. 
\newline

We considered the following sets of initial data. First, we considered a small amplitude sinusoidal perturbation of a homogeneous thermal state. Such a well-controlled system allows to define the criteria that we then use in more non-trivial settings. We continued by 
considering large amplitude Gaussian perturbations of a homogeneous thermal state and finally shockwave solutions.

We proposed three criteria, denoted by A, B and C, that assess different aspects 
of our solutions.
The detailed definitions of these criteria can be found in Section \ref{sec:sinusoidal}, and we summarize them here. They involve measuring the `size' of certain terms in the gradient expansion of the stress tensor in our solutions at a certain time $t$. To do so we will use the $L_1$ norm computed over the spatial domain and times $\{t-\tilde{t},t+\tilde{t}\}$, where $\tilde{t}$ is a characteristic timescale of the problem. Here we use the $L_1$ norm for convenience, but other norms are possible and should give similar results. In practical applications one may need to be quantitative about what `much smaller' means; here we use $10\%$, but this threshold value is arbitrary and in practice it may depend on the initial conditions and/or the details of the fluid under consideration. In future work we will make this precise in the case of holographic fluids, where we have control over the microscopic theory.

{\it Criterion} A. We measure the sizes of the ideal and first order terms on our solutions by evaluating the corresponding expressions in \eqref{eq:tmunu11}. In addition, for each solution, we also measure the size of the second order terms by using the second line of \eqref{constitutive0sheartensor0}.
In this way we compare the ideal, first and second order terms in the gradient expansion and we  check if each term is much smaller than the previous one.  This is the standard criterion in truncated effective field theories. Our criterion A is:  the system is in the effective field theory regime at a given time $t$ if, for all components of the stress tensor, 
 the $L_1$ norm of the shear term is much smaller than the $L_1$ norm the ideal term, and in turn  the $L_1$ norm of the second order terms is much smaller than  the $L_1$ norm of the shear term.

{\it Criterion} B. We measure the size of the $a_1$, $a_2$  terms in \eqref{eq:tmunu11} and compare them with the size of the shear term. The motivation for this criterion is as follows. 
The change of frame terms \eqref{changeframeconformal} are proportional to the lowest order equation of motion, that is, the equations of ideal hydrodynamics 	\eqref{conservationidealexplicit}. Thus, by using the equations of first-order viscous hydrodynamics these terms are, on shell, equivalent to second order terms. Then, if the system is in the effective field theory regime, the frame dependent terms in \eqref{eq:tmunu11} should be much smaller than the shear term, which is frame independent. 
With this motivation, we propose the following criterion: the physics to first order is independent of the chosen causal frame 
at a given time $t$ if, for the spatial components of the stress tensor, the $L_1$ norms of the $a_1$ and $a_2$ terms are much smaller than the $L_1$ norm of the shear term. 

{\it Criterion} C. We consider two evolutions in different (and sufficiently separated, at least in parameter space) causal frames, and a third evolution using ideal hydrodynamics; we compare the difference between the two viscous evolutions with the difference between one viscous evolution and the ideal evolution. 
In this way we compare the effect of changing frame with the effect of including first order viscous terms in the hydrodynamics description.
Our specific criterion is:  the physics to first order is independent on the chosen hydrodynamic frame
at a given time $t$ if  the $L_1$ norm of the difference between the stress tensors obtained from two evolutions performed in different causal frames,  $T^{\text{visc1}}_{\mu\nu}-T^{\text{visc2}}_{\mu\nu}$, is much smaller than the $L_1$ norm of the difference between the stress tensors obtained with ideal hydrodynamics and one of the viscous evolutions,   $T^{\text{visc}}_{\mu\nu}-T^{\text{ideal}}_{\mu\nu}$.
\newline

The main conclusion of this paper is that in all solutions that we have studied, if Criterion A is satisfied, then criteria B and C are are also satisfied; this is a quantitative and precise version of the statement that if the system is in the effective field theory regime, then the physics to first order is independent of the arbitrarily chosen frame.

Thus, we confirm that it is consistent to truncate the hydrodynamic gradient expansion to include only the viscous terms.  
This is a non-trivial result since each of these three criteria assesses a different aspect of what it means for a solution to be `in the effective field theory regime'. One might have thought that a given solution is indeed well described by viscous hydrodynamics only of the three criteria are simultaneously satisfied. We have shown that this is not the case and Criterion A is sufficient to imply to other two.

From the main conclusion and from a practical perspective, 
one could think of the terms $a_1$, $a_2$ as mere regulators of the equations (that make them well posed) and work as if we were in the usual Landau frame, as the choice of frame difference is a higher order effect. However, this is only applicable  when the system is well within the effective field theory regime, and one must be careful with this statement if it is not clear that the system is  in that regime.

One possibility that could happen is that Criterion B is satisfied and the terms $a_1$, $a_2$ are small, but the effects of choosing different frames accumulate over sufficiently long times and they eventually become large. That is, can Criterion B be satisfied and C violated? In our solutions we have always observed that if Criterion B is satisfied, then Criterion C is also satisfied.  

Regarding Criterion B, even if in most cases we have found that when Criterion A is satisfied then Criterion B is also satisfied, we have found some counterexamples; for example see discussion after eq. \eqref{Linear_solution_dispersion_relation_BDNK_k4}. However, these counterexamples only happen when the values of $\{a_1,a_2\}$ are  unreasonable large.
If we restrict the values of $\{a_1,a_2\}$ to be close to saturating the inequalities \eqref{hyperbolicity_conditions_BDNK}, then these counterexamples can be ruled out. Moreover, in these solutions we found that Criterion C is satisfied, even if the $a_1$, $a_2$ terms are not small compared to shear term. 
\newline

In the actual description of the QGP created in heavy-ion collision experiments, when the hydrodynamic codes are initialized, there are situations in which gradients are large. This motivates us to consider solutions in which the system is marginally in the effective field theory regime, that is, Criterion A is almost violated or slightly violated.\footnote{See Figs. \ref{Diference_Ttt_of_two_runs_a1a2_5_a1a2_10}(right), \ref{Comparison_ideal_evolution}(right), \ref{Changing_initial_data}(right), \ref{gaussain_evolution_to_show_size_a1a2}(right) and \ref{Strong_shocks}(top).} We want to test if the physics to first order is independent of the arbitrarily chosen frame even under these circumstances. 
A generic conclusion from our studies is that even if the system is not well within the regime of hydrodynamics, that is, Criterion A is slightly violated, 
the change of frame still does not significantly affect the viscous evolutions compared to the ideal evolution, that is, Criterion C is still satisfied; we also observe that Criterion B is also slightly violated.
Therefore, we conclude that the physics to first order is robust against changes of frame according to Criterion C, even in situations where the system is on the boundary of the regime of applicability of  effective field theory. 

We emphasize that it is convenient in practice to choose hydrodynamic frames that saturate the inequalities \eqref{hyperbolicity_conditions_BDNK}, because it allows to evolve arbitrarily strong shockwaves. Moreover, this choice also guarantees that the field redefinition terms are as small as possible. 
\newline

In the hydrodynamic studies of the QGP, the initial data for the stress tensor is obtained by some prehydrodynamic procedure. If we want to implement the relativistic Navier-Stokes equations in this context, one needs to address the question of constructing initial data in a certain causal frame.
One way to proceed is to diagonalize a given stress tensor and find the data for the evolution variables in the Landau frame; then, use the prescription of effective field theory \eqref{changeframeconformal} to find the corresponding quantities in the causal frame. Therefore, one needs to address the effects of changing frames in the initial data.

To study the effects of changing frames in the initial data, in this paper we have used solutions obtained in a certain frame for a time interval to construct initial data at some $t>0$ in a new frame using \eqref{changeframeconformal}, see  Figs. \ref{Comparison_ideal_evolution} and \ref{gaussain_evolution_to_show_size_a1a2}. The lesson that we have learned from this exercise is that the change of frame in the initial data has a small effect in the subsequent evolution. 
That is, the difference in the two evolutions in different frames due to the change of frame in the initial data is comparable to the difference introduced upon time evolution in different frames when using similar initial data that satisfies the ideal equations. 
In these constructions, the equations of ideal hydrodynamics are nearly satisfied, which implies that the effects of changing frames in the initial data are necessarily small.

We also studied a situation in which the the ideal hydrodynamics equations are far from being satisfied initially, see Fig. 	\ref{Changing_initial_data}; this situation arises, for example, when the time derivatives in the initial data are set to zero. 
We found that this introduces large differences in the initial stress tensor under frame changes and this difference is sustained upon time evolution. To be precise, this difference is large compared to the difference generated solely upon time evolution with initial data that satisfies the ideal equations. 
This can be simply understood because having large error terms in the ideal equations implies that initial data in different frames is significantly different; then Cauchy stability implies that the subsequent evolutions in different frames will also differ significantly.
In practical applications, this situation should not arise since one wants to initialize a hydrodynamic code to describe a system that is in the regime of hydrodynamics, and hence the ideal equations should nearly satisfied.
\newline

We now comment on future directions. 
We have limited our studies to conformal fluids as a natural starting point to perform a detailed analysis of the consequences of using different causal frames. However, physical systems of interest like neutron star mergers or the QGP created in heavy-ion collisions are not conformal, and it would be interesting to extend our analysis to the non-conformal case.
Dynamical evolutions of the relativistic first-order viscous hydrodynamics equations in the non-conformal case have been studied in \cite{Pandya:2022sff}.

An interesting direction for studies of relativistic hydrodynamics is to perform comparisons with microscopic real-time evolutions in quantum field theories. Such comparisons allow to explore quantitatively under which circumstances the effective field theory provides a good description of the microscopic theory. In particular, one would be able to estimate the acceptable threshold values for the different criteria that we have proposed. 
However, obtaining real time evolutions in interacting quantum field theories is in general a very difficult task. In this context, holography turns out to be a very useful tool. 
We can think of a specific holographic theory as a toy model, specific to certain quantum field field theories that are very different from QCD in many aspects, but useful nevertheless to obtain information about the real time dynamics 
in a controlled way; such a computation is not feasible in QCD.  
The qualitative conclusions obtained from this analysis, even if obtained with holography, might apply more generally, and in particular to finite temperature QCD. 
We initiated this line of research in \cite{Bantilan:2022ech} and we will extend these studies in the future.  This paper is part of a long term program that aims to contribute to the implementation of the relativistic Navier-Stokes equations in studies of physical systems of interest such as the QGP created in heavy-ion collisions or astrophysical systems.

\section*{Acknowledgments}
We thank Jorge Casalderrey-Solana, David Mateos and Alexandre Serantes for very useful discussions.
Y.B. is supported by a Maria Zambrano postdoctoral fellowship from the University of Barcelona, by the “Unit of Excellence MdM 2020-2023” award to the Institute of Cosmos Sciences (CEX2019-000918-M), and by
grants PID2019-105614GB-C22, 2021-SGR-872 and PID2022-136224NB-C22, funded by MCIN/AEI/ 10.13039/501100011033/FEDER, UE. 
P.F. was supported by a Royal Society University Research Fellowship (Grant No. UF140319 and URF\textbackslash R\textbackslash 201026) and is currently supported by the STFC Consolidated Grant ST/X000931/1. 

\begin{appendix}

\section{Convergence tests}
\label{sec:AppendixA}

\begin{figure}[thbp]
	\centering
	{\includegraphics[width=0.495\textwidth]{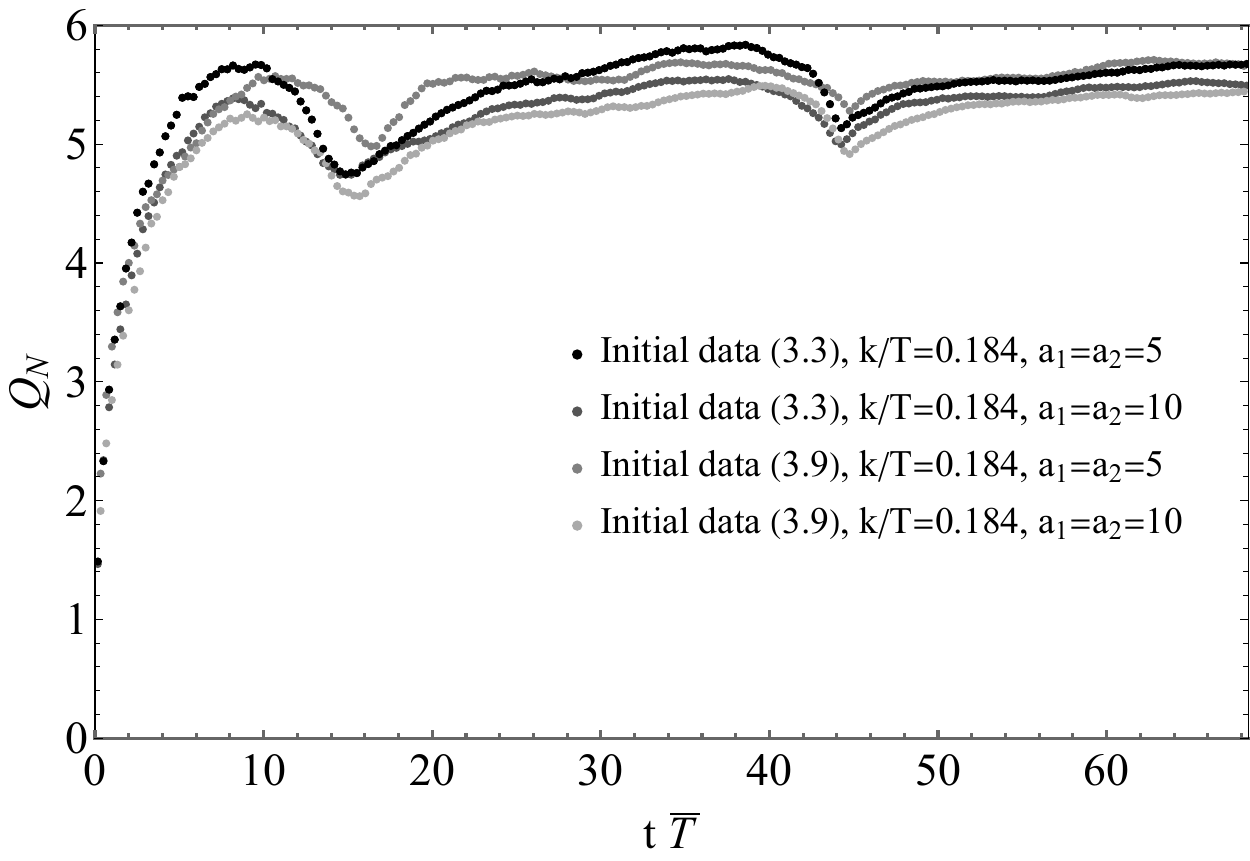}}
	{\includegraphics[width=0.495\textwidth]{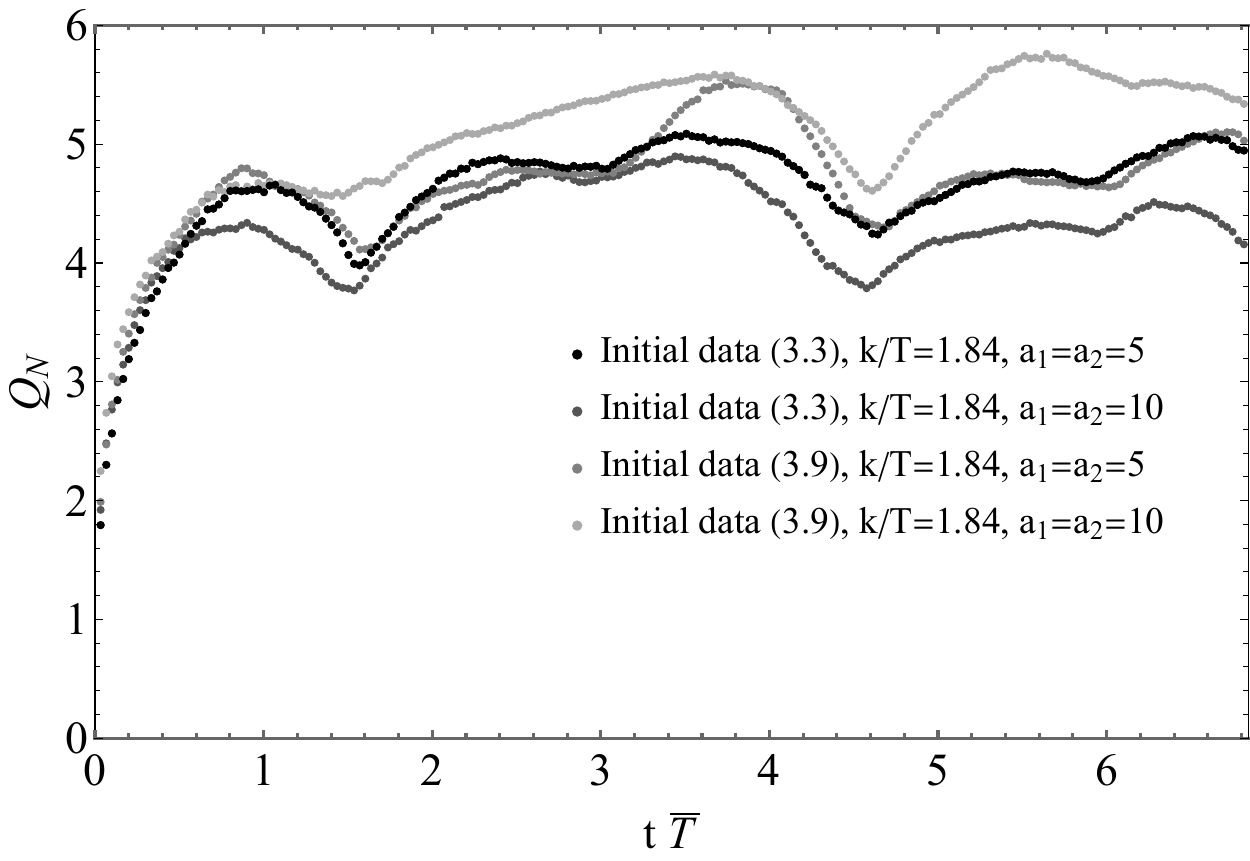}} 
	\caption{Convergence test corresponding to the small amplitude sinusoidal perturbation runs presented in Figs. \ref{Diference_Ttt_of_two_runs_a1a2_5_a1a2_10} and  \ref{Changing_initial_data}, with initial data \eqref{Initial_data_sinusoidal} and \eqref{Initial_data_sinusoidal2_chage_frame} respectively. (Left) Runs with $k/\overline{T}\simeq 0.184$, (right) runs with $k/\overline{T}\simeq 1.84$. 
	The convergence rate approaches 4, or even goes above, as one would expect given our differencing and time integration schemes.}
	\label{Convergence_test_sinusoidals_1}
\end{figure}

\begin{figure}[thbp]
	\centering
	{\includegraphics[width=0.495\textwidth]{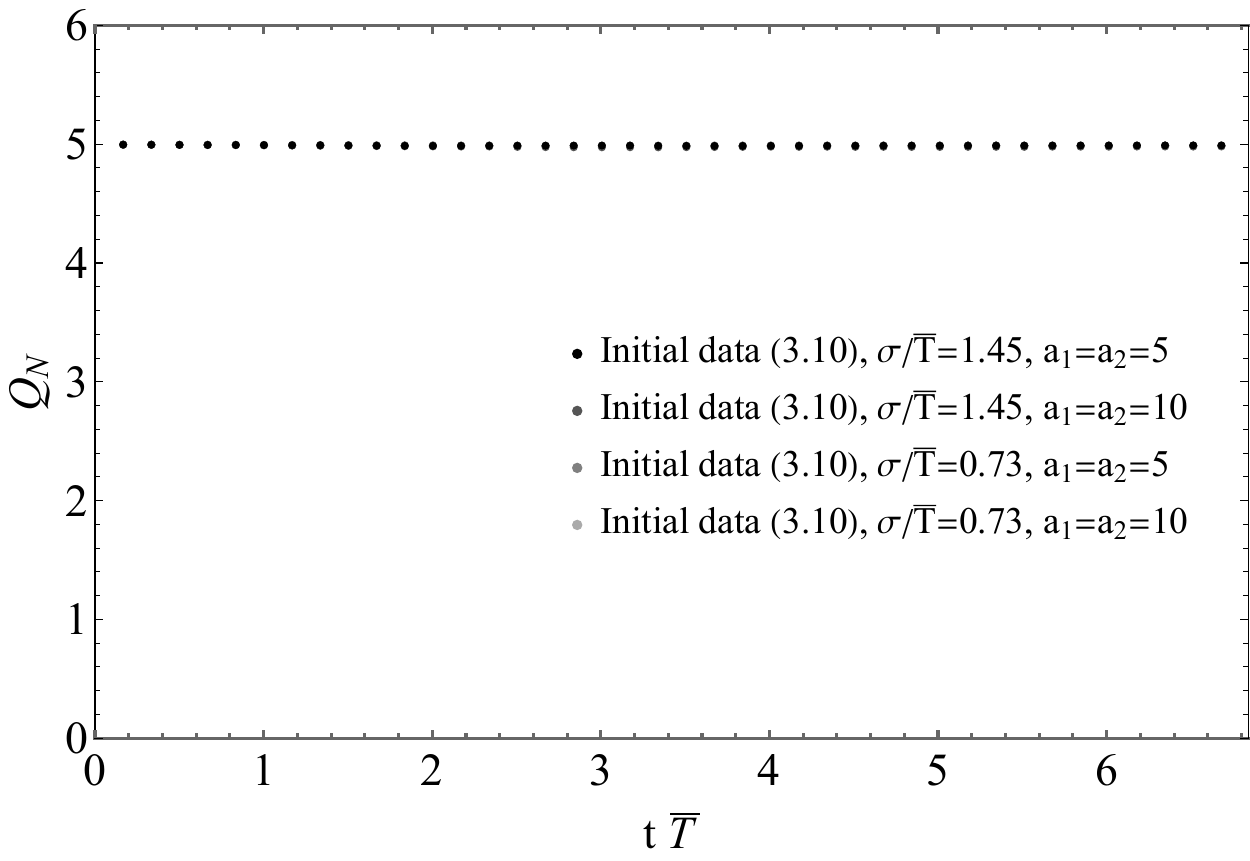}}
	{\includegraphics[width=0.495\textwidth]{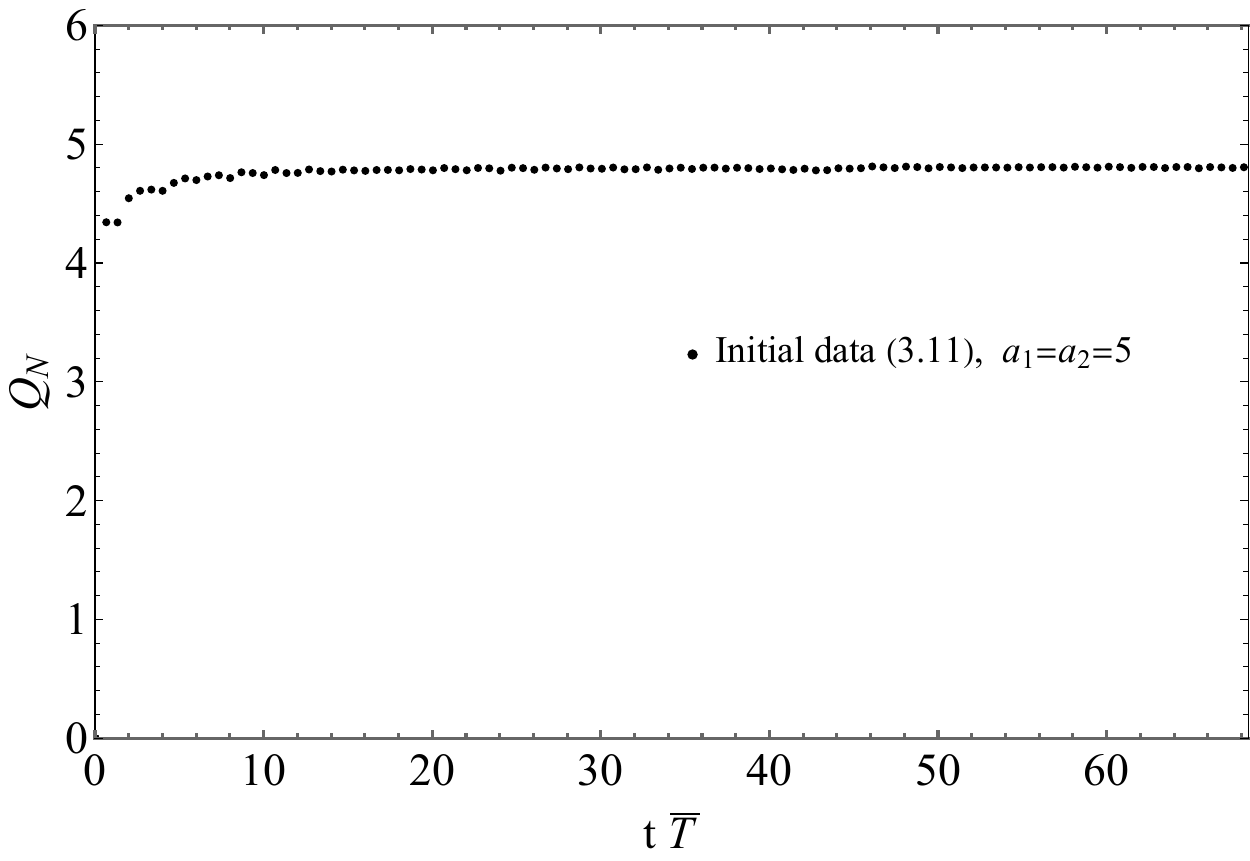}}
	\caption{(Left) Convergence test corresponding to the gaussian perturbation runs presented in Fig. \ref{gaussain_evolution_to_show_size_a1a2} with initial data \eqref{Initial_data_gaussian}. The convergence rate for all the runs is close to 5.
    (Right) Convergence test for the Riemann problem presented in Fig. \ref{Shockwave_profiles} with initial data \eqref{Initial_data_Riemann}.}
	\label{Convergence_test_gaussian_1}
\end{figure}

Our code uses finite difference stencils of order 6 for the spatial derivatives and a fifth order explicit  Runge-Kutta-Nyström Generalized (RKNG) time integration method for the equations written in second order (both in time and space) form. See \cite{Bantilan:2022ech} for more details. In \cite{Bantilan:2022ech} we presented a convergence test showing a convergence factor of 4 or even higher. In this paper we use a 1+1 version of that code, and we
present convergence tests for some representative runs of this paper.
For the runs presented in this paper we typically use $N_x=2048$ points along the spatial dimension and Courant factor 0.1. 
For a set of representative runs, we perform a convergence test for which we use three different resolutions $N_x/4$,$N_x/2$ and $N_x$. In Fig. \ref{Convergence_test_sinusoidals_1} we show convergence tests for the runs presented in Figs. \ref{Diference_Ttt_of_two_runs_a1a2_5_a1a2_10} and \ref{Changing_initial_data}, and in Fig. \ref{Convergence_test_gaussian_1}  convergence tests for the runs presented in Figs. \ref{gaussain_evolution_to_show_size_a1a2} and  \ref{Shockwave_profiles}. In some cases there is an initial transient, but after it in all cases the convergence factor is between 4 and 5, consistent with our method.

\end{appendix}

\bibliographystyle{JHEP}
\bibliography{refs_BDNK}

\end{document}